
\magnification=1200
\hoffset=-.25cm
\voffset=.0cm
\baselineskip=.55cm plus .55mm minus .55mm

%
%
\def\ref#1{\lbrack#1\rbrack}
%
%
%
%
\input amssym.def
\input amssym.tex
%
%
\font\teneusm=eusm10
\font\seveneusm=eusm7
\font\fiveeusm=eusm5
%
%
\font\sf=cmss10
\font\ssf=cmss8
%
%
\font\cps=cmcsc10
%
%
\newfam\eusmfam
\textfont\eusmfam=\teneusm
\scriptfont\eusmfam=\seveneusm
\scriptscriptfont\eusmfam=\fiveeusm
\def\sans#1{\hbox{\sf #1}}
\def\ssans#1{\hbox{\ssf #1}}

\def\proclaim #1. #2\par{\medbreak{\cps #1.\enspace}{\it #2}\par\medbreak}
%
%
%
%

\def\dtr{\hskip 1pt{\rm det}\hskip 1pt}
\def\real{\hskip 1pt{\rm Re}\hskip 1pt}

\def\ker{\hskip 1pt{\rm ker}\hskip 1pt}
\def\coker{\hskip 1pt{\rm coker}\hskip 1pt}
\def\ran{\hskip 1pt{\rm ran}\hskip 1pt}
\def\dom{\hskip 1pt{\rm dom}\hskip 1pt}

\def\tr{\hskip 1pt{\rm tr}\hskip 1pt}

\def\ad{\hskip 1pt{\rm ad}\hskip 1pt}
\def\Ad{\hskip 1pt{\rm Ad}\hskip 1pt}
\def\bit{\hskip 1pt{\rm bit}\hskip 1pt}
\def\Lie{\hskip 1pt{\rm Lie}\hskip 1pt}

\def\Gau{\hskip 1pt{\rm Gau}\hskip 1pt}

\def\Hol{\hskip 1pt{\rm Hol}\hskip 1pt}
\def\Herm{\hskip 1pt{\rm Herm}\hskip 1pt}
\def\Don{\hskip 1pt{\rm Don}\hskip 1pt}
\def\Met{\hskip 1pt{\rm Met}\hskip 1pt}

\def\Aff{\hskip 1pt{\rm Aff}\hskip 1pt}
\def\Conn{\hskip 1pt{\rm Conn}\hskip 1pt}
\def\CF{\hskip 1pt{\rm CF}\hskip 1pt}
\def\ECF{\hskip 1pt{\rm ECF}\hskip 1pt}
\def\HCF{\hskip 1pt{\rm HCF}\hskip 1pt}
\def\HECF{\hskip 1pt{\rm HECF}\hskip 1pt}
\def\SHol{\hskip 1pt{\rm SHol}\hskip 1pt}

\def\DS{{\rm DS}}

%
%
%
%

\hrule\vskip.4cm
\hbox to 16.5 truecm{July 1995   \hfil DFUB 95--17}
\hbox to 16.5 truecm{Version 1  \hfil hep-th/9508054}
\vskip.4cm\hrule\vskip1.5cm
\centerline{\bf DRINFELD--SOKOLOV GRAVITY}
\vskip1cm
\centerline{by}
\vskip.5cm
\centerline{\bf Roberto Zucchini}
\centerline{\it Dipartimento di Fisica, Universit\`a degli Studi di Bologna}
\centerline{\it V. Irnerio 46, I-40126 Bologna, Italy}
\vskip1.5cm\hrule\vskip.9cm
\centerline{\bf Abstract}
\vskip.3cm
\par\noindent
A lagrangian euclidean model of Drinfeld--Sokolov (DS) reduction leading
to general $W$--algebras on a Riemann surface of any genus is presented.
The background geometry is given by the DS principal bundle $K$ associated
to a complex Lie group $G$ and an $SL(2,\Bbb C)$ subgroup $S$. The basic
fields are a hermitian fiber metric $H$ of $K$ and a $(0,1)$ Koszul gauge
field $A^*$ of $K$ valued in a certain negative graded subalgebra $\goth x$
of $\goth g$ related to $\goth s$. The action governing the $H$ and $A^*$
dynamics is the effective action of a DS field theory in the geometric
background specified by $H$ and $A^*$. Quantization of $H$ and $A^*$
implements on one hand the DS reduction and on the other defines a
novel model of $2d$ gravity, DS gravity. The gauge fixing of the DS gauge
symmetry yields an integration on a moduli space of DS gauge equivalence
classes of $A^*$ configurations, the DS moduli space. The model has a
residual gauge symmetry associated to the DS gauge transformations leaving
a given field $A^*$ invariant. This is the DS counterpart of conformal
symmetry. Conformal invariance and certain non perturbative features of the
model are discussed in detail.
\vfill\eject
\vskip.4cm
\item{1.}{\bf Introduction}
\vskip.4cm
\par
In recent years, a considerable amount of work has been devoted to the
study of $W$--algebras \ref{1}. The interest in $W$--algebras stems mainly
from the fact that they are non linear extensions of the Virasoro algebra
appearing as symmetry algebras in certain critical two dimensional
statistical systems as well as in $W$ strings and $W$--gravity
models. The latter in turn are of considerable interest in themselves
as generalizations of ordinary string and gravity models with non standard
values of the critical dimension \ref{2--5}.

The construction of $W$--algebras can be carried out both in a hamiltonian
and in a lagrangian framework. In the former approach \ref{6--12}, based
on the methods of hamiltonian reduction, the currents of a
Wess--Zumino--Novikov--Witten phase space with the standard Kac--Moody
Poisson structure and Virasoro action are subject to a set of conformally
invariant first class constraints corresponding to a certain nilpotent
subalgebra of the relevant symmetry Lie algebra. Upon gauge fixing, the
reduced phase space exhibits a non linear Poisson structure and a Virasoro
action, realizing the $W$--algebra. Quantization is carried out in a
Becchi--Rouet--Stora framework. In the latter approach \ref{10,13},
based on lagrangian local field theory, a certain nilpotent subgroup of the
relevant symmetry group of a Wess--Zumino--Novikov--Witten field theory is
gauged yielding a conformally invariant gauge theory. Quantizing and gauge
fixing \`a la Fadeev--Popov, one gets a quantum field theory whose gauge
invariant operators generate the $W$--algebra. Underlying both approaches is
the existence of an $\goth s\goth l(2)$ subalgebra of the symmetry Lie algebra
defining a halfinteger gradation of the latter\ref{10--12}.

It seems appropriate to test the basic assumptions of such formulations in
new ways and explore the consequences of the results so obtained.
A possible approach in this direction consists in seeing whether
$W$--algebras can be constructed on a topological non trivial world
sheet. In the hamiltonian framework, this has been done in refs. \ref{14}
for Drinfeld--Sokolov lowest weight reductions \ref{15}, where the conformal
properties are manifest. It has not been attempted yet in the
lagrangian framework. This is precisely the aim of this paper.

There are at least two reasons why this is an interesting problem.
First, this is integral part of the programme of constructing the
Polyakov measure \ref{16--19} for $W$--strings and $W$--gravity.
Second, the gauge fixing of the Drinfeld--Sokolov gauge symmetry
leaves in principle a residual integration on the space of
Drinfeld--Sokolov gauge orbits. The existence of such Drinfeld--Sokolov
moduli space is a non trivial feature of Drinfeld--Sokolov lowest weight
reduction which is manifest only in the lagrangian approach.

It is important to appreciate the salient features of the construction
of the present paper by comparing it with earlier lagrangian formulations.
The basic elements of the construction of ref. \ref{10} are a split
simple real Lie group $G$ and an $SL(2,\Bbb R)$ subgroup $S$ of
$G$. To these data, one can associate canonically a halfinteger grading
of  $\goth g$ and a certain negative graded subalgebra $\goth x$
of $\goth g$. One considers then a modified minkowskian $G$
Wess--Zumino--Novikov--Witten model and gauges the subgroup
$X=\exp\goth x$ of $G$. The classical action is
$$\eqalignno{
I^{\rm M}(H,A_-,A_+)
=&~KS^{\rm M}_{\rm WZNW}(H)
+{K\over\pi}\int d^2x\tr\Big[(\partial_+HH^{-1}-t_{+1})A_-&\cr
&\phantom{S^{\rm M}_{\rm WZNW}(H)}
+(H^{-1}\partial_-H-t_{-1})A_+-A_-\Ad HA_+\Big]\vphantom{{1\over\pi}\int},
&(1.1)\cr}
$$
where $\tr$ is the Cartan--Killing form of $\goth g$, the $t_d$ are the
standard generators of $\goth s$ and $K$ is the level. $H$ is the minkowskian
Wess--Zumino field and $dx^+A_++dx^-A_-$ is the minkowskian
$\goth x$ gauge field.  $S^{\rm M}_{\rm WZNW}(H)$ is the customary
minkowskian Wess--Zumino--Novikov--Witten action integrating the variational
identity
$$
\delta S^{\rm M}_{\rm WZNW}(H)={1\over\pi}\int d^2x\tr\Big[\delta HH^{-1}
\partial_-(\partial_+HH^{-1})\Big].
\eqno(1.2)
$$
As recalled above, this field theory yields upon quantization the
$W$--algebra associated to the data $(G,S)$.

On a Riemann surface, one needs a euclidean reformulation of the above.
The basic algebraic data are now a simple complex Lie group $G$
and an $SL(2,\Bbb C)$ subgroup $S$ of $G$. To these data, there is
associated again a grading of $\goth g$ and a negative graded subalgebra
$\goth x$ of $\goth g$. The euclidean version of the action $(1.1)$
should read:
$$\eqalignno{
I^{\rm E}(H,A^*,A)
=&~KS^{\rm E}_{\rm WZNW}(H)
+{K\over\pi}\int d^2z\tr\Big[(\partial HH^{-1}-t_{+1})A^*&\cr
&\phantom{S^{\rm E}_{\rm WZNW}(H)}
+(H^{-1}\bar\partial H-t_{-1})A-A^*\Ad HA\Big].\vphantom{{1\over\pi}\int}
&(1.3)\cr}
$$
$H$ is the euclidean Wess--Zumino field and $dzA+d\bar zA^*$
is the euclidean $\goth x$ gauge field. $S^{\rm E}_{\rm WZNW}(H)$
is the `euclidean Wess--Zumino--Novikov--Witten action' integrating the
variational identity
$$
\delta S^{\rm E}_{\rm WZNW}(H)
={1\over\pi}\int d^2z\tr\Big[\delta HH^{-1}\bar\partial(\partial HH^{-1})\Big].
\eqno(1.4)
$$
Resorting to complex groups is unavoidable when switching from
minkowskian light--cone to euclidean holomorphic geometry. However,
in so doing, I have doubled the number of real field theoretic degrees
of freedom and generated a complex action. To eliminate the spurious
degrees of freedom and have a real positive definite action, one has to
impose on the fields certain reality conditions with respect to a suitable
conjugation. Such consitions are
$$
H=H^\dagger,
\eqno(1.5)
$$
$$
A^*=A^\dagger,
\eqno(1.6)
$$
and $t_d{}^\dagger=t_{-d}$, where $\dagger$ is the compact conjugation of
$\goth g$. This leads to a
reinterpretation of the model with surprising features.

The reality conditions $(1.5)$--$(1.6)$ suggest that $H$ is the fiber
metric for some principal $G$ bundle and that the $(0,1)$
gauge field $A^*$ is the Koszul field corresponding to its holomorphic
structure in the spirit of deformation theory \ref{20}. The
euclidean Wess--Zumino--Novikov--Witten action $S^{\rm E}_{\rm WZNW}(H)$
is then nothing but the Donaldson action first employed by Donaldson in his
studies of Hermitian--Einstein bundles \ref{21}. The principal bundle
in question is the Drinfeld--Sokolov bundle $DS$ discovered in ref. \ref{14}.
$DS$ prescribes the transformation rule of a $\goth g$--valued field
$\Psi(z,\bar z)$ under a coordinate change $z\rightarrow z'$, which reads
$$
\Psi'(z',\bar z')
=\exp\Big(-\ln{\partial z\over\partial z'}\ad t_0\Big)
\exp\Big({\partial\over\partial z'}{\partial z'\over\partial z}
\ad t_{-1}\Big)\Psi(z,\bar z).
\eqno(1.7)
$$
This important relation encapsulates at once the algebraic data $(G,S)$
defining the $W$--algebra and the holomorphic geometry of the underlying
Riemann surface. It also provides a mathematically precise formulation of
Polyakov's ideas of soldering \ref{22}.

This is reminiscent of ordinary gravity \`a la Polyakov \ref{16--19},
where the basic fields are the surface metric $h$ and the Beltrami field
$\mu$ and the effective action $I(h,\mu,\bar\mu)$ exhibits a structure
analogous to the one shown above, the counterpart of the
Wess--Zumino--Novikov--Witten action being the Liouville action.
The resemblance is even more striking when it is realized that
there are field theories whose effective action is a functional of $H$ and
$A^*$ of the form $(1.3)$ with $(1.4)$--$(1.6)$ satisfied. Therefore,
I shall call this euclidean model Drinfeld--Sokolov gravity. After gauge
fixing, the model has a residual gauge symmetry associated to the gauge
transformations leaving the a given Koszul field invariant. This is the
Drinfeld--Sokolov counterpart of conformal symmetry. It also involves
an integration on a non trivial space of Drinfeld--Sokolov gauge orbits.
It must be stressed that the Drinfeld--Sokolov moduli space considered here
is distinct from the $W$--moduli space of ref. \ref{23} and from the moduli
space studied by Hitchin in ref. \ref{24} and later related to quantum
$W$--gravity in ref. \ref{25}.

The plan of the paper is as follows. In sect. 2, the basic notions
concerning the holomorphic and hermitian structures and the symmetries
of the Drinfeld--Sokolov bundle necessary for the understanding of the
following constructions are collected. In sect. 3, the main properties
of Drinfeld--Sokolov field theory are expounded. In sect. 4,
Drinfeld--Sokolov gravity is defined, the  gauge fixing of the
Drinfeld--Sokolov symmetry is illustrated and the formal construction
of the measure is carried out. In sect. 5, the Drinfeld--Sokolov ghost
system is studied in detail. In sect. 6, conformal invariance and
certain non perturbative features of the resulting theory are analyzed
and the remaining unsolved problems are pointed out. Finally, the
appendices explain in great detail the definition of the functional
measures and implementation of the gauge fixing for the interested reader.
\vskip.4cm
\item{2.}{\bf The Drinfeld--Sokolov Bundle}
\vskip.4cm
\par
In the first part this section, I review certain general results concerning
the holomorphic and hermitian geometry of principal bundles on a surface
\ref{26--28}. In the second part, I define the Drinfeld--Sokolov bundle and
analyze its main properties \ref{14,29}.

{\it 1. Holomorphic Structures}

Let $\Sigma$ be a compact Riemann surface of genus $\ell$ with local
holomorphic coordinates $z_a$, where $a$ is a coordinate label. $\Sigma$
is characterized by the holomorphic $1$--cocycle $k$ defined by
$k_{ab}=\partial_a z_b$, where $\partial_a=\partial/\partial z_a$.
In applications, it is necessary to choose a $1$--cocycle square root
of $k$, that is a holomorphic $1$--cocycle
$k^{\otimes{1\over 2}}$ such that $(k^{\otimes{1\over 2}}{}_{ab})^2=k_{ab}$.
For any $j\in\Bbb Z/2$, one can then define the holomorphic $1$--cocycle
$k^{\otimes j}$ by setting $k^{\otimes j}{}_{ab}=
(k^{\otimes{1\over 2}}{}_{ab})^{2j}$. As is well known,
these $1$--cocycles define holomorphic line bundles on $\Sigma$, $k$ and
$k^{\otimes j}$ corresponding to the canonical line bundle and its
tensor powers.

Let $w,\bar w\in\Bbb Z/2$. A conformal field $\psi$ of weights $w,\bar w$
is given as a collection of smooth complex valued maps $\psi_a$ of domain
$\dom z_a$ such that, whenever defined, $\psi_a=k^{\otimes w}\otimes
\bar k^{\otimes\bar w}{}_{ab}\psi_{b}$. The conformal fields $\psi$ of
weights $w,\bar w$ span a infinite dimensional complex linear space
$\CF^{w,\bar w}$.

The spaces $\CF^{w,\bar w}$ and $\CF^{1-w,1-\bar w}$ are dual to each
other. The dual pairing is given by $\langle\phi,\psi\rangle
={1\over\pi}\int_\Sigma d^2z\phi\psi$ for $\psi\in\CF^{w,\bar w}$ and
$\phi\in\CF^{1-w,1-\bar w}$.

The Cauchy--Riemann operator $\bar\partial:\CF^{w,0}\rightarrow\CF^{w,1}$
is locally defined by $(\bar\partial\psi)_a=\bar\partial_a\psi_a$ for
$\psi\in\CF^{w,0}$. The kernel of $\bar\partial$ is the subspace
$\HCF^w$ of holomorphic elements of $\CF^{w,0}$. By the
Riemann--Roch theorem, $\dim\HCF^w-\dim\HCF^{1-w}=(2w-1)(\ell-1)$.

A $(1,0)$ affine connection $\gamma$ is a collection of smooth
complex valued maps $\gamma_a$ of domain $\dom z_a$ such that $\gamma_a=
k_{ab}[\gamma_b+\partial_b\ln k_{ab}]$ whenever defined. $\gamma$ is
characterized by its curvature $f_\gamma$, given locally by
$f_{\gamma a}=\bar\partial_a\gamma_a$. $f_\gamma\in\CF^{1,1}$.
Let $\Aff$ be the family of all $(1,0)$ affine connections $\gamma$.

To any $\gamma\in\Aff$, one can associate the covariant derivative
$\partial_\gamma:\CF^{w,\bar w}\rightarrow
\CF^{w+1,\bar w}$ locally given by $(\partial_\gamma\psi)_a
=(\partial_a-w\gamma_a)\psi_a$ for $\psi\in\CF^{w,\bar w}$.

Let $K$ be a holomorphic $G$--valued $1$--cocycle on $\Sigma$,
where $G$ is a simple complex Lie group. To $K$, one can associate
a smooth principal $G$--bundle $P$ over $\Sigma$ by means
of a well known construction.

A holomorphic structure $\sans s$ is specified by a collection of
smooth $G$--valued maps $V_{\ssans s a}$ of domain $\dom z_a$ such that
there exists a holomorphic $G$--valued $1$--cocycle $K_{\ssans s}$
such that, whenever defined, $V_{\ssans s a}=K_{ab}V_{\ssans s b}
K_{\ssans s ab}{}^{-1}$. Note that the $1$--cocycle $K_{\ssans s}$
characterizes but does not determine the holomorphic structure $\sans s$,
since the map $\sans s\rightarrow K_{\ssans s}$ is many--to--one.
Two holomorphic structures $\sans s_1$ and $\sans s_2$ are said equivalent
if $V_{\ssans s_1 a}=V_{\ssans s_2 a}v_a$ for some holomorphic $G$--valued
function $v_a$ for every $a$. This is indeed an equivalence relation.
Below, I shall not distinguish between equivalent holomorphic structures.
The family of all holomorphic structures of will be denoted by $\Hol$.

Let $\sans s\in\Hol$ and $w,\bar w\in\Bbb Z/2$. An extended
$\sans s$--conformal field $\Psi_{\ssans s}$ of weights $w,\bar w$ is
given as a collection of smooth $\goth g$--valued maps $\Psi_{\ssans s a}$
of domain $\dom z_a$ such that, whenever defined, $\Psi_{\ssans s a}
=k^{\otimes w}{}_{ab}\bar k^{\otimes\bar w}{}_{ab}\Ad K_{\ssans s ab}
\Psi_{\ssans s b}$. The extended $\sans s$--conformal fields $\Psi$ of
weights $w,\bar w$ span a infinite dimensional complex linear space
$\ECF^{w,\bar w}_{\ssans s}$.

The spaces $\ECF^{w,\bar w}_{\ssans s}$ and
$\ECF^{1-w,1-\bar w}_{\ssans s}$ are dual to each other. The dual pairing
is given by $\langle\Phi,\Psi\rangle_{\ssans s}
={1\over\pi}\int_\Sigma d^2z\tr_{\rm ad}\big(\Phi\Psi\big)_{\ssans s}$
for $\Psi_{\ssans s}\in\ECF^{w,\bar w}_{\ssans s}$ and
$\Phi_{\ssans s}\in\ECF^{1-w,1-\bar w}_{\ssans s}$, where $\tr_{\rm ad}$
denotes the Cartan--Killing form of $\goth g$.

The Cauchy--Riemann operator $\bar\partial_{\ssans s}:\ECF^{w,0}_{\ssans s}
\rightarrow\ECF^{w,1}_{\ssans s}$ is locally defined by
$(\bar\partial\Psi)_{\ssans s a}=\bar\partial_a\Psi_{\ssans s a}$. The
kernel of $\bar\partial_{\ssans s}$ is the subspace $\HECF^w_{\ssans s}$
of holomorphic elements of $\ECF^{w,0}_{\ssans s}$. By the
Riemann--Roch theorem, $\dim\HECF^w_{\ssans s}
-\dim\HECF^{1-w}_{\ssans s}=(2w-1)(\ell-1)\dim\goth g$.

A $(1,0)$ $\sans s$--connection $\Gamma_{\ssans s}$ is a collection
of smooth $\goth g$--valued maps $\Gamma_{\ssans s a}$ such that
$\Gamma_{\ssans s a}=k_{ab}[\Ad K_{\ssans s ab}\Gamma_{\ssans s b}
+\partial_bK_{\ssans s ab}K_{\ssans s ab}{}^{-1}]$. The connection
$\Gamma_{\ssans s}$ is characterized by its curvature $F_{\Gamma\ssans s}$
locally given by $F_{\Gamma\ssans s a}=\bar\partial_a\Gamma_{\ssans s a}$.
$F_{\Gamma\ssans s}\in\ECF^{1,1}_{\ssans s}$. Let $\Conn_{\ssans s}$
be the family of all $(1,0)$ $\sans s$--connections $\Gamma_{\ssans s}$.

To any $\gamma\in\Aff$ and $\Gamma_{\ssans s}\in\Conn_{\ssans s}$, one can
associate the covariant derivative
$\partial_{\gamma,\Gamma\ssans s}:\ECF^{w,\bar w}_{\ssans s}\rightarrow
\ECF^{w+1,\bar w}_{\ssans s}$ locally given by
$(\partial_{\gamma,\Gamma}\Psi)_{\ssans s a}
=(\partial_a-w\gamma_a-\ad\Gamma_{\ssans s a})\Psi_{\ssans s a}$.

In applications, the holomorphic structure $\sans s$ is considered as
variable. The dependence on $\sans s$ is then to be studied.

$\Hol$ contains a natural reference holomorphic structure defined by
$V_a=1$ for all $a$. By convention, all geometric objects related to
such structure, such as the holomorphic $1$--cocycle $K$, the extended
conformal fields $\Psi$, the spaces of (holomorphic) extended conformal
fields $\ECF^{w,\bar w}$ and $\HECF^w$, the $(1,0)$ connections $\Gamma$
and their family $\Conn$, etc. will carry no subscript $\sans s$.
In particular, the adjective `conformal' is always understood as
`reference--holomorphic--structure--conformal'.

Let $w,\bar w\in\Bbb Z/2$. A minimal extended conformal field functional
$\Psi$ of weights $w,\bar w$ is a map that associates to any $\sans s\in\Hol$
an element $\Psi_{\ssans s}\in\ECF^{w,\bar w}_{\ssans s}$ in such a way that
the condition $\Psi_a=\Ad V_{\ssans sa}\Psi_{\ssans s a}$ is satisfied
for any $a$. In this way, the dependence of $\Psi_{\ssans s}$ on $\sans s$
is determined entirely by $V_{\ssans s}$. The space of all minimal
extended conformal field functionals
$\Psi$ of weights $w,\bar w$ may thus be identified with $\ECF^{w,\bar w}$
itself.

For any $\Psi\in\ECF^{w,\bar w}$ and $\Phi\in\ECF^{1-w,1-\bar w}$,
$\langle\Phi,\Psi\rangle_{\ssans s}=\langle\Phi,\Psi\rangle$
for $\sans s\in\Hol$.
In this way, the dual pairing $\langle\cdot,\cdot\rangle_{\ssans s}$ of
$\ECF^{w,\bar w}_{\ssans s}$ and $\ECF^{1-w,1-\bar w}_{\ssans s}$
induces a dual pairing $\langle\cdot,\cdot\rangle$ of the spaces of
minimal extended conformal field functionals $\ECF^{w,\bar w}$ and
$\ECF^{1-w,1-\bar w}$.

A minimal $(1,0)$ connection functional $\Gamma$ is a map that associates
to any $\sans s\in\Hol$ an element $\Gamma_{\ssans s}\in\Conn_{\ssans s}$
in such a way that the condition $\Gamma_a=\Ad V_{\ssans sa}
\Gamma_{\ssans s a}+\partial_aV_{\ssans sa}V_{\ssans sa}{}^{-1}$
is satisfied for any $a$. As for minimal extended conformal field
functionals, this condition means that the dependence of
$\Gamma_{\ssans s}$ on $\sans s$ is determined by $V_{\ssans s}$.
The family of minimal $(1,0)$ connection functionals
$\Gamma$ may be identified with $\Conn$ itself.

There exists a parametrization of $\Hol$, the Koszul parametrization
defined next, which is particularly useful in field theoretic applications.

A Koszul field $A^*$ is simply an element of $\ECF^{0,1}$. There is a
one--to--one correspondence between the family of holomorphic structures
$\sans s$ and the family of Koszul fields $A^*$ \ref{20}. The
correspondence, expressed notationally as $\sans s\equiv A^*$, is given
by the relation $A^*{}_a=\bar\partial_aV_{\ssans s a}V_{\ssans s a}{}^{-1}$.
Thus, one may view equivalently $\Hol$ as the manifold formed by all
Koszul fields and cast dependence on $\sans s$ as dependence on $A^*$.
Note that $A^*=0$ for the reference holomorphic structure.

In general, field theoretic expressions are compact when written in
terms of the relevant holomorphic structure $\sans s$. The dependence
on $\sans s$ is however explicit only in the Koszul parametrization
provided one restricts to minimal extended conformal field functionals
and minimal $(1,0)$ connection functionals. The rules for translating
from the first to the second description are the following:
$$
\Psi_{\ssans s}\leftrightarrow\Psi,
\eqno(2.1)
$$
$$
\bar\partial_{\ssans s}\leftrightarrow\bar\partial-\ad A^*,
\eqno(2.2)
$$
$$
F_{\Gamma\ssans s}\leftrightarrow F_\Gamma-\partial_\Gamma A^*,
\eqno(2.3)
$$
$$
\partial_{\gamma,\Gamma\ssans s}\leftrightarrow\partial_{\gamma,\Gamma},
\eqno(2.4)
$$
for $\Psi\in\ECF^{w,\bar w}$, $\gamma\in\Aff$ and $\Gamma\in\Conn$, where
$\sans s\equiv A^*\in\Hol$. If $\Psi\in\ECF^{w,0}$ is such that
$\Psi_{\ssans s}\in\HECF^w_{\ssans s}$, then $(\bar\partial-\ad A^*)\Psi=0$.

{\it 2. Hermitian Structures}

A hermitian surface metric $h$ on $\Sigma$ is a collection of smooth
maps $h_a$ of domain $\dom z_a$ such that $h_a>0$ and
$h_a=k_{ab}\bar k_{ab}h_b$. The hermitian surface metrics $h$ form a
infinite dimensional real functional manifold $\Met$.

Given any metric $h\in\Met$, one can define a Hilbert structure on
$\CF^{w,\bar w}$ by setting
$\langle\psi_1,\psi_2\rangle^{\vphantom{\vee}}_h
={1\over\pi}\int_\Sigma d^2zh^{\otimes 1-w-\bar w}\bar\psi_1\psi_2$
for $\psi_1,\psi_2\in\CF^{w,\bar w}$.

Each metric $h$ is characterized by a $(1,0)$ affine connection $\gamma_h$
locally given by $\gamma_{ha}=\partial_a\ln h_a$. The curvature $f_h$ of
$\gamma_h$ is then given by $f_{ha}=\bar\partial_a\partial_a\ln h_a$.
The covariant derivative of $\gamma_h$ will be denoted by $\partial_h$.

Let $\sans s\in\Hol$ be a holomorphic structure. A $\sans s$--hermitian
fiber metric $H_{\ssans s}$ is defined as a collection of smooth
$G$--valued maps $H_{\ssans s a}$ of domain $\dom z_a$ such that
$H_{\ssans s a}{}^\dagger=H_{\ssans s a}$ and $H_{\ssans s a}=K_{\ssans s ab}
H_{\ssans s b}K_{\ssans s ab}{}^\dagger$, where $\dagger$ denotes
the compact conjugation of $G$. The $\sans s$--hermitian fiber metrics
$H_{\ssans s}$ form an infinite dimensional real manifold $\Herm_{\ssans s}$.

Given metrics $h\in\Met$ and $H_{\ssans s}\in\Herm_{\ssans s}$, one can
define a Hilbert structure on $\ECF^{w,\bar w}_{\ssans s}$ by setting
$\langle\Psi_1,\Psi_2\rangle^{\vphantom{\vee}}_{h,H\ssans s}
={1\over\pi}\int_\Sigma d^2zh^{\otimes 1-w-\bar w}\tr_{\rm ad}
\big(\Ad H\Psi_1{}^\dagger\Psi_2\big)_{\ssans s}$
for $\Psi_{1\ssans s},\Psi_{2\ssans s}\in\ECF^{w,\bar w}_{\ssans s}$.

Each fiber metric $H_{\ssans s}$ is characterized by a $(1,0)$
$\sans s$--connection $\Gamma_{H\ssans s}$ of $K_{\ssans s}$ locally
given by $\Gamma_{H\ssans s a}=\partial_aH_{\ssans s a}
H_{\ssans s a}{}^{-1}$. The curvature $F_{H\ssans s}$ of $\Gamma_{H\ssans s}$
is given by $F_{H\ssans s a}=\bar\partial_a(\partial_aH_{\ssans s a}
H_{\ssans s a}{}^{-1})$.
The covariant derivative associated to a surface
metric $h\in\Met$ and to $H_{\ssans s}$ is
$\partial_{h,H\ssans s}$.

A minimal hermitian fiber metric functional is a map that associates
to each holomorphic structure $\sans s\in\Hol$ a hermitian fiber metric
$H_{\ssans s}\in\Herm_{\ssans s}$ in such a way that
$H_a=V_{\ssans s a}H_{\ssans s a}V_{\ssans s a}{}^\dagger$
holds for any $a$. As for minimal extended conformal field
functionals, this condition means that the dependence of
$H_{\ssans s}$ on $\sans s$ is determined by
$V_{\ssans s}$. Hence, the space of minimal hermitian fiber metric
functionals $H$ may be identified with $\Herm$.

For any $H\in\Herm$ and any two $\Psi_1,\Psi_2\in\ECF^{w,\bar w}$,
$\langle\Psi_1,\Psi_2\rangle^{\vphantom{\vee}}_{h,H\ssans s}
=\langle\Psi_1,\Psi_2\rangle^{\vphantom{\vee}}_{h,H}$ for
$\sans s\in\Hol$. Thus, for a given
minimal hermitian fiber metric functional $H$, the Hilbert
structure $\langle\cdot,\cdot\rangle^{\vphantom{\vee}}_{h,H\ssans s}$
on $\ECF^{w,\bar w}_{\ssans s}$ induces a Hilbert structure
$\langle\cdot,\cdot\rangle^{\vphantom{\vee}}_{h,H}$ on the space of
minimal extended conformal field functionals $\ECF^{w,\bar w}$.

For the curvature $F_H$ and the covariant derivative $\partial_{h,H}$
associated to metrics $h\in\Met$ and $H\in\Herm$, $(2.3)$--$(2.4)$ do not
apply. To express everything in the Koszul parametrization, one has instead
to perform the substitutions
$$
F_{H\ssans  s}\leftrightarrow
F_H-\partial_HA^*-\bar\partial\Ad HA^*{}^\dagger
+[A^*,\Ad HA^*{}^\dagger],
\eqno(2.5)
$$
$$
\partial_{h,H,\ssans s}\leftrightarrow\partial_{h,H}+\ad\Ad HA^*{}^\dagger,
\eqno(2.6)
$$
with $\sans s\equiv A^*\in\Hol$.

{\it 3. The Gauge Group}

A gauge transformation $\alpha$ is a collection of smooth $G$--valued
maps $\alpha_a$ of domain $\dom z_a$ such that, whenever defined,
$\alpha_a=K_{ab}\alpha_bK_{ab}{}^{-1}$. The gauge transformations form
a group $\Gau$ under pointwise multiplication. $\Lie\Gau\cong\ECF^{0,0}$
with the obvious Lie brackets. To $\Gau$, there are associated a few relevant
actions.

$\Gau$ does not act on $\Sigma$ and on the spaces $\CF^{w,\bar w}$ of
conformal fields.

$\Gau$ acts on the family $\Hol$ of holomorphic structures as follows.
If $\alpha\in\Gau$ and $\sans s\in\Hol$, then $\alpha^*\sans s\in\Hol$
is the holomorphic structure specified by $V_{\alpha^*\ssans s a}=
\alpha_aV_{\ssans s a}$. Note that $K_{\alpha^*\ssans s}=K_{\ssans s}$.
The action of $\Gau$ on $\Hol$ is not free. The stability subgroup
${\cal G}(\sans s)$ of a holomorphic structure $\sans s\in\Hol$ in $\Gau$
is formed by all gauge transformations $\eta$ such that
$\eta_{\ssans s a}=V_{\ssans sa}{}^{-1}\eta_aV_{\ssans s a}$ is
holomorphic. In fact, for $\eta\in{\cal G}(\sans s)$, $\eta^*\sans s$
is equivalent to $\sans s$, and hence is not distinguished from the latter.
Note that $\Lie{\cal G}(\sans s)\cong\HECF^0_{\ssans s}$.

Associated to this action is also an action on extended conformal fields
defined as follows. For $\alpha\in\Gau$ and $\Psi_{\ssans s}
\in\ECF^{w,\bar w}_{\ssans s}$, $\alpha^*\Psi_{\alpha^*\ssans s}$ is the
extended conformal field in $\ECF^{w,\bar w}_{\alpha^*\ssans s}$ locally
defined by $\alpha^*\Psi_{\alpha^*\ssans s a}=\Psi_{\ssans s a}$.

The dual pairing $\langle\cdot,\cdot\rangle_{\ssans s}$ of
$\ECF^{w,\bar w}_{\ssans s}$ and $\ECF^{1-w,1-\bar w}_{\ssans s}$ is
covariant under $\Gau$. In fact, $\langle\alpha^*\Phi,\alpha^*\Psi
\rangle_{\alpha^*\ssans s}=\langle\Phi,\Psi\rangle_{\ssans s}$
for $\Psi_{\ssans s}\in\ECF^{w,\bar w}_{\ssans s}$ and
$\Phi_{\ssans s}\in\ECF^{1-w,1-\bar w}_{\ssans s}$.

There is a corresponding action of $\Gau$ on the space of minimal extended
conformal field functionals $\ECF^{w,\bar w}$. For $\alpha\in\Gau$ and
$\Psi\in\ECF^{w,\bar w}$, $\alpha^*\Psi$ is the element of
$\ECF^{w,\bar w}$ locally given by $\alpha^*\Psi_a=\Ad\alpha_a\Psi_a$.
The value $\alpha^*\Psi_{\alpha^*\ssans s}$ of $\alpha^*\Psi$ at the
holomorphic structure $\alpha^*\sans s$ is the result of the action
of $\alpha$ on $\Psi_{\ssans s}$ defined above, as suggested by the
notation.

The dual pairing $\langle\cdot,\cdot\rangle$ of $\ECF^{w,\bar w}$ and
$\ECF^{1-w,1-\bar w}$ is invariant under $\Gau$, i. e. one has
$\langle\alpha^*\Phi,\alpha^*\Psi\rangle=\langle\Phi,\Psi\rangle$
for $\Psi\in\ECF^{w,\bar w}$ and $\Phi\in\ECF^{1-w,1-\bar w}$.

In the Koszul parametrization, the action of $\Gau$ on $\Hol$ translates
into an action on the Koszul field $A^*$. For $\alpha\in\Gau$ and
$A^*\in\Hol$, the action is locally given by $\alpha^*A^*{}_a
=\bar\partial_a\alpha_a\alpha_a{}^{-1}+\Ad\alpha_a A^*{}_a$.
If $\eta\in{\cal G}(\sans s)$ with $\sans s\equiv A^*$, then
the equation $(\bar\partial-\ad A^*)\eta\eta^{-1}=0$
is satisfied.

$\Gau$ is inert on the space of surface metrics $\Met$.

$\Gau$ acts on the hermitian fiber metrics as follows.
For any $\alpha\in\Gau$ and $H_{\ssans s}\in\Herm_{\ssans s}$,
$\alpha^*H_{\alpha^*\ssans s}$ is the element of
$\Herm_{\alpha^*\ssans s}$ locally given by
$\alpha^*H_{\alpha^*\ssans s a}=H_{\ssans s a}$.

It is easy to verify that, for any $h\in\Met$ and any $H_{\ssans s}\in
\Herm_{\ssans s}$,
the Hilbert structure $\langle\cdot,\cdot\rangle_{h,H\ssans s}$ on
$\ECF^{w,\bar w}_{\ssans s}$ defined earlier is $\Gau$ covariant, i. e.
$\langle\alpha^*\Psi_1,\alpha^*\Psi_2
\rangle^{\vphantom{\vee}}_{h,\alpha^*H\alpha^*\ssans s}
=\langle\Psi_1,\Psi_2\rangle^{\vphantom{\vee}}_{h,H\ssans s}$,
for $\Psi_{1\ssans s},\Psi_{2\ssans s}\in\ECF^{w,\bar w}_{\ssans s}$.

There is a corresponding action of $\Gau$ on the space of minimal fiber
metrics functionals $\Herm$. For $\alpha\in\Gau$ and $H\in\Herm$,
$\alpha^*H$ is the element of $\Herm$ locally given by
$\alpha^*H_a=\alpha_aH_a\alpha_a{}^\dagger$.
The value $\alpha^*H_{\alpha^*\ssans s}$ of $\alpha^*H$ at the
holomorphic structure $\alpha^*\sans s$ is the result of the action
of $\alpha$ on $H_{\ssans s}$ defined above.

It is easy to verify that, for any $h\in\Met$ and any $H\in\Herm$,
the Hilbert structure $\langle\cdot,\cdot\rangle_{h,H}$ on
$\ECF^{w,\bar w}$ defined earlier is $\Gau$ invariant, i. e.
$\langle\alpha^*\Psi_1,\alpha^*\Psi_2\rangle^{\vphantom{\vee}}_{h,\alpha^*H}
=\langle\Psi_1,\Psi_2\rangle^{\vphantom{\vee}}_{h,H}$,
for $\Psi_1,\Psi_2\in\ECF^{w,\bar w}$.

In the analysis of symmetries, it is much simpler to proceed at the
infinitesimal level. Let $\Xi$ be the gauge ghost. $\Xi$ is an element of
$\ECF^{0,0}\otimes\bigwedge^1(\Lie\Gau)^\vee$ defining a basis of
$\bigwedge^1(\Lie\Gau)^\vee$. The infinitesimal action of the gauge group
$\Gau$ on field functionals is given be the nilpotent Slavnov operator
$s$, $s^2=0$. From the Maurer--Cartan equations of $\Gau$, one has
$$s\Xi={1\over 2}[\Xi,\Xi].
\eqno(2.7)
$$
Further,
$$
s\psi=0,
\eqno(2.8)
$$
$$
sA^*=\big(\bar\partial-\ad A^*\big)\Xi,
\eqno(2.9)
$$
$$
s\Psi=\ad\Xi\Psi,
\eqno(2.10)
$$
where $\psi\in\CF^{w,\bar w}$, $A^*\in\Hol$ and $\Psi\in\ECF^{w,\bar w}$.

At infinitesimal level, the action $\Gau$ on $\Met$ and $\Herm$ is given by
$$
s\ln h=0,
\eqno(2.11)
$$
$$
sHH^{-1}=\Xi+\Ad H\Xi^\dagger,
\eqno(2.12)
$$
with $h\in\Met$ and $H\in\Herm$.

{\it 4. The Drinfeld--Sokolov Bundle}

The basic data entering in the definition of the Drinfeld--Sokolov
bundle are the following: {\it i}) a simple complex Lie group $G$;
{\it ii}) an $SL(2,\Bbb C)$ subgroup $S$ of $G$ invariant under the
compact conjugation $\dagger$ of $G$; {\it iii})
a Riemann surface $\Sigma$ of genus
$\ell$ with a spinor structure $k^{\otimes {1\over 2}}$.
Let $t_{-1}$, $t_0$, $t_{+1}$ be a set of standard generators
of $\goth s$, so that
$$
[t_{+1},t_{-1}]=2t_0,\quad [t_0,t_{\pm 1}]=\pm t_{\pm 1},
\eqno(2.13)
$$
$$
t_d{}^\dagger=t_{-d}, \quad d=-1,0,+1.
\eqno(2.14)
$$
Then,
$$
K_{ab}=\exp(-\ln k_{ab}t_0)\exp(\partial_a k_{ab}{}^{-1}t_{-1}).
\eqno(2.15)
$$
defines a holomorphic $G$--valued $1$--cocycle $K$ \ref{14}.
This in turn defines a smooth principal $G$--bundle, the
Drinfeld--Sokolov bundle $DS$, whose relevance has been explained
in the introduction.

The Drinfeld--Sokolov bundle has extra structures derived from a special
nilpotent subalgebra $\goth x$ of $\goth g$ associated to $\goth s$.
Such structures will be called Drinfeld--Sokolov and will play an important
role in the following. The reason for this, related to the form of
anomalies, will be explained in detail in the next section.

To the Cartan element $t_0$ of $\goth s$, there is associated a
halfinteger grading of $\goth g$: the subspace $\goth g_m$ of $\goth g$
of degree $m\in \Bbb Z/2$ is the eigenspace of $\ad t_0$ with
eigenvalue $m$. One can further define a
bilinear form $\chi$ on $\goth g$ by $\chi(x,y)=\tr_{\rm ad}(t_{+1}[x,y])$,
$x,y\in \goth g$ \ref{10}. The restriction of $\chi$ to
$\goth g_{-{1\over 2}}$ is non singular. By Darboux theorem,
there is a direct sum decomposition $\goth g_{-{1\over 2}}=
\goth p_{-{1\over 2}}\oplus\goth q_{-{1\over 2}}$ of $\goth g_{-{1\over 2}}$
into subspaces $\goth p_{-{1\over 2}}$ and $\goth q_{-{1\over 2}}$
of the same dimension, which are maximally isotropic and dual
to each other with respect to $\chi$. Set
$$
\goth x=\goth p_{-{1\over 2}}\oplus\bigoplus_{m\leq -1}\goth g_m.
\eqno(2.16)
$$
$\goth x$ is a negative graded nilpotent subalgebra of $\goth g$.

Let $\Hol_{\DS}$ be the family of all holomorphic structures $\sans s$
such that $V_{\ssans s a}$ is $\exp\goth x$--valued for every $a$.
Such structures will be called Drinfeld--Sokolov. For $\sans s\in\Hol_{\DS}$,
$K_{\ssans s ab}=K_{ab}L_{\ssans s ab}$, where $L_{\ssans s ab}$ is a
holomorphic $\exp\goth x$--valued function.

Let $\sans s\in\Hol_{\DS}$ and $w,\bar w\in\Bbb Z/2$. A
Drinfeld--Sokolov extended $\sans s$--conformal field
$\Psi_{\ssans s}$ of weights $w,\bar w$
is an element of $\ECF^{w,\bar w}_{\ssans s}$ such that
$\Psi_{\ssans s a}$ is valued in $\goth x$ for any $a$.
This definition is consistent because of the form of the $1$--cocycle
$K_{\ssans s}$ and the fact that $[t_0,\goth x]\subseteq\goth x$
and $[\goth x,\goth x]\subseteq\goth x$. The Drinfeld--Sokolov fields
$\Psi_{\ssans s}$ of weights $w,\bar w$ span an infinite dimensional
complex linear space $\ECF^{\phantom{\DS}w,\bar w}_{\DS\ssans s}$.
Similarly, a dual Drinfeld--Sokolov extended $\sans s$--conformal
field $\Psi_{\ssans s}$ of weights $w,\bar w$ is an element of
$\ECF^{w,\bar w}_{\ssans s}$ such that $\Psi_{\ssans s a}$ is defined
modulo a $\goth x^\perp$--valued local function for any $a$, where
$\goth x^\perp$ is the orthogonal complement of $\goth x$ with respect to
the Cartan--Killing form $\tr_{\rm ad}$. This definition is also consistent
because of the form of the $1$--cocycle $K_{\ssans s}$ and the fact that
$[t_0,\goth x^\perp]\subseteq\goth x^\perp$ and $[\goth x,\goth x^\perp]
\subseteq\goth x^\perp$. The dual Drinfeld--Sokolov fields $\Psi_{\ssans s}$
of weights $w,\bar w$ span an infinite dimensional complex space
$\ECF^{\vee\hphantom{\rm S}w,\bar w}_{\DS\ssans s}$.

For $\sans s\in\Hol_{\DS}$, the Cauchy--Riemann operator
$\bar\partial_{\ssans s}$ maps $\ECF^{\phantom{\DS}w,0}_{\DS\ssans s}$
into $\ECF^{\phantom{\DS}w,1}_{\DS\ssans s}$. Therefore,
$\bar\partial_{\ssans s}$ defines by restriction
a Cauchy--Riemann operator $\bar\partial_{\rm DS\ssans s}:
\ECF^{\phantom{\DS}w,0}_{\DS\ssans s}\rightarrow
\ECF^{\phantom{\DS}w,1}_{\DS\ssans s}$, the Drinfeld--Sokolov
Cauchy--Riemann operator. In this way, one can consistently define
a notion of holomorphy for Drinfeld--Sokolov extended
$\sans s$--conformal fields.
The subspace of holomorphic elements $\Psi_{\ssans s}$ of
$\ECF^{\phantom{\DS}w,0}_{\DS\ssans s}$ will be denoted by
$\HECF^{\phantom{\DS}w}_{\DS\ssans s}$.
Similarly, $\bar\partial_{\ssans s}$ maps
$\ECF^{\vee\hphantom{\rm S}w,0}_{\DS\ssans s}$ into
$\ECF^{\vee\hphantom{\rm S}w,1}_{\DS\ssans s}$.
So, $\bar\partial_{\ssans s}$ induces a Cauchy--Riemann operator
$\bar\partial^\vee_{\rm DS\ssans s}:
\ECF^{\vee\hphantom{\rm S}w,0}_{\DS\ssans s}
\rightarrow\ECF^{\vee\hphantom{\rm S}w,1}_{\DS\ssans s}$,
the dual Drinfeld--Sokolov Cauchy--Riemann operator.
So, one can consistently define a notion of holomorphy also
for dual Drinfeld--Sokolov extended $\sans s$--conformal fields.
The subspace of holomorphic elements $\Phi_{\ssans s}$ of
$\ECF^{\vee\hphantom{\rm S}w,0}_{\DS\ssans s}$ will be denoted by
$\HECF^{\vee\hphantom{\rm S}w}_{\DS\ssans s}$. There exists an
interesting Drinfeld--Sokolov version of the Riemann--Roch theorem:
$$
\dim\HECF^{\phantom{\DS}w}_{\DS\ssans s}
-\dim\HECF^{\vee\hphantom{\rm S}1-w}_{\DS\ssans s}
=\tr\big[\big((2w-1)1-2\ad t_0\big)p_{\goth x}\big](\ell-1),
\eqno(2.17)
$$
where $p_{\goth x}$ is any projector of $\goth g$ onto $\goth x$ \ref{29}.

The spaces $\ECF^{\phantom{\DS}w,\bar w}_{\DS\ssans s}$ and
$\ECF^{\vee\hphantom{\rm S}1-w,1-\bar w}_{\DS\ssans s}$
are dual to each other. The dual pairing is given by
$\langle\Phi,\Psi\rangle_{\DS\ssans s}
={1\over\pi}\int_\Sigma d^2z\tr_{\rm ad}\big(\Phi\Psi\big)_{\ssans s}$ for
$\Psi\in\ECF^{\phantom{\DS}w,\bar w}_{\DS\ssans s}$ and
$\Phi\in\ECF^{\vee\hphantom{\rm S}1-w,1-\bar w}_{\DS\ssans s}$.
Note that the result of the integration does not depend on the
representative of $\Phi_{\ssans s}$ used.

A Drinfeld--Sokolov $(1,0)$ $\sans s$--connection $\Gamma_{\ssans s}$ is
an element of $\Conn_{\ssans s}$ such that $\Gamma_{\ssans s a}
-{1\over 2}t_{+1}$ is $\goth x^\perp$--valued for every $a$.
This definition is consistent because of the form of the $1$--cocycle
$K_{\ssans s}$ and the fact that $[t_d,\goth x]\subseteq\goth x^\perp$ for
$d=0,-1$, $\goth x\subseteq\goth x^\perp$ and $[\goth x,\goth x^\perp]\subseteq
\goth x^\perp$. If $\Gamma_{\ssans s}$ is Drinfeld--Sokolov, then
$F_{\Gamma \ssans s}=0$ in
$\ECF^{\vee\hphantom{\rm S}1,1}_{\DS\ssans s}$. Let $\Conn_{\DS\ssans s}$
be the family of all Drinfeld--Sokolov $(1,0)$ $\sans s$--connections
$\Gamma_{\ssans s}$.

The reference holomorphic structure is obviously Drinfeld--Sokolov, since
$V_a=1$ is $\exp\goth x$--valued.

Let $w,\bar w\in\Bbb Z/2$. A Drinfeld--Sokolov minimal extended conformal
field functional $\Psi$ of weights $w,\bar w$ is a minimal extended conformal
field functional defined on $\Hol_{\DS}$ and such that, for any
$\sans s\in\Hol_{\DS}$, $\Psi_{\ssans s}\in
\ECF^{\phantom{\DS}w,\bar w}_{\DS\ssans s}$. This definition is
certainly consistent, as the reference holomorphic structure is
Drinfeld--Sokolov, $V_{\ssans s}$ is $\exp\goth x$--valued
and $[\goth x,\goth x]\subseteq\goth x$. The space of Drinfeld--Sokolov
minimal extended conformal field functionals of weights $w,\bar w$ may
clearly be identified with $\ECF^{\phantom{\DS}w,\bar w}_{\DS}$.
Similarly, a dual Drinfeld--Sokolov minimal extended conformal field
functional $\Psi$ of weights $w,\bar w$ is a minimal extended conformal
field functional defined on $\Hol_{\DS}$ and such that, for any
$\sans s\in\Hol_{\DS}$, $\Psi_{\ssans s}\in
\ECF^{\vee\hphantom{\rm S}w,\bar w}_{\DS\ssans s}$.
This definition also is consistent, for the reference holomorphic
structure is Drinfeld--Sokolov, $V_{\ssans s}$ is $\exp\goth x$--valued
and $[\goth x,\goth x^\perp]\subseteq\goth x^\perp$.
The space of dual Drinfeld--Sokolov minimal extended conformal field
functionals of weights $w,\bar w$ may clearly be identified with
$\ECF^{\vee\hphantom{\rm S}w,\bar w}_{\DS}$.

For any $\Psi\in\ECF^{\phantom{\DS}w,\bar w}_{\DS}$ and
$\Phi\in\ECF^{\vee\hphantom{\rm S}1-w,1-\bar w}_{\DS}$,
$\langle\Phi,\Psi\rangle_{\DS\ssans s}=\langle\Phi,\Psi\rangle_{\DS}$
for any $\sans s\in\Hol_{\DS}$.
Therefore, the dual pairing $\langle\cdot,\cdot\rangle_{\DS\ssans s}$
of $\ECF^{\phantom{\DS}w,\bar w}_{\DS\ssans s}$ and
$\ECF^{\vee\hphantom{\rm S}1-w,1-\bar w}_{\DS\ssans s}$
induces a dual pairing $\langle\cdot,\cdot\rangle_{\DS}$ of the
spaces of (dual) Drinfeld--Sokolov minimal extended conformal
field functionals $\ECF^{\phantom{\DS}w,\bar w}_{\DS}$ and
$\ECF^{\vee\hphantom{\rm S}1-w,1-\bar w}_{\DS}$.

A minimal Drinfeld--Sokolov $(1,0)$ connection functional $\Gamma$ is
a minimal $(1,0)$ connection functional defined on $\Hol_{\DS}$
such that, for any $\sans s\in\Hol_{\DS}$, $\Conn_{\ssans s}
\in\Conn_{\DS\ssans s}$. This definition is consistent again because
the reference holomorphic structure is Drinfeld--Sokolov,
$V_{\ssans s}$ is $\exp\goth x$--valued and
the fact that $[t_{+1},\goth x]\subseteq\goth x^\perp$,
$\goth x\subseteq\goth x^\perp$ and $[\goth x,\goth x^\perp]\subseteq
\goth x^\perp$. The space of dual Drinfeld--Sokolov minimal
connection functionals $\Gamma$ may clearly be identified with
$\Conn_{\DS}$.

In the Koszul parametrization, the Drinfeld--Sokolov holomorphic structures
are represented by $\goth x$--valued Koszul fields $A^*$. Such Koszul fields
are also called Drinfeld--Sokolov.

{\it 5. Hermitian Structures of the Drinfeld--Sokolov Bundle}

Let $h\in\Met$ and $H_{\ssans s}\in\Herm_{\ssans s}$ be metrics. The Hilbert
structure $\langle\cdot,\cdot\rangle^{\vphantom{\vee}}_{h,H\ssans s}$
on $\ECF^{w,\bar w}_{\ssans s}$ defines by restriction a Hilbert structure
$\langle\cdot,\cdot\rangle^{\vphantom{\vee}}_{\DS h,H\ssans s}$
on $\ECF^{\phantom{\DS}w,\bar w}_{\DS\ssans s}$.
The Hilbert structure allows one to identify
$\ECF^{\vee\hphantom{\rm S}w,\bar w}_{\DS\ssans s}$ with
$\ECF^{\phantom{\DS}1-w,1-\bar w}_{\DS\ssans s}$.
By definition, the element $\Phi_{h,H\ssans s}
\in\ECF^{\phantom{\DS}1-w,1-\bar w}_{\DS\ssans s}$ corresponding to
$\Phi_{\ssans s}\in\ECF^{\vee\hphantom{\rm S}w,\bar w}_{\DS\ssans s}$
is the unique element of $\ECF^{\phantom{\DS}1-w,1-\bar w}_{\DS\ssans s}$
such that $\langle\Phi|\Psi\rangle^{\vphantom{\vee}}_{\DS\ssans s}
=\langle\Phi_{h,H},\Psi\rangle^{\vphantom{\vee}}_{\DS h,H\ssans s}$
for all $\Psi_{\ssans s}\in\ECF^{\phantom{\DS}1-w,1-\bar w}_{\DS\ssans s}$.
One may now define a Hilbert structure on
$\ECF^{\vee\hphantom{\rm S}w,\bar w}_{\DS\ssans s}$ by setting
$\langle\Phi_1,\Phi_2\rangle^\vee_{\DS h,H\ssans s}
=\langle\Phi_{2h,H},
\Phi_{1h,H}\rangle^{\vphantom{\vee}}_{\DS h,H\ssans s}$.

For any $H\in\Herm$ and any $\Psi_1,\Psi_2\in
\ECF^{\phantom{\DS}w,\bar w}_{\DS}$,
$\langle\Psi_1,\Psi_2\rangle^{\vphantom{\vee}}_{\DS h,H\ssans s}
=\langle\Psi_1,\Psi_2\rangle^{\vphantom{\vee}}_{\DS h,H}$
for $\sans s\in\Hol_{\DS}$. Similarly, for $H\in\Herm$ and
$\Phi_1,\Phi_2\in\ECF^{\vee\hphantom{\rm S}w,\bar w}_{\DS}$,
$\langle\Phi_1,\Phi_2\rangle^\vee_{\DS h,H\ssans s}
=\langle\Phi_1,\Phi_2\rangle^\vee_{\DS h,H}$. Thus, for a given
minimal hermitian fiber metric functional $H$, the Hilbert structures
$\langle\cdot,\cdot\rangle^{\vphantom{\vee}}_{\DS h,H\ssans s}$
on $\ECF^{\phantom{\DS}w,\bar w}_{\DS\ssans s}$ and
$\langle\cdot,\cdot\rangle^\vee_{\DS h,H\ssans s}$
on $\ECF^{\vee\hphantom{\rm S}w,\bar w}_{\DS\ssans s}$
induce Hilbert structures
$\langle\cdot,\cdot\rangle^{\vphantom{\vee}}_{\DS h,H}$
and $\langle\cdot,\cdot\rangle^\vee_{\DS h,H}$ on the spaces of
(dual) Drinfeld--Sokolov minimal extended conformal field
functionals $\ECF^{\phantom{\DS}w,\bar w}_{\DS}$ and
$\ECF^{\vee\hphantom{\rm S}w,\bar w}_{\DS}$, respectively.

{\it 6. The Drinfeld--Sokolov Gauge Group}

The gauge group $\Gau$ does not respect $\Hol_{\DS}$. There is however a
subgroup of $\Gau_{\DS}$ of $\Gau$, the Drinfeld--Sokolov gauge group,
which does. $\Gau_{\DS}$ is formed by those elements $\alpha\in\Gau$
such that $\alpha_a$ is $\exp\goth x$--valued for every $a$.
Clearly, $\Lie\Gau_{\DS}\cong\ECF^{\phantom{\DS}0,0}_{\DS}$.

For any $\sans s\in\Hol_{\DS}$, the stability subgroup
${\cal G}_{\DS}(\sans s)$ of $\sans s$ in $\Gau_{\DS}$ is simply the
intersection ${\cal G}(\sans s)\cap\Gau_{\DS}$. Clearly,
$\Lie{\cal G}_{\DS}(\sans s)\cong\HECF^{\phantom{\DS}0}_{\DS\sans s}$.
$\Lie{\cal G}_{\DS}(\sans s)$ is nilpotent, since $\exp\goth x$ is.

{}From the definition, it is immediate to see that
the action $\alpha^*:\ECF^{w,\bar w}_{\ssans s}
\rightarrow\ECF^{w,\bar w}_{\alpha^*\ssans s}$ associated to
$\alpha\in\Gau_{\DS}$ maps $\ECF^{\phantom{\DS}w,\bar w}_{\DS\ssans s}$
and $\ECF^{\vee\hphantom{\rm S}w,\bar w}_{\DS\ssans s}$ respectively
into $\ECF^{\phantom{\DS}w,\bar w}_{\DS\alpha^*\ssans s}$
and $\ECF^{\vee\hphantom{\rm S}w,\bar w}_{\DS\alpha^*\ssans s}$.

It can also be seen that the dual pairing $\langle\cdot|
\cdot\rangle^{\vphantom{\vee}}_{\DS\ssans s}$ of
$\ECF^{\phantom{\DS}w,\bar w}_{\DS\ssans s}$ and
$\ECF^{\vee\hphantom{\rm S}1-w,1-\bar w}_{\DS\ssans s}$
is $\Gau_{\DS}$ covariant.

Using the fact that $\goth x$ is a subalgebra of
$\goth g$ such that $[\goth x,\goth x^\perp]\subseteq\goth x^\perp$,
it is easy to check that the action of $\Gau_{\DS}$ on $\ECF^{w,\bar w}$
preserves both $\ECF^{\phantom{\DS}w,\bar w}_{\DS}$ and
$\ECF^{\vee\hphantom{\rm S}w,\bar w}_{\DS}$.

The dual pairing $\langle\cdot|\cdot\rangle^{\vphantom{\vee}}_{\DS}$
of $\ECF^{\phantom{\DS}w,\bar w}_{\DS}$ and
$\ECF^{\vee\hphantom{\rm S}1-w,1-\bar w}_{\DS}$
is $\Gau_{\DS}$ invariant.

The Hilbert structures on $\ECF^{\phantom{\DS}w,\bar w}_{\DS\ssans s}$
and $\ECF^{\vee\hphantom{\rm S}w,\bar w}_{\DS\ssans s}$ defined above are
both $\Gau_{\DS}$ covariant.

One can similarly show that
the Hilbert structures on $\ECF^{\phantom{\DS}w,\bar w}_{\DS}$
and $\ECF^{\vee\hphantom{\rm S}w,\bar w}_{\DS}$ are
both $\Gau_{\DS}$ invariant.

To $\Gau_{\DS}$, one can consistently associate a Slavnov operator $s_{\DS}$
and a $\goth x$--valued ghost field $\Xi_{\DS}
\in\ECF^{\phantom{\DS}0,0}_{\DS}\otimes\bigwedge^1(\Lie\Gau_{\DS})^\vee$
obeying $(2.7)$.
$(2.8)$--$(2.10)$ also holds with $A^*$ a Drinfeld--Sokolov Koszul field
and $\Psi$ a (dual) Drinfeld--Sokolov extended conformal field with
$s$ and $\Xi$ replaced by $s_{\DS}$ and $\Xi_{\DS}$. Of course,
$(2.11)$--$(2.12)$ continue to holds with $s$ and $\Xi$ replaced by
$s_{\DS}$ and $\Xi_{\DS}$.

Before completing this section, I shall state the following
conventions. In what follows, {\it when in the same equation there
appear a holomorphic structure $\sans s$ and a Koszul field $A^*$, it is
implicitly assumed, unless otherwise stated, that $\sans s\equiv A^*$}.
Further, {\it all field functionals on $\Hol$ or $\Hol_{\DS}$ are
implicitly assumed, unless otherwise stated, to be minimal field
functionals}.
\vskip.4cm
\item{3.} {\bf Drinfeld--Sokolov Field Theory}
\vskip.4cm
\par
A Drinfeld--Sokolov field theory is a local field theory
whose basic fields are (extended) conformal fields of the Drinfeld--Sokolov
bundle.

The standard classical example to have in mind is the Drinfeld--Sokolov
$B$--$C$ system. The basic fields $B$ and $C$ belong to
${\bf F}\otimes\ECF^{1-w,0}$ and ${\bf F}\otimes\ECF^{w,0}$,
respectively, where $\bf F$ is the fermionic Grassmann algebra.
The action, for a given holomorphic structure $A^*$, is
\footnote{${}^1$}{In the notation of this paper, a functional
$f(X)$ of a complex field $X$ is not necessarily holomorphic. Holomorphy,
when it occurs, will be explicitly stated.}
$$
S(B,C,A^*)
={1\over\pi}\int_\Sigma d^2z
2\real\tr_{\rm ad}(B\bar\partial C)_{\ssans s}.
\eqno(3.1)
$$

In general, the quantization of a Drinfeld--Sokolov field theory requires
the introduction of a hermitian structure $(h,H)\in\Met\times\Herm$ for the
proper definition of the adjoint of the relevant differential operators.
The regularization of the ultraviolet divergencies of the corresponding
functional determinants involves further the use of an ultraviolet cut--off
$\epsilon$. The regularization method which will be applied below
is the so called proper time method \ref{18}. I shall restrict to
Drinfeld--Sokolov field theories for which the bare $\Gau$ invariant bare
effective action $\hat I(h,H,A^*;\epsilon)$ is of the form
$$
\hat I(h,H,A^*;\epsilon)
=-{r\over\pi\epsilon}\int_\Sigma d^2zh
+\bigg[{n\over6\pi}\int_\Sigma d^2zf_h-d_{\ssans s}\bigg]\ln\epsilon
+I_0(h,H,A^*)+O(\epsilon).
\eqno(3.2)
$$
Here $r$, $n$ and $d_{\ssans s}$ are real coefficients.
$\int_\Sigma d^2zf_h$ is the Gauss--Bonnet topological
invariant whose well known value is $2\pi(\ell-1)$.
$I_0(h,H,A^*)$ is a non local functional of
$h$, $H$ and $A^*$ such that
$$
\delta I_0(h,H,A^*)
=-{\kappa_0\over12\pi}\int_\Sigma d^2z\delta\ln hf_h
+{K\over\pi}\int_\Sigma d^2z\tr_{\rm ad}
\big(\delta HH^{-1}F_H\big)_{\ssans s},
\eqno(3.3)
$$
where $\delta$ denotes variation with respect to $h$ and $H$ at fixed
$A^*$ \ref{30}. $\kappa_0$ and $K>0$ are generalized central charges.
The Drinfeld--Sokolov $B$--$C$ system introduced earlier is precisely of
this type with $r=\dim\goth g$, $n=(3w-1)\dim\goth g$, $d_{\ssans s}=
\dim\HECF^w_{\ssans s}$, $\kappa_0=-2(6w^2-6w+1)\dim\goth g$ and $K=1$.

To renormalize the bare effective action, one has to add to it a
counterterm of the form
$$
\Delta\hat I(h,H,A^*;\epsilon)
={r\over\pi\epsilon}\int_\Sigma d^2zh
-\bigg[{n\over6\pi}\int_\Sigma d^2zf_h-d_{\ssans s}\bigg]\ln\epsilon
+\Delta I(h,H,A^*)+O(\epsilon).
\eqno(3.4)
$$
Here, $\Delta I(h,H,A^*)$ is a local but otherwise arbitrary functional
of $h$, $H$ and $A^*$, whose choice defines a renormalization prescription.
The renormalized effective action is thus
$$
I(h,H,A^*)=I_0(h,H,A^*)+\Delta I(h,H,A^*).
\eqno(3.5)
$$
$I_0(h,H,A^*)$ is the renormalized effective action in the
minimal subtraction renormalization scheme.

In what follows, $\Delta I(h,H,A^*)$ is assumed to be independent from $A^*$:
$$
\Delta I(h,H,A^*)=\Delta I(h,H).
\eqno(3.6)
$$
Under this hypothesis, it can be shown that $I(h,H,A^*)$ has the following
structure
$$
I(h,H,A^*)=I(h,H)+L(H,A^*;A)+I_{\rm hol}(A^*;A).
\eqno(3.7)
$$
Here, $A\in\Conn$ is a background $(1,0)$ connection. $I(h,H)$ is the
functional $I(h,H,A^*)$ evaluated at the reference holomorphic structure
$A^*=0$.
$$
L(H,A^*;A)
={K\over\pi}\int_\Sigma d^2z\Big[
2\real\tr_{\rm ad}\big((\Gamma_H-A)A^*\big)
-\tr_{\rm ad}\big(A^*\Ad HA^{*\dagger}\big)\Big].
\eqno(3.8)
$$
$I_{\rm hol}(A^*;A)$ is a non local functional of $A^*$ only depending
on $A$. Next, I shall analyze the properties of the three terms in the
right hand side of $(3.7)$.

In order the counterterm $\Delta\hat I(h,H,A^*;\epsilon)$ to be $\Gau$
invariant, $\Delta I(h,H)$ must satisfy
$$
s\Delta I(h,H)=0.
\eqno(3.9)
$$
In this way, the renormalized effective action $I(h,H,A^*)$ is
$\Gau$ invariant as well. When $(3.9)$ is fulfilled, one has
$$
sI(h,H)={\cal W}(H),
\eqno(3.10)
$$
$$
sL(H,A^*;A)=-{\cal W}(H)-{\cal A}(A^*;A),
\eqno(3.11)
$$
$$
sI_{\rm hol}(A^*;A)={\cal A}(A^*;A),
\eqno(3.12)
$$
where
$$
{\cal W}(H)={K\over\pi}\int_\Sigma d^2z
2\real\tr_{\rm ad}\big(\Xi F_H\big),
\eqno(3.13)
$$
$$
{\cal A}(A^*;A)
=-{K\over\pi}\int_\Sigma d^2z
2\real\tr_{\rm ad}\big(\Xi(F_A-\partial_A A^*)\big)
\eqno(3.14)
$$
are the gauge anomalies.

$I(h,H)$ is a non local functional of $h$ and $H$. Its dependence on
$h$ and $H$ can be analyzed as follows.
The Drinfeld--Sokolov bundle possesses a remarkable property,
the possibility of lifting any surface metric $h\in\Met$ to a fiber metric
$H(h)\in\Herm$. Explicitly, $H(h)$ is given by
$$
H(h)=\exp(-\partial\ln ht_{-1})\exp(-\ln ht_0)\exp(-\bar\partial\ln ht_{+1}).
\eqno(3.15)
$$
This allows one to write $I(h,H)$ as follows.
$$
I(h,H)=I_{\rm conf}(h)+S(h,H)+\Delta I(h,H)-\Delta I(h,H(h)),
\eqno(3.16)
$$
where
$$
I_{\rm conf}(h)=I(h,H(h)),
\eqno(3.17)
$$
$$
S(h,H)=\Omega(H,H(h)).
\eqno(3.18)
$$
Here, for any two $H,H_0\in\Herm$, $\Omega(H,H_0)$ is the Donaldson action
defined by functional path integral
$$
\Omega(H,H_0)
={K\over\pi}\int_{H_0}^H\int_\Sigma d^2z
\tr_{\rm ad}\big(\delta H'H'^{-1}F_{H'}\big).
\eqno(3.19)
$$
The right hand side is independent from the choice of the
functional integration path joining $H_0$ to $H$, since
the functional $1$--form on $\Herm$ integrated is closed and $\Herm$
is clearly contractible. $\Omega(H,H_0)$ can be computed explicitly. The
metric $H\in\Herm$ can be written as $H=\exp\Phi H_0$, where the Donaldson
field $\Phi$ is an element of $\ECF^{0,0}$ such that $\Ad H\Phi^\dagger
=\Phi$. By direct calculation, one then finds
$$
\Omega(H,H_0)
=-{K\over\pi}\int_\Sigma d^2z\tr_{\rm ad}\Big[
\bar\partial\Phi{\exp\ad\Phi-1-\ad\Phi\over(\ad\Phi)^2}
\partial_{H_0}\Phi-\Phi F_{H_0}\Big]
\eqno(3.20)
$$
\ref{31}.

Now, $I_{\rm conf}(h)$ is a non local functional of $h$. Using $(3.3)$
$(3.5)$, $(3.15)$ and $(3.17)$, one can show that
$$
\delta I_{\rm conf}(h)
=-{\kappa_0+\kappa\over 12\pi}\int_\Sigma d^2z\delta\ln hf_h
+\delta\bigg[{\lambda\over\pi}\int_\Sigma d^2zh^{-1}f_h{}^2
+\Delta I(h,H(h))\bigg],
\eqno(3.21)
$$
where
$$
\kappa=-12K\tr_{\rm ad}(t_0{}^2),
\eqno(3.22)
$$
$$
\lambda=-2K\tr_{\rm ad}(t_0{}^2).
\eqno(3.23)
$$
If
$$
\Delta I(h,H)={\lambda_0\over\pi}\int_\Sigma d^2zh^{-1}f_h{}^2,
\eqno(3.24)
$$
where $\lambda_0$ is some constant, then $(3.21)$ simplifies into
$$
\delta I_{\rm conf}(h)
=-{\kappa_0+\kappa\over 12\pi}\int_\Sigma d^2z\delta\ln hf_h
+{\lambda_0+\lambda\over\pi}\delta\int_\Sigma d^2zh^{-1}f_h{}^2.
\eqno(3.25)
$$
A counterterm $\Delta I(h,H,A^*)$ for which $(3.24)$ holds is given by
the right hand side of $(3.24)$ itself and clearly satisfies both
$(3.6)$ and $(3.9)$.
Setting $\lambda_0=-\lambda$, $I_{\rm conf}(h)$ becomes the renormalized
effective action of a conformal field theory of conformal central charge
$\kappa_{\rm conf}=\kappa_0+\kappa$. Note that {\it the shift $\kappa$ given
by $(3.22)$ is precisely the classical central charge of the classical
$W$--algebras associated to the pair $(G,S)$, if $K$ is interpreted as the
Wess--Zumino--Novikov--Witten level}. For a generic value of $\lambda_0$,
one obtains a more general renormalized effective action with a
$\int\sqrt h R_h{}^2$ term yielding a model of induced $2d$ gravity of the
same type as that considered in refs. \ref{32--33}.

The functional $S(h,H)$ is local. In fact, the Donaldson field $\Phi(h,H)$
relevant here, given by
$$
\exp\Phi(h,H)=HH(h)^{-1},
\eqno(3.26)
$$
is clearly a local functional of $h$ and $H$ and $\Omega(H,H_0)$, given by
$(3.20)$, is a local functional of $\Phi$ and $H_0$.

{}From the above discussion, it follows that {\it the suitably renormalized
effective action $I(h,H)$ differs from the conformal effective action
$I_{\rm conf}(h)$ by a local functional of $h$ and $H$. In particular, the
$H$ dependence is local}.

{}From $(3.8)$, it is apparent that $L(H,A^*;A)$, the interaction term of $H$
and $A^*$, is local.

$I_{\rm hol}(A^*;A)$ is the real part of a holomorphic functional of $A^*$
and $A$ \ref{30}. Holomorphic factorization is an important feature of the
model which however will not be discussed in this paper. Its independence
from $H$ is crucial.

One has thus reached the following important conclusion. {\it The full
suitably renormalized $\Gau$ invariant effective action $I(h,H,A^*)$ is a
local functional of $H$}.

An important observation, related to the analysis of ref. \ref{10},
is the following. If one restricts to Drinfeld--Sokolov holomorphic
structures $A^*\in\Hol_{\DS}$ and to Drinfeld--Sokolov background
connections $A\in\Conn_{\DS}$, then the functionals $L(H,A^*;A)$ and
$I_{\rm hol}(A^*;A)$ are independent from $A$. Further, under the action of
Drinfeld--Sokolov gauge group $\Gau_{\DS}$, one has relations analogous to
$(3.10)$--$(3.12)$, with $s$, ${\cal W}(H)$ and ${\cal A}(A^*;A)$ replaced
by $s_{\DS}$, ${\cal W}_{\DS}(H)$ and ${\cal A}_{\DS}(A^*;A)$, respectively,
where ${\cal W}_{\DS}(H)$ and ${\cal A}_{\DS}(A^*;A)$ are given by
$(3.13)$--$(3.14)$ with $\Xi$ substituted by $\Xi_{\DS}$. In this case,
however, one has
$$
{\cal A}_{\DS}(A^*;A)=0,\quad A\in\Conn_{\DS},~A^*\in\Hol_{\DS}
\eqno(3.27)
$$
identically by $(2.16)$. Henceforth, {\it it is assumed that
$A\in\Conn_{\DS}$}.
\vskip.4cm
\item{4.} {\bf Drinfeld--Sokolov Gravity}
\vskip.4cm
\par
In Polyakov's approach to two dimensional gravity, the functional integration
over all smooth metrics on the string world sheet is reduced into an
integration over the conformal factor of the metric $h$ and on the Beltrami
field $\mu$. The action governing the quantum dynamics of such fields
is the diffeomorphism invariant effective action of a conformal field theory.

In many respects, the quantization of Drinfeld--Sokolov gravity parallels
that of ordinary two dimensional gravity. One integrates over all fiber
metrics $H$ of $\Herm$ and on all Drinfeld--Sokolov Koszul fields
$A^*$ of $\Hol_{\rm DS}$. The action of such fields is the $\Gau_{\DS}$
invariant bare effective action $\hat I(h,H,A^*)$ of a Drinfeld--Sokolov
field theory of the type described in sect. 3. The partition function is
thus of the form
$$
{\cal Z}_\Theta(h)
=\int_{\Herm\times\Hol_{\DS}}{(DH)\otimes(DA^*)
\over{\rm vol}(\Gau_{\rm DS})}\hat\Theta(h,H,A^*)\exp\hat I(h,H,A^*),
\eqno(4.1)
$$
where $\hat\Theta(h,H,A^*)$ is some bare $\Gau_{\DS}$--invariant insertion.
This is of course a rather formal expression whose precise meaning is to be
defined. The relation of this quantization prescription with earlier
approaches, in particular with that of ref. \ref{10}, has been discussed
in the introduction.

The basic configuration space is the cartesian product $\Herm\times\Hol_{\DS}$
carrying the action of $\Gau_{\DS}$ described in sect. 2.
To gauge fix, one has to transform the functional
integral on $\Herm\times\Hol_{\DS}$ into one on a configuration space
containing, roughly speaking, a factor $\Gau_{\DS}$ by computing the jacobian
of the corresponding functional change of variables.

To properly carry out the gauge fixing, it is necessary to define a good
moduli space of Drinfeld--Sokolov holomorphic structures modulo the action
of the Drinfeld--Sokolov gauge group and characterize the stability group
of Drinfeld--Sokolov holomorphic structures. This requires a notion of
stability. A thorough geometric investigation of this issue is beyond the
scope of this paper. Nevertheless, it is still possible to make an educated
guess about these geometric structures by the following argument.

As well known, every stable holomorphic structure is simple and the
family $\SHol$ of stable holomorphic structures is dense in $\Hol$
and invariant under the action of the gauge group $\Gau$ \ref{27--28}.
Here, the relevant holomorphic structures are those of $\Hol_{\DS}$
and the relevant symmetry group is the Drinfeld--Sokolov gauge group
$\Gau_{\DS}$. No holomorphic structure $\sans s\in\Hol_{\DS}$ is stable
in the customary sense. It is however reasonable to assume by analogy
that, for any reasonable definition of Drinfeld--Sokolov stability, a
Drinfeld--Sokolov stable holomorphic structure should be Drinfeld--Sokolov
simple and that the family $\SHol_{\DS}$ of Drinfeld--Sokolov stable
holomorphic structures should be dense in $\Hol_{\DS}$ and invariant
under the action of the Drinfeld--Sokolov gauge group $\Gau_{\DS}$.
Recall that a holomorphic structure $\sans s\in\Hol$ is simple if
the subgroup ${\cal G}(\sans s)$ of $\sans s$--holomorphic gauge
transformations of $\Gau$ is trivial \ref{27--28}, a condition
equivalent to the vanishing of the space $\HECF^0_{\ssans s}$, since
$\Lie{\cal G}(\sans s)\cong\HECF^0_{\ssans s}$. Similarly, a
holomorphic structure $\sans s\in\Hol_{\DS}$ is said Drinfeld--Sokolov
simple if ${\cal G}_{\DS}(\sans s)$ has minimal dimension, or,
equivalently, if the space $\HECF^{\phantom{\DS}0}_{\DS\ssans s}$ has
minimal dimension, since $\Lie{\cal G}_{\DS}(\sans s)\cong
\HECF^{\phantom{\DS}0}_{\DS\ssans s}$.

In analogy to the ordinary moduli space, the Drinfeld--Sokolov moduli space
${\cal M}_{\DS}$ will be defined as the quotient $\SHol_{\DS}/\Gau_{\DS}$.
${\cal M}_{\DS}$ is a finite dimensional complex manifold.

For $\sans s$ varying in $\SHol_{\DS}$, the groups ${\cal G}_{\DS}(\sans s)$
are all isomorphic to the same complex Lie group ${\cal G}_{\DS}$. In fact
they all are of the form $\exp\HECF^{\phantom{\DS}0}_{\DS\ssans s}$, where
the spaces $\HECF^{\phantom{\DS}0}_{\DS\ssans s}$ are all valued in the
same nilpotent subalgebra of $\goth x$ of $\goth g$ and can be continuously
deformed into one another by continuously varying $\sans s$ in
$\SHol_{\DS}$. ${\cal G}_{\DS}$ is nilpotent, since $\exp\goth x$ is.

In this paper, {\it it will be assumed that $\dim\HCF^{{1\over 2},0}=0$}.
This holds for an even spinor structure and for a generic holomorphic
structure of $\Sigma$. It is merely a technically simplifying hypothesis
with a very nice consequence. If the assumption is fulfilled, all
holomorphic structures are Drinfeld--Sokolov simple. This is no longer true
in the generic situation, where even the reference holomorphic structure
characterized by the $1$--cocycle $(2.15)$ may fail to be Drinfeld--Sokolov
simple \ref{29}.

A method for computing the dimensions of ${\cal G}_{\DS}$ and
${\cal M}_{\DS}$ exploiting the Drinfeld--Sokolov simplicity
has been presented in \ref{29}. They are given by
$$
\dim{\cal G}_{\DS}
=\cases{0,\vphantom{\big[} &if $\ell=0$,\cr
\dim\goth x_{\rm int}, \vphantom{\big[} &if $\ell=1$,\cr
\dim\goth g_{-1}-\tr\big[\big(2\ad t_0+1\big)p_{\goth x}\big](\ell-1),
\vphantom{\big[} &if $\ell\geq 2$,\cr}
\eqno(4.2)
$$
$$
\dim{\cal M}_{\DS}
=\cases{-\tr\big[\big(2\ad t_0+1\big)p_{\goth x}\big],\vphantom{\big[}
&if $\ell=0$,\cr
\dim\goth x_{\rm int}, \vphantom{\big[} &if $\ell=1$,\cr
\dim\goth g_{-1},\vphantom{\big[} &if $\ell\geq 2$,\cr}
\eqno(4.3)
$$
where $p_{\goth x}$ is any projector of $\goth g$ onto
$\goth x$ and $\goth x_{\rm int}=\bigoplus_{m\in\Bbb Z,m\leq -1}\goth g_m$.

The relevant configuration space is properly $\Herm\times\SHol_{\DS}$
A natural parametrization of $\Herm\times\SHol_{\DS}$ is provided by
$$
H(\tilde H,\alpha)=\alpha^*\tilde H=\alpha \tilde H\alpha^\dagger,
\eqno(4.4)
$$
$$
A^*(t,\alpha)=\alpha^*A^*(t)
=\bar\partial\alpha\alpha^{-1}+\Ad\alpha A^*(t),
\eqno(4.5)
$$
where $t\in{\cal M}_{\DS}$, $\tilde H\in\Herm$, $\alpha\in\Gau_{\DS}$,
and $t\in{\cal M}_{\DS}\rightarrow A^*(t)\in\SHol_{\DS}$ is a fiducial
gauge slice. The parametrization
possesses a ${\cal G}_{\DS}$--symmetry as follows
from the following argument. Any two elements
$(t,\tilde H,\alpha)$ and $(t',\tilde H',\alpha')$ of
${\cal M}_{\DS}\times\Herm\times\Gau_{\DS}$ have the same image under
$(4.4)$--$(4.5)$ if and only if $t'=t$ and $\tilde H'=
\eta \tilde H\eta^\dagger$ and $\alpha'=
\alpha\eta^{-1}$ for some $\eta\in{\cal G}_{\DS}(\sans s_t)$
with $\sans s_t\equiv A^*(t)$, since ${\cal G}_{\DS}(\sans s_t)$ is
the subgroup of $\Gau_{\DS}$ leaving $A^*(t)$ invariant. Now, for fixed
$t\in{\cal M}_{\DS}$, the maps
$$
\eta^*\tilde H=\eta\tilde H\eta^\dagger,
\eqno(4.6)
$$
$$
\vphantom{\alpha}^\eta\alpha=\alpha\eta^{-1},
\eqno(4.7)
$$
with $\eta\in{\cal G}_{\DS}(\sans s_t)$, define an action of
${\cal G}_{\DS}(\sans s_t)$ on $\Herm\times\SHol_{\DS}$.
The action $(4.6)$--$(4.7)$ is free and is a symmetry of $(4.4)$--$(4.5)$.
Since ${\cal G}_{\DS}\cong{\cal G}_{\DS}(\sans s_t)$ for
any $t$, it is a ${\cal G}_{\DS}$ symmetry. One can then construct the space
${\cal M}_{\DS}\times(\Herm\times\Gau_{\DS})/{\cal G}_{\DS}(\sans s_\cdot)
=\prod_{t\in{\cal M}_{\DS}}\{t\}\times
((\Herm\times\Gau_{\DS})/{\cal G}_{\DS}(\sans s_t))$. This
provides the realization of the configuration space relevant
for the implementation of the gauge fixing.

The second realization is rather unwieldy, because the meaning of
the functional integration on a functional manifold of the form
$(\Herm\times\Gau_{\DS})/{\cal G}_{\DS}(\sans s_t)$
for fixed $t\in{\cal M}_{\DS}$ is not quite clear.
One way of solving this problem consists in transforming
the integration on such
functional manifold into an integration on $\Herm\times\Gau_{\DS}$
with a residual unfixed gauge symmetry corresponding to
${\cal G}_{\DS}(\sans s_t)$. To do this, one employs
the obvious isomorphism $\Herm\times\Gau_{\DS}\cong((\Herm\times\Gau_{\DS})
/{\cal G}_{\DS}(\sans s_t))\times{\cal G}_{\DS}$, where the action of
${\cal G}_{\DS}(\sans s_t)$ on $\Herm\times\Gau_{\DS}$ is given by
$(4.6)$--$(4.7)$. Upon choosing a group isomorphism of
$\zeta(\cdot;t):{\cal G}_{\DS}\rightarrow{\cal G}_{\DS}(\sans s_t)$
of ${\cal G}_{\DS}$ onto ${\cal G}_{\DS}(\sans s_t)$,
the isomorphisms is explicitly given by
$$
H(\tilde H,g)=\zeta(g;t)^*\tilde H=\zeta(g;t)\tilde H\zeta(g;t)^\dagger,
\eqno(4.8)
$$
$$
\omega(\alpha,g)=\vphantom{\alpha}^{\zeta(g;t)}\alpha=\alpha\zeta(g;t)^{-1},
\eqno(4.9)
$$
where $(\tilde H,\alpha)$ varies in a slice of $\Herm\times\Gau_{\DS}$
representing the quotient $(\Herm\times\Gau_{\DS})
/{\cal G}_{\DS}(\sans s_t))$ and $g\in{\cal G}_{\DS}$.

The definition of the functional measures on the relevant field spaces and
the computation of the jacobians relating such measures is carried out
by means certain formal prescriptions outlined below. It is
important to realize that such prescriptions serve only the purpose
of producing and justifying heuristicly a definition of the measure
of the gauge fixed partition ${\cal Z}_\Theta(h)$ and should not
in any way be interpreted as a means of proving theorems about
an otherwise well defined field theoretic model.

To any complex Hilbert space $\cal H$ with inner product $\langle\cdot,
\cdot\rangle$, there is associated a real Hilbert space ${\cal H}^{\rm r}$
with inner product $\langle\cdot,\cdot\rangle^{\rm r}$. ${\cal H}^{\rm r}$
is just ${\cal H}$ seen as a real vector space by restricting the numerical
field from $\Bbb C$ to $\Bbb R$. $\langle x_1,x_2\rangle^{\rm r}
=2\real\langle x_1,x_2\rangle$ for $x_1,x_2\in{\cal H}^{\rm r}={\cal H}$.
In particular, $\Vert x\Vert^{{\rm r} 2}=2\Vert x\Vert^2$.

To any real Hilbert space $\cal H$, there is associated a translation
invariant functional measure $(Dx)$ normalized so that $\int_{\cal H}(Dx)
\exp\big(-{1\over 2}\Vert x\Vert^2\big)=1$.

If $\cal F$ is a real Hilbert manifold, then, for any $f\in{\cal F}$, the
tangent space $T_f{\cal F}$ of $\cal F$ at $f$ is a Hilbert space with
norm $\Vert\delta f\Vert_{|f}$ and measure $(D\delta f)_{|f}$. This
defines a measure $(Df)_{|f}$ on $\cal F$ by identifying $(Df)_{|f}$ with
$(D\delta f)_{|f}$ at $f$. In general, $(Df)_{|f}$ is not
translation invariant, depending explicitly on $f$.

If $\cal F$ and $\cal E$ are Hilbert manifolds and $\varphi:{\cal F}
\rightarrow{\cal E}$ is an invertible map, then $\cal F$ is a
parameter space for $\cal E$ and it is possible to transform
functional integration on $\cal E$ with measure $(De)_{|e}$
into functional integration on $\cal F$ with measure $(Df)_{|f}$.
To this end, one needs the jacobian relation
$(D\varphi(f))_{|\varphi(f)}=[\det(\delta\varphi(f))]^{1\over 2}
(Df)_{|f}$, where, for any $f\in{\cal F}$, $\delta\varphi(f):
T_f{\cal F}\rightarrow T_{\varphi(f)}{\cal E}$ is the tangent map
of $\varphi$ at $f$.

Applying the above formal recipes, one can define real Hilbert
structures on $\Herm$, $\SHol_{\DS}$, $\Gau_{\DS}$, ${\cal M}_{\DS}$
and ${\cal G}_{\DS}$ and obtain in this way the corresponding functional
measures $(DH)_{h|H}$, $(DA^*)_{H|A^*}$, $(D\alpha)_{h,H|\alpha}$,
$(Dt)_{|t}$ and $(Dg)_{|g}$. The measures depend on a background
surface metric $h\in\Met$ and on a fiber metric $H\in\Herm$ through
the underlying Hilbert structures. $h$ is fixed whereas $H$ is chosen
to be the varying metric integrated over in the functional integral.
Using these basic Hilbert structures and functional measures, one can
define real Hilbert structures on the derived field spaces defined above,
obtain the corresponding functional measures, implement the gauge fixing in
the partition function and computing the resulting functional jacobians.
The details of this analysis are rather technical and have been lumped in
app. A for the interested reader. Here, I shall limit myself to illustrate
the result.

By varying $(4.5)$ with respect to $\alpha$ and taking $(2.2)$
into account, it appears that the Drinfeld--Sokolov ghost kinetic
operator is $\bar\partial_{\DS\ssans s_t}$ acting on
$\ECF^{\phantom{\DS}0,0}_{\DS\ssans s_t}$. Hence, the Drinfeld--Sokolov
Fadeev--Popov determinant is something like
$\dtr(\bar\partial_{\DS\ssans s}{}^\star\bar\partial_{\DS\ssans s})$
for a Drinfeld--So\-ko\-lov holomorphic structure $\sans s\in\Hol_{\DS}$.
This notation is a little bit too formal.
First, the adjoint $\bar\partial_{\DS\ssans s}{}^\star$ of
$\bar\partial_{\DS\ssans s}$ is defined with respect to
the Hilbert structures $\langle\cdot,\cdot
\rangle^{\vphantom{\vee}}_{\DS h,H\ssans s}$
$\ECF^{\phantom{\DS}0,0}_{\DS\ssans s}$ and
$\ECF^{\phantom{\DS}0,1}_{\DS\ssans s}$ corresponding to
the fixed background surface metric $h$ and the varying fiber metric
$H$. Secondly, the ghost kinetic operator $\bar\partial_{\DS\ssans s}$
has zero eigenvalues which have to be removed from the determinant.
Hence, the Drinfeld--Sokolov Fadeev--Popov determinant should properly be
$\dtr'_{h,H}(\bar\partial_{\DS\ssans s}{}^\star
\bar\partial_{\DS\ssans s})$, where the dependence on the metrics $h$ and
$H$ and the removal of the zero eigenvalues are
explicitly stated. The resulting functional of $h$, $H$ and $\sans s$ is
essentially the bare ghost effective action once the zero modes and
comodes of $\bar\partial_{\DS\ssans s}$ are properly taken care of.
Let $\{e_i(\sans s)_{\ssans s}|i=1,\cdots,\dim{\cal G}_{\DS}\}$ be a basis
of $\ker\bar\partial_{\DS\ssans s}$. Since $\bar\partial_{\DS\ssans s}$
is defined independently from any choice of hermitian structure, the
$e_i(\sans s)_{\ssans s}$ can be chosen independent from $h$ and $H$.
Let $\{f^j(\sans s)_{\ssans s}|j=1,\cdots,\dim{\cal M}_{\DS}\}$
be a basis of $\coker\bar\partial_{\DS\ssans s}$. This is defined here
as the annihilator of $\ran\bar\partial_{\DS\ssans s}$ in
$\ECF^{\vee\hphantom{\rm S}1,0}_{\DS\ssans s}$ under the dual pairing
$\langle\cdot|\cdot\rangle^{\vphantom{\vee}}_{\DS\ssans s}$.
$\bar\partial_{\DS\ssans s}$ being defined independently of any choice of
hermitian structure, the $f^j(\sans s)_{\ssans s}$ can also be chosen
independent from $h$ and $H$. The bare effective action
$\hat I^{\rm gh}(h,H,A^*)$ is
$$
\hat I^{\rm gh}(h,H,A^*)
=\ln\bigg[{\dtr'_{h,H}(\bar\partial_{\DS\ssans s}{}^\star
\bar\partial_{\DS\ssans s})\over
\det M^{\vphantom{\vee}}_{h,H\ssans s}(e(\sans s))
\det M^\vee_{h,H\ssans s}(f(\sans s))}\bigg],
\eqno(4.10)
$$
where
$$
M^{\vphantom{\vee}}_{h,H\ssans s}(e(\sans s))_{ij}
=\langle e_i(\sans s),e_j(\sans s)\rangle^{\vphantom{\vee}}_{\DS h,H\ssans s},
\quad i,j=1,\cdots,\dim{\cal G}_{\DS},
\eqno(4.11)
$$
$$
M^\vee_{h,H\ssans s}(f(\sans s))^{kl}
=\langle f^k(\sans s),f^l(\sans s)\rangle^\vee_{\DS h,H\ssans s},
\quad k,l=1,\cdots,\dim{\cal M}_{\DS}.
\eqno(4.12)
$$
are the Gramian matrices of the bases $\{e_i(\sans s)_{\ssans s}\}$ and
$\{f^j(\sans s)_{\ssans s}\}$.

Below, I shall make some reasonable assumptions on the gauge slice function
$A^*(t)$ and the group isomorphism $\zeta(g;t)$. Though they are not
strictly necessary for the formal manipulations of functional integrals
required by the gauge fixing, they are such to guarantee the holomorphic
factorization on ${\cal M}_{\DS}$ of all finite dimensional factors entering
in the measure of the gauge fixed partition function ${\cal Z}_{\Theta}(h)$,
a property known to hold in ordinary string theory which one would like to
keep also in the present context.

As first assumption, the gauge slice function $t\rightarrow A^*(t)$ is
assumed to be analytic:
$$
\bar\partial_tA^*(t)=0.
\eqno(4.13)
$$
It is not known to me whether it is possible to find a gauge slice
function $A^*(t)$ globally holomorphic on ${\cal M}_{\DS}$. In general,
$A^*(t)$ may develop singularities on a submanifold of ${\cal M}_{\DS}$
of non zero codimension, where $A^*(t)$ fails to be transverse to the
action of the gauge group $\Gau_{\DS}$ on $\SHol_{\DS}$. The
singularities may eventually entail divergencies in the modular
integration.

$(4.13)$ implies that the family of elliptic operators
$t\rightarrow\bar\partial_{\DS\ssans s_t}$ is complex analytic.
So, setting $e_i(t)=e_i(\sans s_t)$ and $f^j(t)=f^j(\sans s_t)$,
one also has $\bar\partial_te_i(t)=0$ and
$\bar\partial_t f^j(t)=0$.

For fixed $t\in{\cal M}_{\DS}$, define
$$
\sigma^*_j(t)=\partial_{t^j}A^*(t),\quad
j=1,\cdots,\dim{\cal M}_{\DS}.
\eqno(4.14)
$$
Since $\SHol_{\DS}\subseteq\ECF^{\phantom{\DS}0,1}_{\DS}$,
$\sigma^*_j(t)\in\ECF^{\phantom{\DS}0,1}_{\DS}$. The $\sigma^*_j(t)$
are analytic, since $A^*(t)$ is. They are also
linearly independent, since $A^*(t)$ defines a gauge slice,
except perhaps on the submanifold of ${\cal M}_{\DS}$ where
$A^*(t)$ is singular. Using the $\sigma^*_j(t)$, one can build
the matrix
$$
F(t,f)^i_j=\langle f^i(t)|\sigma^*_j(t)
\rangle^{\vphantom{\vee}}_{\DS},\quad
i,j=1,\cdots,\dim{\cal M}_{\DS}.
\eqno(4.15)
$$
$F(t,f)$ is analytic on ${\cal M}_{\DS}$.

As second assumption, the map $\zeta(g;t)$ is assumed
to be analytic in both arguments:
$$
\zeta(g;t)^{-1}\bar\partial_g\zeta(g;t)=0,
\eqno(4.16)
$$
$$
\zeta(g;t)^{-1}\bar\partial_t\zeta(g;t)=0.
\eqno(4.17)
$$
As a function of $t$, $\zeta(g;t)$ may develop singularities
on some submanifold of ${\cal M}_{\DS}$ of non zero codimension, where
$\zeta(g;t)$ fails to be a group isomorphism.

For fixed $t\in{\cal M}_{\DS}$, define
$$
\tau_i(t)=\zeta(1;t)^{-1}\partial_{g^i}\zeta(1;t),\quad
i=1,\cdots,\dim{\cal G}_{\DS}.
\eqno(4.18)
$$
$\tau_i(t)\in\ECF^{\phantom{\DS}0,0}_{\DS}$, since
$\Lie{\cal G}_{\DS}(\sans s_t)\subseteq\ECF^{\phantom{\DS}0,0}_{\DS}$.
The $\tau_i(t)$ are analytic, since $\zeta(g;t)$ is. They are also linearly
independent, since $\zeta(g;t)$ is a group isomorphism, except perhaps on
the submanifold of ${\cal M}_{\DS}$ where $\zeta(g;t)$ is singular.
Away from that submanifold, they span $\Lie{\cal G}_{\DS}(\sans s_t)
\cong\ker\bar\partial_{\DS\ssans s_t}$. One then picks vectors
$\{\tau^{\vee i}(t)|i=1,\cdots,\dim{\cal G}_{\DS}\}$ in
$\ECF^{\vee\hphantom{\rm S}1,1}_{\DS}$ defining a basis dual
to $\{\tau_i(t)|i=1,\cdots,\dim{\cal G}_{\DS}\}$ with respect to the
dual pairing $\langle\cdot|\cdot\rangle^{\vphantom{\vee}}_{\DS}$
and depending analytically on $t$. Using the $\tau^{\vee i}(t)$, one can
build the matrix
$$
E(t,e)^i_j=\langle\tau^{\vee i}(t)|e_j(t)
\rangle^{\vphantom{\vee}}_{\DS},\quad
i,j=1,\cdots,\dim{\cal G}_{\DS}.
\eqno(4.19)
$$
$E(t,e)$ does not depend on the choice of the $\tau^{\vee i}(t)$. $E(t,e)$
is clearly analytic on ${\cal M}_{\DS}$.

Let $\nu(g)$ be a left invariant positive
$(\dim{\cal G}_{\DS},\dim{\cal G}_{\DS})$ form on ${\cal G}_{\DS}$.
Hence, $L_f{}^*\nu(g)=\nu(g)$, for any $f\in{\cal G}_{\DS}$. Using $\nu(g)$,
one can define the volume $v_\nu=\int_{{\cal G}_{\DS}}(Dg)_{|g}\nu(g)$
of ${\cal G}_{\DS}$. This is actually divergent, as ${\cal G}_{\DS}$ is non
compact. The gauge fixed partition function ${\cal Z}_\Theta(h)$ reads
$$\eqalignno{
{\cal Z}_\Theta(h)
=&~\int_{{\cal M}_{\DS}}(Dt)_{|t}\big|\det F(t,f)\det E(t,e)\big|^2
{\nu(1)\over v_\nu}\int_{\Herm}(DH)_{h|H}&\cr
\times&~\hat\Theta(h,H,A^*(t))
\exp\big(\hat I(h,H,A^*(t))+\hat I^{\rm gh}(h,H,A^*(t))\big).
\hskip1cm
&(4.20)\cr}
$$
The denominator $v_\nu(t)$ reflects the residual unfixed ${\cal G}_{\DS}$
gauge symmetry, as mentioned earlier. In fact, $\Theta(h,H,A^*)$,
$\hat I(h,H,A^*)$ and $\hat I^{\rm gh}(h,H,A^*)$ are ${\cal G}_{\DS}(\sans s)$
invariant as functionals of $H$, the former two by $\Gau_{\DS}$ invariance,
the latter as a consequence of $(4.10)$--$(4.12)$ and the nilpotence of
${\cal G}_{\DS}(\sans s)$. By $(4.14)$--$(4.15)$, the measure is a
$(\dim{\cal M}_{\DS},\dim{\cal M}_{\DS})$ form on ${\cal M}_{\DS}$ so that
the $t$ integration is well defined. From $(4.10)$--$(4.12)$, $(4.14)$ and
$(4.19)$, it is immediate to see that the measure is independent from the
choice of the bases $\{e_i(t)\}$ and $\{f^j(t)\}$. Gauge invariance ensures
the measure is independent from the choice of the gauge slice $A^*(t)$. It
may also be shown that it is independent from the choice of the group
isomorphism $\zeta(g;t)$. The measure is also independent
from the choice of $\nu$, since left invariance entails that $\nu$ is
determined up to a positive constant. Finally, the measure is independent
from the choice of the coordinates of ${\cal G}_{\DS}$ at $1$, provided
of course one uses the same coordinates for the $\tau(t)_i$ and
$\nu(1)$.

The contribution of the Drinfeld--Sokolov ghosts has a functional
integral representation. Let $\bf G$ be the ghost Grassmann algebra.
The ghost fields are $\beta\in{\bf G}\otimes(\Lie\Gau_{\DS})^\vee$ and
$\gamma\in{\bf G}\otimes\Lie\Gau_{\DS}$. The isomorphisms
$(\Lie\Gau_{\DS})^\vee\cong\ECF^{\vee\hphantom{\rm S}1,0}_{\DS}$
and $\Lie\Gau_{\DS}\cong\ECF^{\phantom{\DS}0,0}_{\DS}$ allow one
to construct the appropriate ghost functional measures
$(D\beta)_{H|\beta}$ and $(D\gamma)_{h,H|\gamma}$.
The Drinfeld--Sokolov ghost action is
$$
S(\beta,\gamma,A^*)
=2\real\langle\beta|\bar\partial_{\DS}\gamma\rangle_{\DS\ssans s}
={1\over\pi}\int_\Sigma d^2z2\real\tr_{\rm ad}
\big(\beta\bar\partial_{\DS}\gamma\big)_{\ssans s}.
\eqno(4.21)
$$
Then, by standard functional techniques, one can show that
$$\eqalignno{
|\det&F(t,f)\det E(t,e)|^2
\exp\hat I^{\rm gh}(h,H,A^*(t))&\cr
=&\int_{{\bf G}\otimes(\Lie\Gau_{\DS})^\vee\times
{\bf G}\otimes\Lie\Gau_{\DS}}
(D\beta)_{H|\beta}\otimes(D\gamma)_{h,H|\gamma}
\exp\big(-S(\beta,\gamma,A^*(t))\big)
&\cr
&\hphantom{\int_{{\bf G}\otimes(\Lie\Gau_{\DS})^\vee\times
{\bf G}\otimes\Lie\Gau_{\DS}}}
\times\Big|\prod_i\langle\beta|\sigma^*_i(t)\rangle^{\vphantom{\vee}}_{\DS}
\prod_j\langle\tau^{\vee j}(t)|\gamma\rangle^{\vphantom{\vee}}_{\DS}\Big|^2.
&(4.22)\cr}
$$

The formal similarities with the construction of the Polyakov measure for
ordinary strings are evident \ref{16--19}. A detailed study of the
Drinfeld--Sokolov ghost system is now in order.
\vskip.4cm
\item{5.} {\bf The Drinfeld--Sokolov Ghost System}
\vskip.4cm
\par
The study of Drinfeld--Sokolov ghost effective action is problematic.
For any Drinfeld--Sokolov holomorphic structure $\sans s\in\SHol_{\DS}$,
the Cauchy--Riemann operator $\bar\partial_{\DS\ssans s}$ acts on the
Drinfeld--Sokolov space $\ECF^{\phantom{\DS}0,0}_{\DS\ssans s}$. However,
the hermitian structure is defined in terms of a metric $H_{\ssans s}\in
\Herm_{\ssans s}$, which does not respect the $\goth x$--valuedness of the
Drinfeld--Sokolov fields, since, for $\Psi_{\ssans s}\in
\ECF^{\phantom{\DS}0,0}_{\DS\ssans s}$, $(\Ad H\Psi^\dagger)_{\ssans s}$
in not $\goth x$--valued in general. This renders the application
of standard field theoretic techniques to the study of the Drinfeld-Sokolov
ghost system impossible. This problem has been solved in a general
context in ref. \ref{29} by using the method of local projectors which
now I shall briefly recall.

Given a metric $H_{\ssans s}\in\Herm_{\ssans s}$, one can introduce the
orthogonal projector $\varpi(H)_{\ssans s}$ of $\ECF^{w,\bar w}_{\ssans s}$
onto $\ECF^{\phantom{\DS}w,\bar w}_{\DS\ssans s}$ with Hilbert structures
corresponding to $H_{\ssans s}$ defined in sect. 2. $\varpi(H)_{\ssans s}$
is given as a collection of local maps $\varpi(H)_{\ssans s a}$ valued
in the endomorphisms of $\goth g$ with range $\goth x$ such that
$\varpi(H)_{\ssans s a}=\Ad\Ad K_{\ssans s ab}\varpi(H)_{\ssans s b}$
whenever defined and that $\varpi(H)_{\ssans s}{}^2=\varpi(H)_{\ssans s}$
and $\big(\Ad H\varpi(H){}^\dagger\Ad H^{-1}\big)_{\ssans s}
=\varpi(H)_{\ssans s}$, where $\varpi(H){}^\dagger$ is the pointwise
adjoint of $\varpi(H)$ with respect to the hermitian inner product on
$\goth g$ defined by $(x,y)=\tr_{\rm ad}(x^\dagger y)$ for
$x,y\in\goth g$.

Recall that the Cauchy Riemann operator $\bar\partial_{\ssans s}$ maps
$\ECF^{\phantom{\DS}w,}_{\DS\ssans s}$ into
$\ECF^{\phantom{\DS}w,1}_{\DS\ssans s}$. It can be shown that this implies
that $\varpi(H)_{\ssans s}$ obeys the relation
$$
\big(\bar\partial\varpi(H)\varpi(H)\big)_{\ssans s}=0,
\eqno(5.1)
$$
Projectors $\varpi(H)_{\ssans s}$ satisfying $(5.1)$ were introduced
earlier in the mathematical literature in the analysis of
Hermitian--Einstein and Higgs bundles \ref{34--35}.

The dependence of $\varpi(H)_{\ssans s}$ is minimal in the sense explained
in sect. 2, i. e. $\varpi(H)_a=\Ad V_{\ssans sab}\varpi(H)_{b\ssans s}
\Ad V_{\ssans sab}{}^{-1}$.

The independence of the range of $\varpi(H)$ from $H$ implies that
$$
\delta\varpi(H)\varpi(H)=0.
\eqno(5.2)
$$
By combining $\Ad H$ hermiticity of $\varpi(H)$ and $(5.2)$, one obtains
$$
\delta\varpi(H)
=-\varpi(H)\ad(\delta HH^{-1})(1-\varpi(H)).
\eqno(5.3)
$$
This identity is a functional differential equation constraining the
dependence of $\varpi(H)$ on $H$ and shows that $\varpi(H)$ is a local
functional of $H$.

Let $H_0$ be a reference fiber metric in $\Herm$. As explained in sect. 3.,
any other fiber metric $H\in\Herm$ can be written as $H=\exp\Phi H_0$,
where the Donaldson field $\Phi$ is an element of $\ECF^{0,0}$ such that
$\Ad H\Phi^\dagger=\Phi$. Using $(5.3)$, it is straightforward to show that
$\varpi(H)$ has a local Taylor expansion in $\Phi$ of the form
$$
\varpi(H)=\sum_{r=0}^\infty{1\over r!}\varpi^{(r)}(\Phi,H_0),
\eqno(5.4)
$$
where, for each $r\geq 0$, $\varpi^{(r)}(\Phi,H_0)$ transforms as $\varpi(H)$
under coordinate changes and is a homogeneous degree $r$ polynomial in $\Phi$:
$$
\eqalignno{
\varpi^{(0)}(\Phi,H_0)
=&~\varpi(H_0),
&\cr
\varpi^{(1)}(\Phi,H_0)=&~
-\varpi(H_0)\ad\Phi(1-\varpi(H_0)),
&\cr
\varpi^{(2)}(\Phi,H_0)=
&~\varpi(H_0)\ad\Phi(1-2\varpi(H_0))\ad\Phi(1-\varpi(H_0)),
&\cr
\varpi^{(3)}(\Phi,H_0)=
&~\varpi(H_0)
\big[\ad\Phi(3\varpi(H_0)-1)\ad\Phi(1-\varpi(H_0))\ad\Phi&\cr
&+\ad\Phi(2-3\varpi(H_0))\ad\Phi\varpi(H_0)\ad\Phi\big]
(1-\varpi(H_0)),
&\cr
&\vdots.
&(5.5)\cr}
$$

It is not difficult to show that the projector $\varpi(H(h))$
corresponding to the metric $H(h)$ given in $(3.15)$ is given by
$$
\varpi(H(h))
=\exp(-\partial\ln h\ad t_{-1})p_{\goth x}\exp(\partial\ln h\ad t_{-1}),
\eqno(5.6)
$$
where $p_{\goth x}$ is the orthogonal projector of $\goth g$ onto $\goth x$
with respect to the hermitian inner product $(\cdot,\cdot)$ of $\goth g$
defined above.

Next, consider the $\Gau_{\DS}$ invariant unrenormalized
Drinfeld--Sokolov ghost
effective action $\hat I^{\rm gh}(h,H,A^*)$ with $A^*\in\SHol_{\DS}$ a
Drinfeld--Sokolov holomorphic structure. Because of the unboundedness of the
ghost kinetic operator $\bar\partial_{\DS\ssans s}$,
$\hat I^{\rm gh}(h,H,A^*)$ suffers ultraviolet divergencies which have
to be regularized by means of an ultraviolet cut--off $\epsilon$. As in sect.
3, I shall adopt here proper time regularization \ref{18}.
Next, I shall analyze the main properties of this effective action.
It turns out that the Drinfeld--Sokolov ghost system is not
a Drinfeld--Sokolov field theory of the type discussed
in sect. 3. In spite of this, it shares many of the qualitative
features of a Drinfeld--Sokolov field theory, as is shown below.

Using the methods of \ref{29}, it can be seen that
$\hat I^{\rm gh}(h,H,A^*;\epsilon)$ has the following expansion as
$\epsilon\rightarrow 0$:
$$\eqalignno{
&\hat I^{\rm gh}(h,H,A^*;\epsilon)
=-{r^{\rm gh}\over\pi\epsilon}\int_\Sigma d^2zh+
\bigg[{-r^{\rm gh}\over 6\pi}\int_\Sigma d^2zf_h
&\cr
&
+{1\over 2\pi}\int_\Sigma d^2z
\tr\big((\ad F_H+\bar\partial\partial_H\varpi(H))\varpi(H)\big)_{\ssans s}
-q\bigg]\ln\epsilon+I^{\rm gh}_0(h,H,A^*)+O(\epsilon).
&(5.7)\cr}
$$
Here, $r^{\rm gh}=\dim\goth x$ and $q=\dim{\cal G}_{\DS}$.
$\partial_H\varpi(H)=\partial\varpi(H)-[\ad\Gamma_H,\varpi(H)]$.
The first two terms of the coefficient of $\ln\epsilon$ are topological
invariants. In fact,
$\int_\Sigma d^2zf_h=2\pi(\ell-1)$ is the Guass--Bonnet invariant, already
encountered in sect. 3, and
$\int_\Sigma d^2z\tr\big((\ad F_H+\bar\partial\partial_H\varpi(H))\varpi(H)
\big)_{\ssans s}=-2\pi\tr[\ad t_0p_{\goth x}](\ell-1)$
is the Chern--Weil invariant of $DS$, where $p_{\goth x}$ is defined
below $(5.6)$. $I^{\rm gh}_0(h,H,A^*)$ is a non local functional of
$h$, $H$ and $A^*$ such that
$$
\eqalignno{
&\delta I^{\rm gh}_0(h,H,A^*)
={r^{\rm gh}\over 6\pi}\int_\Sigma d^2z\delta\ln hf_h
&\cr
&-{1\over 2\pi}\int_\Sigma d^2z\Big[\delta\ln h
\tr\big((\ad F_H+\bar\partial\partial_H\varpi(H))\varpi(H)\big)_{\ssans s}
+\tr\big(\ad(\delta HH^{-1})\varpi(H)\big)_{\ssans s}f_h\Big]&\cr
&+{1\over\pi}\int_\Sigma d^2z
\tr\big(\ad(\delta HH^{-1})(\ad F_H+\bar\partial\partial_H\varpi(H))
\varpi(H)\big)_{\ssans s}.
&(5.8)\cr}
$$
where $\delta$ denotes variation with respect to $h$ and $H$ at fixed $A^*$.

To renormalize the bare effective action $\hat I^{\rm gh}(h,H,A^*;\epsilon)$,
one has to add to it a counterterm of the form
$$
\eqalignno{
&\Delta\hat I^{\rm gh}(h,H,A^*;\epsilon)
={r^{\rm gh}\over\pi\epsilon}\int_\Sigma d^2zh-
\bigg[{-r^{\rm gh}\over 6\pi}\int_\Sigma d^2zf_h&\cr
&
+{1\over 2\pi}\int_\Sigma d^2z
\tr\big((\ad F_H+\bar\partial\partial_H\varpi(H))\varpi(H)\big)_{\ssans s}
-q\bigg]\ln\epsilon+\Delta I^{\rm gh}(h,H,A^*)+O(\epsilon),~~
&(5.9)\cr}
$$
Here, $\Delta I^{\rm gh}(h,H,A^*)$ is a local but otherwise arbitrary
functional of $h$, $H$ and $A^*$, whose choice defines a renormalization
prescription, as in Drinfeld--Sokolov field theory. The renormalized
effective action is thus
$$
I^{\rm gh}(h,H,A^*)
=I^{\rm gh}_0(h,H,A^*)+\Delta I^{\rm gh}(h,H,A^*).
\eqno(5.10)
$$

Below, $\Delta I^{\rm gh}(h,H,A^*)$ is assumed to be independent from
$A^*$:
$$
\Delta I^{\rm gh}(h,H,A^*)=\Delta I^{\rm gh}(h,H).
\eqno(5.11)
$$
It can be shown that, if this condition is fulfilled,
$I^{\rm gh}(h,H,A^*)$ has the following structure
$$
I^{\rm gh}(h,H,A^*)=I^{\rm gh}(h,H)+L^{\rm gh}(H,A^*;A,\rho)
+I^{\rm gh}_{\rm hol}(A^*;A,\rho).
\eqno(5.12)
$$
Here, $A\in\Conn_{\DS}$ is a background Drinfeld--Sokolov
$(1,0)$ connection. $\rho$ is a background local
projector on $\goth x$. In analogy to $\varpi(H)$, $\rho$ is given as a
collection of maps $\rho_a$ valued in the endomorphisms of $\goth g$ with
range $\goth x$ such that $\rho_a=\Ad\Ad K_{ab}\rho_b$ whenever defined and
that $\rho{}^2=\rho$. $I^{\rm gh}(h,H)$ is the functional
$I^{\rm gh}(h,H,A^*)$ evaluated at the reference holomorphic structure
$A^*=0$.
$$\eqalignno{
L^{\rm gh}(H,A^*;A,\rho)
=&~{1\over\pi}\int_\Sigma d^2z\Big[
2\real\tr\big((\varpi(H)\ad\Gamma_H-\rho\ad A)\ad A^*\big)&\cr
&~\phantom{{1\over\pi}\int_\Sigma d^2z\Big[
2\real\tr\big((\varpi(H)}
-\tr\big(\ad A^*\varpi(H)\ad\Ad HA^{*\dagger}\big)\Big].
&(5.13)\cr}
$$
Using $(2.15)$ and the fact that $\goth x$ is a nilpotent subalgebra
of $\goth g$ such that $[t_d,\goth x]\subseteq\goth x$ for $d=0,-1$,
it is straightforward to verify that the integrand
belongs to $\CF^{1,1}$ so that the integration can be carried
out. $I^{\rm gh}_{\rm hol}(A^*;A,\rho)$ is a non local functional of
$A^*$ depending on $A$ and $\rho$. Next, I shall study the
properties of the three contributions in the right hand side of
$(5.12)$.

In order the counterterm $\Delta\hat I^{\rm gh}(h,H,A^*;\epsilon)$ to be
$\Gau_{\DS}$ invariant, $\Delta I^{\rm gh}(h,H)$ must satisfy
$$
s_{\DS}\Delta I^{\rm gh}(h,H)=0.
\eqno(5.14)
$$
This ensures that the renormalized effective action $I^{\rm gh}(h,H,A^*)$ is
also $\Gau_{\DS}$ invariant. Under this assumption, one has
$$
s_{\DS}I^{\rm gh}(h,H)
={\cal W}^{\rm gh}_{\DS}(H),
\eqno(5.15)
$$
$$
s_{\DS}L^{\rm gh}(H,A^*;A,\rho)
=-{\cal W}^{\rm gh}_{\DS}(H)-{\cal A}^{\rm gh}_{\DS}(A^*;A,\rho),
\eqno(5.16)
$$
$$
s_{\DS}I^{\rm gh}_{\rm hol}(A^*;A,\rho)
={\cal A}^{\rm gh}_{\DS}(A^*;A,\rho),
\eqno(5.17)
$$
where
$$
{\cal W}^{\rm gh}_{\DS}(H)={1\over\pi}\int_\Sigma d^2z
2\real\tr\big(\ad\Xi_{\DS}\bar\partial(\varpi(H)\ad\Gamma_H)\big),
\eqno(5.18)
$$
$$
{\cal A}^{\rm gh}_{\DS}(A^*;A,\rho)
=-{1\over\pi}\int_\Sigma d^2z
2\real\tr\big(\ad\Xi_{\DS}(\bar\partial(\rho\ad A)
-\partial(\ad A^*\rho)+[\rho\ad A,\ad A^*\rho])\big)
\eqno(5.19)
$$
are the ghost gauge anomalies. Using $(2.15)$ and the properties of
$\goth x$ recalled below $(5.13)$, it is straightforward to verify
that the integrand belongs to $\CF^{1,1}$ so that the integration can be
carried out. As a check, I have verified that the restriction of
${\cal A}^{\rm gh}_{\DS}(A^*;A,\rho)$ to $\Lie{\cal G}_{\DS}(\sans s)$
vanishes as it should.

$I^{\rm gh}(h,H)$ is a non local functional of $h$ and $H$. Its dependence on
$h$ and $H$ can be analyzed as follows. Using the fiber metric $H(h)$ defined
in $(3.15)$, one has
$$
I^{\rm gh}(h,H)=I^{\rm gh}_{\rm conf}(h)+S^{\rm gh}(h,H)+F(h,H,H(h))
+\Delta I^{\rm gh}(h,H)-\Delta I^{\rm gh}(h,H(h)),
\eqno(5.20)
$$
where
$$
I^{\rm gh}_{\rm conf}(h)=I^{\rm gh}(h,H(h)),
\eqno(5.21)
$$
$$
S^{\rm gh}(h,H)=\Omega^{\rm gh}(h,H(h)).
\eqno(5.22)
$$
Here, for any two metrics $H,H_0\in\Herm$, $\Omega^{\rm gh}(H,H_0)$
is the Drinfeld--Sokolov generalization of the Donaldson action
defined by functional path integral
$$
\Omega^{\rm gh}(H,H_0)
={1\over\pi}\int_{H_0}^H\int_\Sigma d^2z
\tr\big(\ad(\delta H'H'^{-1})(\ad F_{H'}+\bar\partial\partial_{H'}\varpi(H'))
\varpi(H')\big).
\eqno(5.23)
$$
$F(h,H,H_0)$ is the functional
$$
F(h,H,H_0)
={-1\over 2\pi}\int_{H_0}^H\int_\Sigma d^2z\tr\big(
\ad(\delta H'H'^{-1})\varpi(H')\big))f_h.
\eqno(5.24)
$$
The right hand sides of $(5.23)$ and $(5.24)$
are both independent from the choice of the
functional integration path joining $H_0$ to $H$, since
the functional $1$--forms on $\Herm$ integrated are closed and $\Herm$
is contractible. This can easily be verified using $(5.1)$ and $(5.3)$.
$\Omega^{\rm gh}(H,H_0)$ can be computed in terms of the Donaldson field
$\Phi$ of $H$ relative to $H_0$ by using the local Taylor expansion
$(5.4)$--$(5.5)$ of $\varpi(H)$. The result is
$$
\Omega^{\rm gh}(H,H_0)
=-{1\over\pi}\int_\Sigma d^2z
\tr\Big[K^*(\Phi,\bar\partial\Phi,H_0)\ad\partial_{H_0}\Phi
-D(\Phi,H_0)\ad F_{H_0}+T(\Phi,H_0)\Big],
\eqno(5.25)
$$
where
$$\eqalignno{
D(\Phi,H_0)
=&\sum_{m=0}^\infty{1\over(m+1)!}\sum_{n=0}^m{m\choose n}
\varpi^{(m-n)}(\Phi,H_0)\ad\Phi\varpi^{(n)}(\Phi,H_0),
&(5.26)\cr
T(\Phi,H_0)
=&\sum_{m=0}^\infty{1\over(m+1)!}\sum_{n=0}^m{m\choose n}
\partial_{H_0}\varpi^{(m-n)}(\Phi,H_0)\ad\Phi
\bar\partial\varpi^{(n)}(\Phi,H_0),~~
&(5.27)\cr
K^*(\Phi,\bar\partial\Phi,H_0)
=&\sum_{m=0}^\infty{1\over(m+2)!}\sum_{n=0}^m{m+1\choose n}
(-\ad\ad\Phi)^{m-n}&\cr
\times&\sum_{k=0}^n{n\choose k}\bar\partial
\big(\varpi^{(n-k)}(\Phi,H_0)\ad\Phi\big)\varpi^{(k)}(\Phi,H_0).
&(5.28)\cr}
$$
By a similar and simpler calculation, one finds
$$
F(h,H,H_0)
={-1\over 2\pi}\int_\Sigma d^2z\tr J(\Phi,H_0)f_h,
\eqno(5.29)
$$
where
$$
J(\Phi,H_0)=\sum_{r=0}^\infty{1\over(r+1)!}\ad\Phi\varpi^{(r)}(\Phi,H_0).
\eqno(5.30)
$$

Now, $I^{\rm gh}_{\rm conf}(h)$ is a non local functional of $h$. By using
$(3.15)$, $(5.6)$ and $(5.8)$, one can obtain the variational relation
obeyed by $I^{\rm gh}_{\rm conf}(h)$. This can be written in rather explicit
form, because of the simple dependence of $H(h)$ and $\varpi(H(h))$ on
$h$. By a somewhat lengthy but straightforward calculation, one finds
$$
\delta I^{\rm gh}_{\rm conf}(h)=
-{\kappa^{\rm gh}\over 12\pi}\int_\Sigma d^2z \delta\ln h f_h
+\delta\bigg[{\lambda^{\rm gh}\over\pi}\int_\Sigma d^2zh^{-1}f_h{}^2
+\Delta I^{\rm gh}(h,H(h))\bigg],
\eqno(5.31)
$$
where
$$
\kappa^{\rm gh}=-2\tr\big[\big(6(\ad t_0)^2+6\ad t_0+1\big)p_{\goth x}\big],
\eqno(5.32)
$$
$$
\lambda^{\rm gh}=-\tr\big(\ad t_{+1}\ad t_{-1}p_{\goth x}\big).
\eqno(5.33)
$$
If
$$
\Delta I^{\rm gh}(h,H)=\Delta'I^{\rm gh}(h,H)
+{\lambda^{\rm gh}_0\over\pi}\int_\Sigma d^2zh^{-1}f_h{}^2,
\eqno(5.34)
$$
where $\lambda^{\rm gh}_0$ is some constant and $\Delta'I^{\rm gh}(h,H)$
is a local functional of $h$ and $H$ such that
$$
\Delta'I^{\rm gh}(h,H(h))=0,
\eqno(5.35)
$$
$(5.31)$ becomes simply
$$
\delta I^{\rm gh}_{\rm conf}(h)=
-{\kappa^{\rm gh}\over 12\pi}\int_\Sigma d^2z \delta\ln h f_h
+{\lambda^{\rm gh}_0+\lambda^{\rm gh}\over\pi}
\delta\int_\Sigma d^2zh^{-1}f_h{}^2.
\eqno(5.36)
$$
A counterterm $\Delta I^{\rm gh}(h,H,A^*)$ for which $(5.34)$
holds is given by the right hand side of $(5.34)$ with
$\Delta'I^{\rm gh}(h,H)$ satisfying $(5.14)$ and $(5.35)$
and clearly satisfies both $(5.11)$ and $(5.14)$.
Choosing $\lambda^{\rm gh}_0=-\lambda^{\rm gh}$ yields a renormalized
effective action $I^{\rm gh}_{\rm conf}(h)$ describing a conformal field
theory of central charge $\kappa^{\rm gh}_{\rm conf}=\kappa^{\rm gh}$.
{\it This is precisely the central charge of the Drinfeld--Sokolov ghost
system of the $W$--algebra associated to the pair $(G,S)$ as computed with
the methods of hamiltonian reduction and conformal field theory} \ref{10}
\footnote{${}^2$}{The odd looking sign of the mid term in the right hand
side of $(5.32)$ is due to the fact that $\goth x$ is negative graded.}.
For a generic value of $\lambda_0$, one obtains a renormalized effective
action with a $\int\sqrt h R_h{}^2$ term yielding a model of induced $2d$
gravity of the same type as that considered in refs. \ref{32-33}, as
in sect. 3.

The functional $S^{\rm gh}(h,H)$ and $F(h,H,H(h))$ are local. In fact,
the Donaldson field relevant here is $\Phi(h,H)$, defined in $(3.26)$.
{}From $(5.6)$, the locality of $\Phi(h,H)$ as a functional of $h$ and $H$
and eqs. $(5.25)$--$(5.30)$ showing that $\Omega^{\rm gh}(H,H_0)$ and
$F(h,H,H_0)$ are local functionals of $\Phi$ and $H_0$, the statement is
evident.

{}From the above discussion, it follows that {\it the suitably renormalized
Drinfeld--Sokolov ghost effective action $I^{\rm gh}(h,H)$ differs from the
conformal effective action $I^{\rm gh}_{\rm conf}(h)$ by a local functional of
$h$ and $H$. In particular, the $H$ dependence is local}.

{}From $(5.13)$, it appears that $L^{\rm gh}(H,A^*;A,\rho)$, the interaction
term of $H$ and $A^*$, is local.

It is also likely, though no proof is available at present, that
$I^{\rm gh}_{\rm hol}(A^*;A,\rho)$ is the real part of a holomorphic
functional of $A^*$ and $A$ and $\rho$, entailing holomorphic
factorization. Its crucial property, however, is its independence from
$H$.

One has thus reached the following important conclusion. {\it The full
suitably renormalized $\Gau_{\DS}$ invariant Drinfeld--Sokolov ghost
effective action $I^{\rm gh}(h,H,A^*)$ is a local functional of $H$}.

One could choose $\Delta'I^{\rm gh}(h,H)=0$ above. There is however a
different more interesting choice, namely
$$
\Delta'I^{\rm gh}(h,H)=-F(h,H,H(h)).
\eqno(5.37)
$$
Using $(5.3)$ and $(3.15)$, one can show that
$$\eqalignno{
\delta\Delta'I^{\rm gh}(h,H)
=&~{1\over 2\pi}\int_\Sigma d^2z\Big[\delta\ln h
\tr\big((\ad F_H+\bar\partial\partial_H\varpi(H))\varpi(H)\big)&\cr
&\phantom{{1\over 2\pi}\int_\Sigma d^2z\Big[\delta\ln h
\tr\big(}
+\tr\big(\ad(\delta HH^{-1})\varpi(H)\big)f_h\Big]&\cr
-&~{1\over 2\pi}\int_\Sigma d^2z\Big[\delta\ln h
\tr\big((\ad F_{H_0}+\bar\partial\partial_{H_0}\varpi(H_0))\varpi(H_0)\big)&\cr
&\phantom{{1\over 2\pi}\int_\Sigma d^2z\Big[\delta\ln h
\tr\big(}
+\tr\big(\ad(\delta H_0H_0{}^{-1})\varpi(H_0)\big)f_h\Big]\Big|_{H_0=H(h)}.
&(5.38)\cr}
$$
Hence, the counterterm $\Delta'I^{\rm gh}(h,H)$ has the nice property
of cancelling the mid term of $(5.8)$ separating the $\delta\ln h$ and
$\delta HH^{-1}$ terms in $\delta I^{\rm gh}(h,H,A^*)$.
\vskip.4cm
\item{6.} {\bf Conformal Invariance}
\vskip.4cm
\par
Let us go back to eq. $(4.20)$ providing the expression of the gauge
fixed partition function ${\cal Z}_\Theta(h)$. Here, I shall assume that
the insertion $\hat\Theta(h,H,A^*)$ contains only the counterterms necessary
to absorb the ultraviolet divergencies of the bare effective actions
$\hat I(h,H,A^*)$ and $\hat I^{\rm gh}(h,H,A^*)$. Thus, $\hat\Theta(h,H,A^*)$
has the structure
$$
\hat\Theta(h,H,A^*)
=\exp\big(\Delta\hat I(h,H,A^*)+\Delta\hat I^{\rm gh}(h,H,A^*)\big)
\hat\theta(h,A^*),
\eqno(6.1)
$$
where $\Delta\hat I(h,H,A^*)$ and $\Delta\hat I^{\rm gh}(h,H,A^*)$
are given by $(3.4)$ and $(5.9)$ in the proper time regularization scheme
and $\hat\theta(h,A^*)$ is a $\Gau_{\DS}$ invariant functional of $h$ and
$A^*$. Then, after cancellation of matter and ghost ultraviolet divergencies,
$(4.20)$ may be written as
$$
{\cal Z}(h)=\int_{{\cal M}_{\DS}}(Dt)_{|t}
\big|\det F(t,f)\det E(t,e)\big|^2{\nu(1)\over v_\nu}
\hat\theta(h,A^*(t)){\cal Z}^{\rm herm}(h,A^*(t)),
\eqno(6.2)
$$
where
$$
{\cal Z}^{\rm herm}(h,A^*)
=\int_{\Herm}(DH)_{h|H}\exp I^{\rm tot}(h,H,A^*),
\eqno(6.3)
$$
$$
I^{\rm tot}(h,H,A^*)=I(h,H,A^*)+I^{\rm gh}(h,H,A^*).
\eqno(6.4)
$$
The problem to tackle next is the study of the partition
function ${\cal Z}^{\rm herm}(h,A^*)$. By the discussion of sects. 3 and 5,
the underlying $H$ field theory is local.

Before proceeding, an important remark is in order. Using the results
of sect. 3 of ref. \ref {29}, it is easy to show that, for fixed $\sans s
\equiv A^*\in\SHol_{\DS}$, the action $I^{\rm tot}(h,H,A^*)$ is
invariant under the subgroup ${\cal G}_{\DS}'(\sans s)$ of
$\exp\goth n_{\goth x}$--valued elements of ${\cal G}(\sans s)$, where
$\goth n_{\goth x}$ is the normalizer $\goth x$. ${\cal G}_{\DS}'(\sans s)$
is larger than ${\cal G}_{\DS}(\sans s)$.
For varying $\sans s\in\SHol_{\DS}$, the groups ${\cal G}_{\DS}'(\sans s)$
are all isomorphic to the same complex Lie group ${\cal G}_{\DS}'$
containing ${\cal G}_{\DS}$. Therefore, even after formally
dividing by the volume $v_\nu$ of ${\cal G}_{\DS}$, the partition function
${\cal Z}^{\rm herm}(h,A^*)$ is still divergent. This problem can be solved
either by insertions that break the extra gauge symmetry or by further
gauge fixing. The following analysis of conformal invariance
is not affected by this.

In the method used here, the $H$ functional integration is viewed as the
integration on a suitable manifold of classical $H$ configurations times
the functional integration on the quantum $H$ fluctuations around each of
the corresponding $H$ vacua.

The classical action for the $H$ field is $I^{\rm tot}(h,H,A^*)$.
The classical $H$ equation obtained from $I^{\rm tot}(h,H,A^*)$ is
$$
\Big[F_H+K^{-1}\Pi(H)\big(\ad F_H+\bar\partial\partial_H\varpi(H)\big)
\varpi(H)\Big]_{\ssans s}=0.
\eqno(6.5)
$$
Here, $\Pi(H)_{\ssans s}$ is defined as follows. Consider the real vector
space of local fields $\tilde X_{\ssans s}$ valued in the endomorphisms
of $\goth g$ such that $\tilde X_{\ssans s a}=\Ad\Ad K_{\ssans s ab}
\tilde X_{\ssans s b}$ and that $(\Ad\Ad H\tilde X^\dagger)_{\ssans s}
=\tilde X_{\ssans s}$, equipped with the pointwise Hilbert norm
$\tilde X_{\ssans s}\rightarrow\tr(\tilde X^2)_{\ssans s}$. Let
$\tilde\Pi(H)_{\ssans s}$ be the orthogonal projector of such space onto
its subspace of elements $\tilde X_{\ssans s}$ of the form
$\tilde X_{\ssans s}=\ad X_{\ssans s}$ for some local field $X_{\ssans s}$
such that $X_{\ssans s a}=\Ad K_{\ssans s ab}X_{\ssans s b}$
and that $(\Ad HX^\dagger)_{\ssans s}=X_{\ssans s}$. $\tilde
\Pi(H)_{\ssans s}$ is a field valued in the endomorphisms of space of
endomorphisms of $\goth g$ such that
$\tilde \Pi(H)_{\ssans s a}=\Ad\Ad\Ad K_{\ssans sab}
\tilde \Pi(H)_{\ssans s b}$ and depending locally on $H_{\ssans s}$ since
the $H_{\ssans s}$ hermiticity condition is local. Since $\goth g$ is
simple, the adjoint representation $\ad$ is faithful so that $\ad^{-1}$ is
defined. By definition, $\Pi(H)_{\ssans s}=\ad^{-1}\tilde\Pi(H)_{\ssans s}$.
$(6.5)$ is easily obtained by using the variational identities $(3.3)$,
$(5.8)$ and $(5.38)$.
I do not have any proof that eq. $(6.5)$ admit solutions. I shall assume
anyway that solutions exists.

Eq. $(6.5)$ does not contain the surface metric $h$. It is therefore
conformally invariant. This is a consequence of the renormalization
prescription of the Drinfeld--Sokolov ghost sector used corresponding
to the choice $(5.37)$ of the finite part of the ghost counterterm.

The general solution of eq. $(6.5)$ is a function $H_{\rm cl}(n;\sans s)$
depending on $\sans s$ of a set of parameters $n$ varying in some finite
dimensional real manifold $\cal N$. The $n$ label the different solutions.
For fixed $\sans s$, the metrics $H_{\rm cl}(n;\sans s)$ span a finite
dimensional submanifold $\Herm_{\rm cl}(\sans s)$ of $\Herm$.

Since $I^{\rm tot}(h,H,A^*)$ is ${\cal G}_{\DS}'(\sans s)$ invariant,
if $\eta\in{\cal G}_{\DS}'(\sans s)$ and $H\in\Herm_{\rm cl}(\sans s)$,
then also $\eta^*H\in\Herm_{\rm cl}(\sans s)$.
So, the space of solutions of eq. $(6.5)$ for fixed $\sans s$ is
${\cal G}_{\DS}'(\sans s)$ invariant. There exists therefore a free action
$n\rightarrow\vphantom{n}^gn$, $g\in{\cal G}_{\DS}'$, of ${\cal G}_{\DS}'$
on $\cal N$ such that $H_{\rm cl}(\vphantom{n}^gn;\sans s)=\zeta'(g;\sans s)^*
H_{\rm cl}(n;\sans s)$ for some isomorphism $\zeta'(\cdot;\sans s):
{\cal G}'_{\DS}\rightarrow{\cal G}'_{\DS}(\sans s)$.

To carry out the functional integration of the $H$ quantum fluctuations
around the classical vacua, one needs a fibration $\varphi(\cdot;\sans s):
\Herm\rightarrow{\cal N}$ depending parametrically on a
holomorphic structure $\sans s$. The fibration yields a parametrization of
$\Herm$ of the form
$$
H(\Phi,n;\sans s)=\exp\Phi H_{\rm cl}(n;\sans s),
\eqno(6.6)
$$
where $n\in{\cal N}$ and $\Phi\in\ECF^{0,0}$ with
$\Ad H_{\rm cl}(n;\sans s)\Phi{}^\dagger=\Phi$ subject to the constraint
that $\exp\Phi H_{\rm cl}(n;\sans s)\in\varphi^{-1}(n;\sans s)$.
Such Donaldson fields $\Phi$ form a real manifold obviously
isomorphic to $\varphi^{-1}(n;\sans s)H_{\rm cl}(n;\sans s)^{-1}$.

The fibration $\varphi(n;\sans s)$ must have the following
properties. For any $n\in{\cal N}$ and any $H\in
\varphi^{-1}(n;\sans s)$, $T_H\Herm=T_H\varphi^{-1}(n;\sans s)\oplus
{\cal H}_H(n;\sans s)$, where ${\cal H}_H(n;\sans s)$
is some subspace of $T_H\Herm$ of dimension equal to that of
$\cal N$ and the direct sum is orthogonal with respect to the
Hilbert structure in $\Herm$ (cf. app. A). Further,
$T_{H(\Phi,n;\ssans s)}\varphi^{-1}(n;\sans s)
=\exp(\ad\Phi/2)T_{H_{\rm cl}(n;\ssans s)}\varphi^{-1}(n;\sans s)$
and ${\cal H}_{H(\Phi,n;\ssans s)}(n;\sans s)
=\exp(\ad\Phi/2){\cal H}_{H_{\rm cl}(n;\ssans s)}(n;\sans s)$.
Finally, one has ${\cal H}_{H_{\rm cl}(n;\ssans s)}(n;\sans s)
=T_{H_{\rm cl}(n;\ssans s)}\Herm_{\rm cl}(\sans s)$.

The fibration $\varphi(n;\sans s)$ must also be ${\cal G}_{\DS}'$
covariant, i. e. $\varphi^{-1}(\vphantom{n}^gn;\sans s)=
\zeta'(g;\sans s)^*\varphi^{-1}(n;\sans s)$ for any
$g\in{\cal G}_{\DS}'$. This implies the ${\cal G}_{\DS}'$
covariance of the parametrization $(6.6)$, being
$H(\zeta'(g;\sans s)^*\Phi,\vphantom{n}^gn)=\zeta'(g;\sans s)^*H(\Phi,n)$.
One must also have that
$T_{\zeta'(g;\ssans s)^*H}\varphi^{-1}(\vphantom{n}^gn;\sans s)
=\zeta'(g;\sans s)^*T_H\varphi^{-1}(n;\sans s)$ and
${\cal H}_{\zeta'(g;\ssans s)^*H}(\vphantom{n}^gn;\sans s)
=\zeta'(g;\sans s)^*{\cal H}_H(n;\sans s)$.

One clearly has the isomorphism $\Herm\cong{\cal N}\times
\varphi^{-1}(\cdot;\sans s)$, where ${\cal N}\times
\varphi^{-1}(\cdot;\sans s)=\prod_{n\in{\cal N}}\{n\}\times
\varphi^{-1}(n;\sans s)$. One can use the isomorphism to transform
the functional integration on $\Herm$ into one on ${\cal N}\times
\varphi^{-1}(\cdot;\sans s)$. To this end, one has to provide
$\cal N$ and each $\varphi^{-1}(n;\sans s)$ with the appropriate
real Hilbert structure and construct the corresponding functional
measures $(Dn)_{|n}$ and $(D\Phi)_{h,H_{\rm cl}(n;\ssans s)|\Phi}$.
Details may be found in app. B.

Using the fibration $\varphi(\cdot;\sans s)$, the partition
function ${\cal Z}^{\rm herm}(h,A^*)$ can be written as
$$
{\cal Z}^{\rm herm}(h,A^*)
=\int_{{\cal N}}(Dn)_{|n}\exp I^{\rm tot}(h,H_{\rm cl}(n;\sans s),A^*)
{\cal Z}^{\rm herm}_{\rm qu}(h,A^*;n),
\eqno(6.7)
$$
where
$$\eqalignno{
{\cal Z}^{\rm herm}_{\rm qu}(h,A^*;n)
=&~[\det J(h,A^*;n)]^{1\over 2}
\int_{{\varphi^{-1}}(n;\ssans s)H_{\rm cl}(n;\ssans s)^{-1}}
(D\Phi)_{h,H_{\rm cl}(n;\ssans s)|\Phi}&\cr
&~\hphantom{[\det J(h,A^*;n)]^{1\over 2}}
\times\exp I^{\rm tot}_{\rm qu}(\exp\Phi H_{\rm cl}(n;\sans s),A^*;n),
&(6.8)\cr}
$$
$$\eqalignno{
J(h,A^*;n)_{rs}=
&~\langle\partial_{n^r}H_{\rm cl}(n;\sans s)H_{\rm cl}(n;\sans s)^{-1},
\partial_{n^s}H_{\rm cl}(n;\sans s)H_{\rm cl}(n;\sans s)^{-1}
\rangle^{\vphantom{\vee}}_{h,H_{\rm cl}(n;\ssans s)},&\cr
&~\quad r,s=1,\cdots,\dim{\cal N},
&(6.9)\cr}
$$
$$
I^{\rm tot}_{\rm qu}(H,A^*;n)
=I^{\rm tot}(h,H,A^*)-I^{\rm tot}(h,H_{\rm cl}(n;\sans s),A^*).
\eqno(6.10)
$$
$I^{\rm tot}_{\rm qu}(H,A^*;n)$ is the quantum fluctuation action.
The independence of $I^{\rm tot}_{\rm qu}(H,A^*;n)$ from $h$ follows
straightforwardly from $(6.4)$, $(3.5)$, $(5.10)$, $(3.3)$, $(3.24)$,
$(5.8)$, $(5.34)$ and $(5.38)$. Details about the derivation of this
formula are provided in app. B. $(6.7)$ may be cast in more suggestive form
as follows.

Define
$$\eqalignno{
V^{\rm tot}(h,H,A^*;A,\rho)
=&~I^{\rm tot}(h,H,A^*)
-I_{\rm hol}(A^*;A)-I^{\rm gh}_{\rm hol}(A^*;A,\rho)&\cr
=&~I(h,H)+I^{\rm gh}(h,H)+L(H,A^*;A)+L^{\rm gh}(H,A^*;A,\rho)
&(6.11)\cr}
$$
(cf. eqs. $(6.4)$, $(3.7)$ and $(5.12)$). Now, for a fixed $A^*$, one can
impose the constraint $\delta V^{\rm tot}(h,H,A^*;A,\rho)/
\delta A^*=0$ on the solutions of eq. $(6.5)$. This can be written in the
form
$$
\Big[\Gamma_H-A
+K^{-1}\Pi^{\rm c}(H)(\varpi(H)\ad\Gamma_H-\rho\ad A)\Big]_{\ssans s}=0
\quad {\rm in}~\ECF^{\vee\hphantom{\rm S}1,0}_{\DS\ssans s}.
\eqno(6.12)
$$
Here, $\Pi^{\rm c}(H)$ is defined similarly to $\Pi(H)$
below $(6.5)$, by considering instead the complex vector space of local
fields $\tilde Z_{\ssans s}$ valued in the endomorphisms of $\goth g$ such
that $\tilde Z_{\ssans s a}=\Ad\Ad K_{\ssans s ab}\tilde Z_{\ssans s b}$
equipped with the pointwise Hilbert norm $\tilde Z_{\ssans s}
\rightarrow\tr(\Ad\Ad H\tilde Z^\dagger\tilde Z)_{\ssans s}$.
The above equation depends on the background fields $A$ and $\rho$
at order $O(K^{-1})$, except when the grading of $\goth g$ induced
by $\goth s$ is integer. Below, I assume that, for any $\sans s\in
\SHol_{\DS}$, there are common solutions of the dynamical equation
$(6.5)$ and the constraint $(6.12)$ at least for some choice $A_0$ and
$\rho_0$ of the backgrounds. I further assume that such solutions are
of the form $H_{\rm cl}(n;\sans s)$ for $n$ varying in some submanifold
${\cal N}_{\DS}$ of $\cal N$.

$V^{\rm tot}(h,H,A^*;A,\rho)$ is ${\cal G}_{\DS}'(\sans s)$ invariant,
as $I^{\rm tot}(h,H,A^*)$, $I_{\rm hol}(A^*;A)$ and
$I^{\rm gh}_{\rm hol}(A^*;A,\rho)$ are. Hence, if
$\eta\in{\cal G}_{\DS}'(\sans s)$ and $H$ satisfies $(6.12)$,
then also $\eta^*H$ does. So, ${\cal N}_{\DS}$ is invariant under the
action of ${\cal G}_{\DS}'$ on $\cal N$ defined earlier.

Consider the classical action $I^{\rm tot}(h,H_{\rm cl}(n;\sans s),A^*)$.
If $n\in{\cal N}_{\DS}$, $H_{\rm cl}(n;\sans s)$ satisfies both $(6.5)$
and $(6.12)$. Then, by $(6.4)$, $(3.7)$, $(5.12)$, $(6.5)$ and $(6.12)$,
the functional $V^{\rm tot}(h,H_{\rm cl}(n;\sans s),A^*)$ is
independent from $A^*$. Thus, one can evaluate it by setting $A^*=0$.
{}From here, using $(6.11)$, $(3.8)$, $(5.13)$,
$(3.16)$, $(3.24)$, $(5.20)$, $(5.34)$ and $(5.37)$, one finds
$$
I^{\rm tot}(h,H_{\rm cl}(n;\sans s),A^*)
=I^{\rm tot}_{\rm conf}(h)+\Delta I^{\rm tot}_{\rm conf}(h;n)
+I^{\rm tot}_{\rm hol}(A^*;A_0,\rho_0),
\eqno(6.13)
$$
where
$$
I^{\rm tot}_{\rm conf}(h)=I_{\rm conf}(h)+I^{\rm gh}_{\rm conf}(h),
\eqno(6.14)
$$
$$
\Delta I^{\rm tot}_{\rm conf}(h;n)
=S(h,H_{\rm cl}(n))+S^{\rm gh}(h,H_{\rm cl}(n)),
\eqno(6.15)
$$
$$
I^{\rm tot}_{\rm hol}(A^*;A,\rho)=
I_{\rm hol}(A^*;A,\rho)+I^{\rm gh}_{\rm hol}(A^*;A,\rho),
\eqno(6.16)
$$
$S(h,H_{\rm cl}(n))$ and $S^{\rm gh}(h,H_{\rm cl}(n))$ being given
$(3.18)$ and $(5.22)$ and
$H_{\rm cl}(n)$ being $H_{\rm cl}(n;\sans s)$ evaluated at the reference
holomorphic structure. By $(3.25)$ and $(5.36)$, {\it
$I^{\rm tot}_{\rm conf}(h)$
is the effective action of a conformal field theory of central charge
$\kappa^{\rm tot}_{\rm conf 0}=\kappa_0+\kappa+\kappa^{\rm gh}$, where
$\kappa$ and $\kappa^{\rm gh}$ are given respectively by $(3.22)$ and
$(5.32)$}. $\Delta I^{\rm tot}_{\rm conf}(h;n)$ is a local functional of
$h$, since the two terms in the right hand side of $(6.15)$ are,
as is explained in sects. 3 and 5.

By the classical $H$ equation $(6.5)$,
$I^{\rm tot}(h,H_{\rm cl}(n;\sans s),A^*)$ is constant as a
function of $n$ on each connected component ${\cal N}_i$ of $\cal N$.
Thus, it may be evaluated at any point $n_i\in{\cal N}_i\cap{\cal N}_{\DS}$,
which I assume to be non empty. Then, on account of $(6.13)$,
$(6.7)$ may be written as
$$
\eqalignno{
{\cal Z}^{\rm herm}(h,A^*)
=&~\sum_i\exp\big(I^{\rm tot}_{\rm conf}(h)
+\Delta I^{\rm tot}_{\rm conf}(h;n_i)
+I^{\rm tot}_{\rm hol}(A^*;A_0,\rho_0)\big)&\cr
\times&~\int_{{\cal N}_i}(Dn)_{|n}{\cal Z}^{\rm herm}_{\rm qu}(h,A^*;n).
&(6.17)\cr}
$$
Next, one has to study the partition function
${\cal Z}^{\rm herm}_{\rm qu}(h,A^*;n)$, but before doing that a few
important remarks are in order.

Eqs. $(6.5)$ and $(6.12)$ are rather complicated because of the
Drinfeld--Sokolov ghost contributions proportional to $K^{-1}$. In the
limit $K\rightarrow\infty$, however, the ghosts decouple and they
simplify considerably. Calling $H_\infty$ the corresponding $H$
configuration, the equations become
$$
\big(F_{H_\infty}\big)_{\ssans s}=0,
\eqno(6.18)
$$
$$
\big(\Gamma_{H_\infty}-A\big)_{\ssans s}=0
\quad {\rm in}~\ECF^{\vee\hphantom{\rm S}1,0}_{\DS\ssans s}.
\eqno(6.19)
$$
So, $H_\infty$ is a flat fiber metric such that $\Gamma_{H_\infty}$ is
Drinfeld--Sokolov, since $A$ is.
Equations of this form were found in \ref{10} on a minkowskian
cylindrical world sheet and shown to be equivalent to the non abelian
Toda equations associated to the pair $(G,S)$. On a euclidean
topologically non trivial world sheet, however, one has to take into
account further constraints coming from global definedness and non singularity.
One then finds that the above equations admit solutions $H_\infty$
of Toda type at genus $\ell=0$. For instance, $H(h_{\rm cc})$, where
$h_{\rm cc}$ is the constant curvature surface metric with
$-2h_{\rm cc}{}^{-1}f_{h_{\rm cc}}=1$ and $H(h)$ is given by $(3.15)$,
satisfies $(6.18)$--$(6.19)$ for the reference holomorphic structure.
In higher genus there still are
solutions of Toda type but ones which are hermitian with respect to a non
compact conjugation of the Lie algebra $\goth g$ \ref{36--37}. The use
of the compact conjugation $\dagger$ however cannot be avoided since
positivity of the various Hilbert structures in the construction of the
measures is indispensable. If the Toda solutions are the only solutions
available, then it will be necessary to introduce some type of insertion in
the $H$ functional integral providing extra terms in the classical equations
compensating for the problem. Unfortunately, very little is known at present
about these equations on a Riemann surface.

The partition function ${\cal Z}^{\rm herm}_{\rm qu}(h,A^*;n)$ can be
computed to leading order in a semiclassical expansion with expansion
parameter $\hbar\equiv K^{-1}$. To this end, one rescales the Donaldson
field $\Phi$ into $K^{-{1\over 2}}\Phi^0$ and expands in powers
of $K^{-{1\over 2}}$. In so doing, one must take into account that
the classical solution $H_{\rm cl}(n;\sans s)$, the fibration
$\varphi(n,\sans s)$ and the functional measure
$(D\Phi)_{h,H_{\rm cl}(n;\ssans s)|\Phi}$, also, depend on $K$.

Below, it is assumed that the metric $H_{\rm cl}(n;\sans s)$ has a
well defined limit $H_{\rm cl\infty}(n;\sans s)$ in $\Herm$ as
$K\rightarrow+\infty$ for every $n\in{\cal N}$ satisfying $(6.18)$ and
that such $K\rightarrow+\infty$ solutions span a submanifold
$\Herm_{\rm cl\infty}(\sans s)$ of $\Herm$.

It can be seen that, in the limit $K\rightarrow+\infty$, one has
$(D\Phi)_{h,H_{\rm cl}(n;\ssans s)|\Phi}
=z_K(h)[1+O(K^{-1})](D\Phi^0_\infty)_{h,H_{\rm cl\infty}(n;\ssans s)}$.
Here, $\Phi^0_\infty$ varies in $\Don(H_{\rm cl\infty}(n;\ssans s))$,
where $\Don(H_{\rm cl\infty})$ is the space of Donaldson fields
$\Phi^0_\infty\in\ECF^{0,0}$ satisfying
$\Ad H_{\rm cl\infty}\Phi^{0\dagger}_\infty=\Phi^0_\infty$ and orthogonal
in $\ECF^{0,0{\rm r}}$ with Hilbert structure $\langle\cdot,\cdot
\rangle^{\rm r}_{h,H_{\rm cl\infty}}$ to the kernel of the operator
$\Delta_{H_{\rm cl\infty}\ssans s}=-\big(\bar\partial
\partial_{H_{\rm cl\infty}}\big)_{\ssans s}$. This follows from the
properties of the fibration and the fact that the tangent vectors
$\delta H_{\infty}H_{\infty}{}^{-1}$ to $\Herm_{\rm cl\infty}(\sans s)$
at $H_{\rm cl\infty}$ satisfy $\big(\Delta_{H_{\rm cl\infty}}
(\delta H_{\infty}H_{\infty}{}^{-1})\big)_{\ssans s}=0$.
$(D\Phi^0_\infty)_{h,H_{\rm cl\infty}}$ is the translation invariant
measure on $\Don(H_{\rm cl\infty})$ obtained from the obvious real Hilbert
structure.
$z_K(h)=\big[\dtr_{h,H_{\rm cl\infty}(n;\ssans s)}(K^{-1}1)\big]^{1\over 2}$
is a constant arising because the different normalization
of the fields $\Phi$ and $\Phi^0$ related by $\Phi=K^{-{1\over 2}}\Phi^0$.
It depends on $h$ because of the $h$ dependence of the measure.

Proceeding in this way, one finds
$$\eqalignno{
{\cal Z}^{\rm herm}_{\rm qu}(h,A^*;n)
=&~z_K(h)
[\det J_\infty(h,A^*;n)]^{1\over 2}\int_{\Don(H_{\rm cl\infty}(n;\ssans s))}
(D\Phi^0_\infty)_{h,H_{\rm cl\infty}(n;\ssans s)}&\cr
\times&~
\exp\big(-S^{\rm tot}_{\rm qu\infty}(\Phi^0_\infty,A^*;H_{\rm cl\infty}
(n;\sans s))\big)\big[1+O(K^{-1})\big],
&(6.20)\cr}
$$
where $J_\infty(h,A^*;n)$ is given by $(6.8)$ with
$H_{\rm cl}(n;\sans s)$ replaced by $H_{\rm cl\infty}(n;\sans s)$ and
$$
S^{\rm tot}_{\rm qu\infty}(\Phi^0_\infty,A^*;H_{\rm cl\infty})
={1\over 2\pi}\int_\Sigma d^2z\tr_{\rm ad}\big(\Phi^0_\infty
\Delta_{H_{\rm cl\infty}}\Phi^0_\infty\big)_{\ssans s}
\eqno(6.21)
$$
is the Gaussian fluctuation action.
The effective action $\hat I^{\rm herm}_{\rm qu}(h,A^*;n)=\ln
{\cal Z}^{\rm herm}_{\rm qu}(h,A^*;n)$ is therefore
$$
\hat I^{\rm herm}_{\rm qu}(h,A^*;n)=
-{1\over 2}\ln\bigg[{\dtr'_{h,H_{\rm cl\infty}(n;\ssans s)}
\Delta_{h,H_{\rm cl\infty}(n;\ssans s)}\over
\det J_\infty(h,A^*;n)}\bigg]+\ln z_K(h)+O(K^{-1}),
\eqno(6.22)
$$
where $\Delta_{h,H_{\rm cl\infty}\ssans s}
=h^{-1}\Delta_{H_{\rm cl\infty}\ssans s}$.
Here, I shall use again proper time regularization scheme.
Then, the effective action becomes dependent on the proper time cut off
$\epsilon$. Taking into account that the vectors
$\partial_{n^r}H_{\rm cl\infty}(n;\sans s)H_{\rm cl\infty}(n;\sans s)^{-1}$
span $\ker\Delta_{h,H_{\rm cl\infty}(n;\ssans s)}$, one finds, using standard
heat kernel techniques,
$$\eqalignno{
\hat I^{\rm herm}_{\rm qu}(h,A^*;n;\epsilon)
=&~{(1-\ln K)\dim\goth g\over 2\pi\epsilon}
\int_\Sigma d^2z h
+{1\over 2}\bigg[{\dim\goth g\over 6\pi}
\int_\Sigma d^2z f_h
+\dim{\cal N}\bigg]\ln\epsilon&\cr
+&~W_{\rm conf}(h)+\Lambda(A^*;n)
+\ln K^{c_\ell}
+O(\epsilon)+O(K^{-1}),&\cr
c_\ell=&~{1\over 6}\dim\goth g(\ell-1)+{1\over 2}\dim{\cal N}
\vphantom{\int}.
&(6.23)\cr}
$$
Here, $W_{\rm conf}(h)$ is a non local functional of $h$ such that
$$
\delta W_{\rm conf}(h)
=-{\dim\goth g\over 12\pi}\int_\Sigma d^2z\delta hf_h.
\eqno(6.24)
$$
$\Lambda(A^*;n)$ is a non local functional of $A^*$ depending on $n$.
The ultraviolet divergencies can be cancelled by adding to the bare
effective action the counterterm
$$\eqalignno{
\Delta\hat I^{\rm herm}_{\rm qu}(h,A^*;\epsilon)
=&~-{(1-\ln K)\dim\goth g\over 2\pi\epsilon}
\int_\Sigma d^2z h
-{1\over 2}\bigg[{\dim\goth g\over 6\pi}
\int_\Sigma d^2z f_h
+\dim{\cal N}\bigg]\ln\epsilon&\cr
+&~O(\epsilon)+O(K^{-1}).
&(6.25)\cr}
$$
This must be independent from $n$, since the divergent terms of
$\ln{\cal Z}^{\rm herm}(h,A^*)$ depend only on $h$ and $A^*$.
The renormalized effective action is thus
$$\eqalignno{
I^{\rm herm}_{\rm qu}(h,A^*;n)
=&~\hat I^{\rm herm}_{\rm qu}(h,A^*;n;\epsilon)
+\Delta\hat I^{\rm herm}_{\rm qu}(h,A^*;\epsilon)&\cr
=&~W_{\rm conf}(h)+\Lambda(A^*;n)
+\ln K^{c_\ell}+O(K^{-1}).
&(6.26)\cr}
$$
{}From $(6.24)$, the variation of $I^{\rm herm}_{\rm qu}(h,A^*;n)$
with respect to $h$ at fixed $A^*$ and $n$ is
$$
\delta I^{\rm herm}_{\rm qu}(h,A^*;n)
=-{\kappa^{\rm herm}_{\rm conf}\over 12\pi}\int_\Sigma d^2z\delta hf_h
+O(K^{-1}),
\eqno(6.27)
$$
where
$$
\kappa^{\rm herm}_{\rm conf}=\dim\goth g+O(K^{-1}).
\eqno(6.28)
$$
It appears from here that, {\it to order $O(K^0)$, the renormalized effective
action $I^{\rm herm}_{\rm qu}(h,A^*;n)$ is that of a conformal field
theory of central charge $\kappa^{\rm herm}_{\rm conf}$ given by
$(6.28)$. This is in agreement with the exact result obtained
by conformal field theory techniques for the Wess--Zumino--Novikov--Witten
model}
$$
\kappa^{\rm herm}_{\rm conf}={K\dim\goth g\over K+c^\vee},
\eqno(6.29)
$$
where $c^\vee$ is the dual Coxeter number.
It remains to be seen if the agreement continues to hold at higher
orders in $K^{-1}$, though physical intuition would seem to suggest
so since the short distance structure of Drinfeld--Sokolov gravity
is essentially the same as that of the Wess--Zumino--Novikov--Witten
model.

{}From $(6.2)$, $(6.17)$ and $(6.26)$, choosing
$$
\hat\theta(h,A^*)
=\exp\Delta\hat I^{\rm herm}_{\rm qu}(h,A^*)\big[1+O(K^{-1})\big],
\eqno(6.30)
$$
where $\Delta\hat I^{\rm herm}_{\rm qu}(h)$ is given by
$(6.25)$ in the proper time regularization scheme, one has
$$\eqalignno{
{\cal Z}(h)=&~K^{c_\ell}\sum_i\exp\big(I^{\rm tot}_{\rm conf}(h)
+\Delta I^{\rm tot}_{\rm conf}(h;n_i)+W_{\rm conf}(h)\big)&\cr
\times&~\int_{{\cal M}_{\DS}}(Dt)_{|t}\big|\det F(t,f)\det E(t,e)\big|^2
\exp I^{\rm tot}_{\rm hol}(A^*(t);A_0,\rho_0){\nu(1)\over v_\nu}&\cr
\times&~\int_{{\cal N}_i}(Dn)_{|n}\exp\Lambda (A^*(t);n)\big[1+O(K^{-1})\big].
&(6.31)\cr}
$$
This is the final form of the partition function. To order $O(K^0)$,
conformal invariance is manifest.

Several issues remain to be investigated. The analysis expounded is to
some extent formal due to the lack of detailed geometric information
about the Drinfeld--Sokolov moduli space ${\cal M}_{\DS}$, the
Drinfeld--Sokolov stability group ${\cal G}_{\DS}$ and the parameter
space ${\cal N}$. A thorough investigation of these spaces is desirable.
Also, the holomorphic structure on the Riemann surface $\Sigma$ has
been kept fixed throughout. One may try to deform the complex structure and
study the resulting effects in the framework of deformation theory using
the Beltrami parametrization. Such deformations should be a special subset
of more general deformations parametrized by generalized Beltrami
differentials \ref{38--39}. The study of this matter requires a
better understanding of $W$ geometry, which at present is lacking.
This issue is also related to that of the analysis of the $\Gau_{\DS}$
invariant content of the model. In fact, the generalized Beltrami
differentials should be the sources of a suitable basis of $\Gau_{\DS}$
invariant operators including the energy momentum tensor. At this level,
$W$ symmetries are expected to emerge.
\vskip.4cm
\par\noindent
{\bf Acknowledgements.} I wish to voice my gratitude to E. Aldrovandi,
F. Bastianelli, M. Bauer, G. Falqui, S. Lazzarini and R. Stora for helpful
discussions.
\vskip.4cm
{\bf Appendix A.}
\vskip.4cm
\par
In this appendix, I shall provide the basic details about the derivation
of the measure $(4.20)$. The notation used here is the same as that
defined in sect. 4. I also set $q=\dim{\cal G}_{\DS}$ and
$m=\dim{\cal M}_{\DS}$.

Let us construct the basic Hilbert manifolds. All such Hilbert manifolds
are real, though as ordinary manifolds, they may be complex. Below,
$h\in\Met$ is a generic surface metric on $\Sigma$, which will be kept
fixed throughout.

Consider first $\Herm$. For any $H\in\Herm$, the tangent space $T_H\Herm$
is the subspace of $\ECF^{0,0{\rm r}}$ spanned by the elements
$\delta HH^{-1}$ such that $\Ad H(\delta HH^{-1})^\dagger=\delta HH^{-1}$
and equipped with the Hilbert structure
${1\over 2}\langle\cdot,\cdot\rangle_{h,H}^{\rm r}$. The factor
$1\over 2$ is conventional. Hence, one has
$\Vert\delta HH^{-1}\Vert_{h|H}{}^2=\Vert\delta HH^{-1}\Vert_{h,H}{}^2$,
where the norm in the right hand side is that of $\ECF^{0,0}$.
In this way, $\Herm$ becomes a real Hilbert manifold.

Next, consider $\SHol_{\DS}$. For any $A^*\in\SHol_{\DS}$, the tangent
space $T_{A^*}\SHol_{\DS}$ is just $\ECF^{\phantom{\DS}0,1{\rm r}}_{\DS}$
with the Hilbert structure $\langle\cdot,\cdot\rangle_{\DS h,H}^{\rm r}$
depending on a fiber metric $H\in\Herm$. This is actually independent from
$h$. Denoting by $\delta A^*$ a generic element of $T_{A^*}\SHol_{\DS}$, one
has $\Vert\delta A^*\Vert_{H|A^*}{}^2=2\Vert\delta A^*\Vert_{\DS h,H}{}^2$,
the norm in the right hand side being that of $\ECF^{\phantom{\DS}0,1}_{\DS}$.
In this way, $\SHol_{\DS}$ becomes a real Hilbert manifold.

Next, consider $\Gau_{\DS}$. For any $\alpha\in\Gau_{\DS}$, the tangent
space $T_\alpha\Gau_{\DS}$ is just $\ECF^{\phantom{\DS}0,0{\rm r}}_{\DS}$
with the Hilbert structure $\langle\cdot,\cdot\rangle_{\DS h,H}^{\rm r}$
depending on a fiber metric $H\in\Herm$. A generic element of
$T_\alpha\Gau_{\DS}$ is of the from $\alpha^{-1}\delta\alpha$. One
thus has $\Vert\alpha^{-1}\delta\alpha\Vert_{h,H|\alpha}{}^2
=2\Vert\alpha^{-1}\delta\alpha\Vert_{\DS h,H}{}^2$, the norm in the right
hand side being that of $\ECF^{\phantom{\DS}0,0}_{\DS}$. In this way,
$\Gau_{\DS}$ too becomes a real Hilbert manifold. Because of the form of
the tangent vectors, the Hilbert manifold structure defined is left
invariant.

Consider now ${\cal M}_{\DS}$. For any $t\in{\cal M}_{\DS}$,
$T_t{\cal M}_{\DS}$ is just $(\Bbb C^m)^{\rm r}$
with the standard euclidean inner product
$\langle\cdot,\cdot\rangle^{\rm r}$. So,
$\Vert\delta t\Vert_{|t}{}^2=2|\delta t|^2$
for $\delta t\in T_t{\cal M}_{\DS}$. ${\cal M}_{\DS}$ becomes thus
a real Hilbert manifold of dimension $2m$.

Finally, consider ${\cal G}_{\DS}$. For any $g\in{\cal G}_{\DS}$,
$T_g{\cal G}_{\DS}$ is just $(\Bbb C^q)^{\rm r}$
with the standard euclidean inner product
$\langle\cdot,\cdot\rangle^{\rm r}$. So,
$\Vert\delta g\Vert_{|g}{}^2=2|\delta g|^2$
for $\delta g\in T_g{\cal G}_{\DS}$. In this way, ${\cal G}_{\DS}$
becomes a real $2q$ dimensional Hilbert manifold.

The first problem to tackle is the definition of the Hilbert manifold
structures of the two realizations $\Herm\times\SHol_{\DS}$ and
${\cal M}_{\DS}\times(\Herm\times\Gau_{\DS})/{\cal G}_{\DS}(\sans s_\cdot)$
of the configuration space.

$\Herm\times\SHol_{\DS}$ can be given naturally the structure of
real Hilbert manifold as follows. For and $(H,A^*)\in\Herm\times\SHol_{\DS}$,
$T_{(H,A^*)}\Herm\times\SHol_{\DS}\cong T_H\Herm\oplus
T_{A^*}\SHol_{\DS}$ with the Hilbert norm
$$
\Vert\delta HH^{-1}\oplus\delta A^*\Vert_{h,H|(H,A^*)}{}^2
=\Vert\delta HH^{-1}\Vert_{h|H}{}^2
+\Vert\delta A^*\Vert_{H|A^*}{}^2.
\eqno(A.1)
$$

Providing ${\cal M}_{\DS}\times(\Herm\times\Gau_{\DS})/
{\cal G}_{\DS}(\sans s_\cdot)$ with a
Hilbert manifold structure is slightly trickier
because of the quotient by the action of ${\cal G}_{\DS}(\sans s_\cdot)$.
For any $(t,\tilde H,\alpha)\in
{\cal M}_{\DS}\times(\Herm\times\Gau_{\DS})/{\cal G}_{\DS}(\sans s_\cdot)$,
one has $T_{(t,\tilde H,\alpha)}{\cal M}_{\DS}
\times(\Herm\times\Gau_{\DS})/{\cal G}_{\DS}(\sans s_\cdot)
\cong T_t{\cal M}_{\DS}\oplus((T_{\tilde H}\Herm$ $\oplus T_\alpha\Gau_{\DS})
/T\bit_{(\tilde H,\alpha)}(1;t)\Lie{\cal G}_{\DS}(\sans s_t))$.
$\bit_{(\tilde H,\alpha)}(\cdot;t):{\cal G}_{\DS}(\sans s_t)\rightarrow
\Herm\times\Gau_{\DS}$ is the orbit map associated to
the ${\cal G}_{\DS}(\sans s_t)$ action $(4.6)$--$(4.7)$. Its tangent map
$T\bit_{(\tilde H,\alpha)}(1;t)$ maps $\Lie{\cal G}_{\DS}(\sans s_t)$
into the subspace of $T_{\tilde H}\Herm
\oplus T_\alpha\Gau_{\DS}$ spanned by the vectors of the form
$(\delta\eta+\Ad\tilde H\delta\eta^\dagger)\oplus(-\delta\eta)$
with $\delta\eta\in\Lie{\cal G}_{\DS}(\sans s_t)$.
This follows from the linearization of $(4.6)$--$(4.7)$.
The tangent space can be given a Hilbert
structure as follows. One equips $T_t{\cal M}_{\DS}\oplus T_{\tilde H}\Herm
\oplus T_\alpha\Gau_{\DS}$ with the Hilbert norm
$$
\Vert\delta t\oplus\delta\tilde H\tilde H^{-1}\oplus
\alpha^{-1}\delta\alpha\Vert_{h,\tilde H|(t,\tilde H,\alpha)}{}^2
=\Vert\delta t\Vert_{|t}{}^2
+\Vert\delta \tilde H\tilde H^{-1}\Vert_{h|\tilde H}{}^2
+\Vert\alpha^{-1}\delta\alpha\Vert_{h,\tilde H|\alpha}{}^2.
\eqno(A.2)
$$
Then, one has the identification
$T_{(t,\tilde H,\alpha)}{\cal M}_{\DS}
\times(\Herm\times\Gau_{\DS})/{\cal G}_{\DS}(\sans s_\cdot)
\cong T_t{\cal M}_{\DS}\oplus((T_{\tilde H}\Herm\oplus T_\alpha\Gau_{\DS})
\ominus T\bit_{(\tilde H,\alpha)}(1;t)\Lie{\cal G}_{\DS}(\sans s_t))$.
The right hand side carries the Hilbert structure induced by that of
$T_t{\cal M}_{\DS}\oplus T_{\tilde H}\Herm\oplus T_\alpha\Gau_{\DS}$.
In this way, ${\cal M}_{\DS}
\times(\Herm\times\Gau_{\DS})/{\cal G}_{\DS}(\sans s_\cdot)$
becomes a real Hilbert manifold. The above construction is independent from
the choice of the representative $(\tilde H,\alpha)$ of the corresponding
equivalence class modulo the ${\cal G}_{\DS}(\sans s_t)$ action
$(4.6)$--$(4.7)$. Indeed, different choices
lead to unitarily equivalent realizations of the Hilbert tangent space,
as is straightforward to check.

One has to compute now the jacobian $J(t,h,\tilde H)$ of the map
$(4.4)$--$(4.5)$ relating the functional measures
of $\Herm\times\SHol_{\DS}$ and ${\cal M}_{\DS}\times
(\Herm\times\Gau_{\DS})/{\cal G}_{\DS}(\sans s_\cdot)$:
$$\eqalignno{
&\big(D\delta H(\tilde H,\alpha)H(\tilde H,\alpha)^{-1}
\big)_{h|H(\tilde H,\alpha)}\otimes
\big(D\delta A^*(t,\alpha)\big)_{H(\tilde H,\alpha)|A^*(t,\alpha)}
&\cr
=&~J(t,h,\tilde H)\Big[(D\delta t)_{|t}\otimes
\big(D\delta\tilde H\tilde H^{-1}\big)_{h|\tilde H}
&\cr
&~~\hskip4cm\otimes\big(D\alpha^{-1}\delta\alpha\big)_{h,\tilde H|\alpha}\Big]
\Big|_{(T\bit_{(\tilde H,\alpha)}(1;t)\Lie{\cal G}_{\DS}(\ssans s_t))^\perp}.
&(A.3)\cr}
$$
By explicit calculation, one finds
$$
J(t,h,\tilde H)=2^q\Delta(t,h,\tilde H)\dtr P(t,\tilde H),
\eqno(A.4)
$$
where
$$
\Delta(t,h,\tilde H)
=\dtr'_{h,\tilde H}\big((\bar\partial-\ad A^*(t)){}^\star
(\bar\partial-\ad A^*(t))\big)
\eqno(A.5)
$$
is the functional determinant of the Laplacian associated to the operator
$(\bar\partial-\ad A^*(t)):T_1\Gau_{\DS}\rightarrow T_{A^*(t)}\SHol_{\DS}$
with the given Hilbert structures with the zero eigenvalues removed and
$$
P(t,\tilde H)_{i,j}
=\langle\partial_{t^i} A^*(t),p(t,\tilde H)
\partial_{t^j} A^*(t)\rangle_{h,\tilde H},\quad i,j=1,\cdots,m
\eqno(A.6)
$$
where $p(t,\tilde H)$ is the orthogonal projector on $\coker(\bar\partial-
\ad A^*(t))$ in $T_{A^*(t)}\SHol_{\DS}$. All determinants are taken on the
complex field.
\par\noindent
{\it Proof}.
The calculation of the jacobian requires to begin with
the computation of the tangent map of the map $(4.4)$--$(4.5)$. This is
given by
$$
\delta H(\tilde H,\alpha)H(\tilde H,\alpha)^{-1}
=\Ad\alpha(\delta\tilde H\tilde H^{-1}+\alpha^{-1}\delta\alpha
+\Ad\tilde H(\alpha^{-1}\delta\alpha)^\dagger),
\eqno(A.7)
$$
$$
\delta A^*(t,\alpha)
=\Ad\alpha((\bar\partial-\ad A^*(t))(\alpha^{-1}\delta\alpha)
+\delta_tA^*(t)).
\eqno(A.8)
$$
as follows from a simple variational calculation. The Hilbert structure of
the tangent bundle of ${\cal M}_{\DS}\times(\Herm\times\Gau_{\DS})/
{\cal G}_{\DS}(\sans s_\cdot)$ may be disentangled by means of the
following orthogonal decomposition:
$$\eqalignno{
&~T_t{\cal M}_{\DS}\oplus((T_{\tilde H}\Herm\oplus T_\alpha\Gau_{\DS})
\ominus T\bit_{(\tilde H,\alpha)}(1;t)\Lie{\cal G}_{\DS}(\sans s_t))
&\cr
=&~T_t{\cal M}_{\DS}
\oplus
(T_{\tilde H}\Herm\ominus T\bit_{\tilde H}(1;t)\Lie{\cal G}_{\DS}(\sans s_t))
&\cr
&~\oplus
(T_\alpha\Gau_{\DS}
\ominus T\bit_\alpha(1;t)\Lie{\cal G}_{\DS}(\sans s_t))
\oplus E_t.
&(A.9)\cr}
$$
Here, $\bit_{\tilde H}(\cdot;t):{\cal G}_{\DS}(\sans s_t)\rightarrow\Herm$
is the orbit map associated to the ${\cal G}_{\DS}(\sans s_t)$ action
$(4.6)$ on $\Herm$. Its tangent map $T\bit_{\tilde H}(1;t)$ maps
$\Lie{\cal G}_{\DS}(\sans s_t)$ into the subspace of $T_{\tilde H}\Herm$
spanned by the vectors of the form $\delta\eta+\Ad\tilde H\delta\eta^\dagger$
with $\delta\eta\in\Lie{\cal G}_{\DS}(\sans s_t)$. Similarly,
$\bit_\alpha(\cdot;t):{\cal G}_{\DS}(\sans s_t)\rightarrow\Gau_{\DS}$
is the orbit map associated to the ${\cal G}_{\DS}(\sans s_t)$ action
$(4.7)$ on $\Gau_{\DS}$. Its tangent map
$T\bit_\alpha(1;t)$ maps $\Lie{\cal G}_{\DS}(\sans s_t)$
into the subspace of $T_\alpha\Gau_{\DS}$ spanned by the vectors
$\delta\eta\in\Lie{\cal G}_{\DS}(\sans s_t)$. $E_t$ is the subspace of
$T_{\tilde H}\Herm\oplus T_\alpha\Gau_{\DS}$
spanned by the vectors of the form
$(\delta\eta+\Ad\tilde H\delta\eta^\dagger)\oplus\delta\eta$
with $\delta\eta\in\Lie{\cal G}_{\DS}(\sans s_t)$,
where a sign difference in the second component with respect to the vectors
spanning $T\bit_{(\tilde H,\alpha)}(1;t)\Lie{\cal G}_{\DS}(\sans s_t)$
is to be noticed. Hence, for any $\delta\tilde H\tilde H^{-1}\in
T_{\tilde H}\Herm$ and $\alpha^{-1}\delta\alpha\in T_\alpha\Gau_{\DS}$,
one has the decompositions
$$
\delta\tilde H\tilde H^{-1}
=\delta\tilde H\tilde H^{-1}{}_\perp+\delta\eta
+\Ad\tilde H\delta\eta^\dagger,
\eqno(A.10)
$$
$$
\alpha^{-1}\delta\alpha
=\alpha^{-1}\delta\alpha_\perp+\delta\eta,
\eqno(A.11)
$$
where $\delta\eta\in\Lie{\cal G}_{\DS}(\sans s_t)$,
$\delta\tilde H\tilde H^{-1}{}_\perp\in
T_{\tilde H}\Herm\ominus T\bit_{\tilde H}(1;t)
\Lie{\cal G}_{\DS}(\sans s_t)$ and
$\alpha^{-1}\delta\alpha_\perp\in T_\alpha\Gau_{\DS}$
$\ominus T\bit_\alpha(1;t)\Lie{\cal G}_{\DS}(\sans s_t)$.
By substituting $(A.10)$--$(A.11)$ into $(A.7)$--$(A.8)$ and the result into
$(A.1)$, one finds
$$\eqalignno{
&~\Vert\delta H(\tilde H,\alpha)H(\tilde H,\alpha)^{-1}
\oplus\delta A^*(t,\alpha)
\Vert_{h,H(\tilde H,\alpha)|(H(\tilde H,\alpha),A^*(t,\alpha))}{}^2&\cr
=&~\Vert\delta\tilde H\tilde H^{-1}{}_\perp+\alpha^{-1}\delta\alpha_\perp
+\Ad\tilde H(\alpha^{-1}\delta\alpha_\perp)^\dagger
\Vert_{h|\tilde H}{}^2&\cr
+&~\Vert(\bar\partial-\ad A^*(t))
(\alpha^{-1}\delta\alpha_\perp+\alpha^{-1}\delta\alpha_\perp(t,\tilde H))
\Vert_{\tilde H|A^*(t)}{}^2&\cr
+&~\Vert p(t,\tilde H)\delta_t A^*(t)\Vert_{\tilde H|A^*(t)}{}^2
+2\Vert(\delta\eta+\Ad\tilde H\delta\eta^\dagger)\oplus\delta\eta
\Vert_{h,\tilde H|(\tilde H,1)}{}^2.
&(A.12)\cr}
$$
Here, $\alpha^{-1}\delta\alpha_\perp(t,\tilde H)$
is some element of $T_\alpha\Gau_{\DS}
\ominus T\bit_\alpha(1;t)\Lie{\cal G}_{\DS}(\sans s_t)$
depending on $t$ and $\tilde H$ whose explicit expression will not matter.
In deducing $(A.12)$, one exploits the fact that the Cartan Killing form
$\tr_{\rm ad}$ vanishes on $\goth x$ because of the nilpotence of $\goth x$.
One also uses the fact that $\Lie{\cal G}_{\DS}(\sans s_t)
\subseteq\HECF^{\phantom{\DS}0}_{\DS\ssans s_t}$ so that, for
$\delta\eta\in\Lie{\cal G}_{\DS}(\sans s_t)$,
$(\bar\partial-\ad A^*(t))\delta\eta=0$. Using the jacobian relation
$(A.3)$, the normalization condition for the measures and $(A.12)$,
it is straightforward to obtain $(A.4)$--$(A.6)$. {\it QED}

The jacobian $J(t,h,\tilde H)$ is a positive $(m,m)$ form on
${\cal M}_{\DS}$. It does not depend on $\alpha$, a consequence
of $\Gau_{\DS}$ gauge invariance. It can be shown that, for any
$\eta\in{\cal G}_{\DS}(\sans s_t)$,
$J(t,h,\vphantom{H}^\eta\tilde H)=J(t,h,\tilde H)$,
i. e. $J(t,h,\tilde H)$ is invariant under the action $(4.5)$ of
${\cal G}_{\DS}(\sans s_t)$ on $\Herm$. This is expected on general
grounds as a consequence of the ${\cal G}_{\DS}$ symmetry of the
parametrization $(4.4)$--$(4.5)$.

Next, one has to define the Hilbert manifold structure of the
isomorphic spaces $\Herm\times\Gau_{\DS}$ and
$((\Herm\times\Gau_{\DS})/{\cal G}_{\DS}(\sans s_t))\times{\cal G}_{\DS}$.

$\Herm\times\Gau_{\DS}$ has an obvious structure of real Hilbert manifold.
For $(H,\omega)\in\Herm\times\Gau_{\DS}$, $T_{(H,\omega)}
\Herm\times\Gau_{\DS}\cong T_H\Herm\oplus T_\omega\Gau_{\DS}$ equipped
with the Hilbert norm
$$
\Vert\delta HH^{-1}\oplus
\omega^{-1}\delta\omega\Vert_{h,H|(H,\omega)}{}^2
=\Vert\delta HH^{-1}\Vert_{h|H}{}^2
+\Vert\omega^{-1}\delta\omega\Vert_{h,H|\omega}{}^2.
\eqno(A.13)
$$

$((\Herm\times\Gau_{\DS})/{\cal G}_{\DS}(\sans s_t))\times{\cal G}_{\DS}$
can also be given a structure of Hilbert manifold.
For any $(\tilde H,\alpha,g)\in
((\Herm\times\Gau_{\DS})/{\cal G}_{\DS}(\sans s_t))\times{\cal G}_{\DS}$,
$T_{(\tilde H,\alpha,g)}
((\Herm\times\Gau_{\DS})/{\cal G}_{\DS}(\sans s_t))\times{\cal G}_{\DS}\cong
((T_{\tilde H}\Herm\oplus T_\alpha\Gau_{\DS})
/T\bit_{(\tilde H,\alpha)}(1;t)\Lie{\cal G}_{\DS}(\sans s_t))\oplus
T_g{\cal G}_{\DS}$. One equips $T_{\tilde H}\Herm\oplus T_\alpha\Gau_{\DS}
\oplus T_g{\cal G}_{\DS}$ with the Hilbert norm
$$
\Vert\delta\tilde H\tilde H^{-1}\oplus\alpha^{-1}\delta\alpha
\oplus\delta g\Vert_{h,\tilde H|(\tilde H,\alpha,g)}{}^2
=\Vert\delta\tilde H\tilde H^{-1}\Vert_{h|\tilde H}{}^2
+\Vert\alpha^{-1}\delta\alpha\Vert_{h,\tilde H|\alpha}{}^2
+\Vert\delta g\Vert_{|g}{}^2.
\eqno(A.14)
$$
Then, one has the identification $T_{(\tilde H,\alpha,g)}
((\Herm\times\Gau_{\DS})/{\cal G}_{\DS}(\sans s_t))\times{\cal G}_{\DS}\cong
((T_{\tilde H}\Herm\oplus T_\alpha\Gau_{\DS})
\ominus T\bit_{(\tilde H,\alpha)}(1;t)\Lie{\cal G}_{\DS}(\sans s_t))
\oplus T_g{\cal G}_{\DS}$. This above construction is independent
up to unitary equivalence from the choice of the representative
$(\tilde H,\alpha)$ of the corresponding equivalence class modulo the
${\cal G}_{\DS}(\sans s_t)$ action $(4.6)$--$(4.7)$.

One has now to compute the jacobian $K(t,g,h,\tilde H)$ of the map
$(4.8)$--$(4.9)$ relating the functional measures on
$\Herm\times\Gau_{\DS}$ and $((\Herm\times\Gau_{\DS})/
{\cal G}_{\DS}(\sans s_t))\times{\cal G}_{\DS}$. One has
$$\eqalignno{
&\big(D\delta H(\tilde H,g)H(\tilde H,g)^{-1}
\big)_{h|H(\tilde H,g)}
\otimes
\big(D\omega(\alpha,g)^{-1}\delta\omega(\alpha,g)
\big)_{h,H(\tilde H,g)|\omega(\alpha,g)}
&\cr
=&~K(t,g,h,\tilde H)\Big[
\big(D\delta\tilde H\tilde H^{-1}\big)_{h|\tilde H}
\big(D\alpha^{-1}\delta\alpha\big)_{h,\tilde H|\alpha}
&\cr
&~~~\hskip5cm\otimes(D\delta g)_{|g}\Big]
\Big|_{(T\bit_{(\tilde H,\alpha)}(1;t)
\Lie{\cal G}_{\DS}(\ssans s_t))^\perp}.
&(A.15)\cr}
$$
The expression obtained is
$$
K(t,g,h,\tilde H)
=2^q\dtr Q(t,g,h,\tilde H),
\eqno(A.16)
$$
where
$$
Q(t,g,h,\tilde H)_{i,j}
=\langle\zeta(g;t)^{-1}\partial_{g^i}\zeta(g;t),
\zeta(g;t)^{-1}\partial_{g^j}\zeta(g;t)
\rangle_{h,\tilde H},\quad i,j=1\cdots,q
\eqno(A.17)
$$
and the determinant is taken on the complex field.
\par\noindent
{\it Proof}.
The tangent map of the parametrization $(4.8)$--$(4.9)$ is given by
$$\eqalignno{
&~\delta H(\tilde H,g)H(\tilde H,g)^{-1}&\cr
=&~\Ad\zeta(g;t)(\delta\tilde H\tilde H^{-1}
+\zeta(g;t)^{-1}\delta_g\zeta(g;t)
+\Ad\tilde H(\zeta(g;t)^{-1}\delta_g\zeta(g;t))^\dagger),
&(A.18)\cr}
$$
$$
\omega(\alpha,g)^{-1}\delta\omega(\alpha,g)
=\Ad\zeta(g;t)(\alpha^{-1}\delta\alpha
-\zeta(g;t)^{-1}\delta_g\zeta(g;t)).
\eqno(A.19)
$$
By substituting $(A.18)$--$(A.19)$ into $(A.13)$, one obtains
$$\eqalignno{
&\Vert\delta H(\tilde H,g)H(\tilde H,g)^{-1}\oplus
\omega(\alpha,g)^{-1}\delta\omega(\alpha,g)
\Vert_{h,H(\tilde H,g)|(H(\tilde H,g),\omega(\alpha,g))}{}^2&\cr
=&~\Vert\delta\tilde H\tilde H^{-1}\Vert_{h|\tilde H}{}^2
+\Vert\alpha^{-1}\delta\alpha\Vert_{h,\tilde H|\alpha}{}^2
+2\Vert\zeta(g;t)^{-1}\delta_g\zeta(g;t)\Vert_{h,\tilde H|\zeta(g;t)}{}^2.
&(A.20)\cr}
$$
Using the jacobian relation $(A.15)$, the normalization condition of the
measures and $(A.20)$, it is straightforward to obtain $(A.16)$--$(A.17)$.
{\it QED}

The jacobian $K(t,g,h,\tilde H)$ is $(q,q)$ form on ${\cal G}_{\DS}$.
Its independence from $\alpha$ is a consequence of the left invariance of
the measure on $T_\alpha\Gau_{\DS}$. From $(A.17)$, it is apparent that
$L_f{}^*K(t,g,h,\tilde H)=K(t,g,h,\tilde H)$ for any $f\in{\cal G}_{\DS}$,
i. e. $K(t,g,h,\tilde H)$ is left invariant. Under the right action of
${\cal G}_{\DS}$, one has instead
$R_f{}^*K(t,g,h,\tilde H)=K(t,g,h,\zeta(f;t)^*\tilde H)$.

Now, all elements required for the implementation of the gauge fixing
procedure are available. Consider a $\Gau_{\DS}$--invariant functional
$\Theta(h,H,A^*)$. Hence, for any $\alpha\in\Gau_{\DS}$,
$\Theta(h,\alpha^*H,\alpha^*A^*)=\Theta(h,H,A^*)$.
The functional integral
$$
{\cal J}_\Theta(h)=\int_{\Herm\times\SHol_{\DS}}
(DH)_{h|H}\otimes(DA^*)_{H|A^*}\Theta(h,H,A^*)
\eqno(A.21)
$$
is thus divergent because of the $\Gau_{\DS}$ invariance of the
integrand. The problem to solve next is the factorization
of the divergent gauge volume.

On account of the isomorphism $(4.4)$--$(4.5)$ of $\Herm\times\SHol_{\DS}$
and ${\cal M}_{\DS}\times(\Herm\times\Gau_{\DS})/
{\cal G}_{\DS}(\sans s_\cdot)$, the jacobian relation $(A.3)$
and the $\Gau_{\DS}$ invariance of $\Theta(h,\tilde H,A^*)$, one has
$$\eqalignno{
{\cal J}_\Theta(h)
=&\int_{{\cal M}_{\DS}}(Dt)_{|t}
\int_{(\Herm\times\Gau_{\DS})/{\cal G}_{\DS}(\ssans s_t)}
(D\tilde H)_{h|\tilde H}\otimes
(D\alpha)_{h,\tilde H|\alpha}&\cr
&\phantom{\int_{{\cal M}_{\DS}}(Dt)_t
\int_{(\Herm\times\Gau_{\DS})/{\cal G}_{\DS}(\ssans s_t)}(}
\times J(t,h,\tilde H)\Theta(h,\tilde H,A^*(t)).
&(A.22)\cr}
$$
Because of the quotient by ${\cal G}_{\DS}(\sans s_t)$, it is not
possible to factor out the gauge volume yet. This requires a few extra
steps.

Define
$$
v(t,h,H)=\int_{{\cal G}_{\DS}}(Dg)_{|g}K(t,g,h,H).
\eqno(A.23)
$$
$v(t,h,H)$ is actually divergent since ${\cal G}_{\DS}$ is
a non compact group. However, formally, by the form of the right
${\cal G}_{\DS}$ action on $K(t,g,h,\tilde H)$,
$v(t,h,\vphantom{H}^\eta H)=v(t,h,H)$ for any
$\eta\in{\cal G}_{\DS}(\sans s_t)$, i. e.
$v(t,h,H)$ is ${\cal G}_{\DS}(\sans s_t)$ invariant.
The infinite volume of the gauge group is
$$
V(h,H)=\int_{\Gau_{\DS}}(D\omega)_{h,H|\omega}.
\eqno(A.24)
$$
Now, from the isomorphism $(4.8)$--$(4.9)$ of $\Herm\times\Gau_{\DS}$ and
$((\Herm\times\Gau_{\DS})/{\cal G}_{\DS}(\sans s_t))\times{\cal G}_{\DS}$,
using the jacobian relation $(A.15)$ and the ${\cal G}_{\DS}(\sans s_t)$
invariance of $J(t,h,\tilde H)$, the $\Gau_{\DS}$ invariance of
$\Theta(h,\tilde H,A^*)$, $(A.23)$ and the ${\cal G}_{\DS}(\sans s_t)$
invariance of $v(t,h,H)$, one has
$$\eqalignno{
&~\int_{\Herm}(DH)_{h|H}
{V(h,H)\over v(t,h,H)}J(t,h,H)\Theta(h,H,A^*(t))
&\cr
=&~\int_{\Herm\times\Gau_{\DS}}
(DH)_{h|H}\otimes(D\omega)_{h,H|\omega}
{1\over v(t,h,H)}J(t,h,H)\Theta(h,H,A^*(t))
&\cr
=&~\int_{(\Herm\times\Gau_{\DS})/{\cal G}_{\DS}(\ssans s_t)}
(D\tilde H)_{h|\tilde H}\otimes(D\alpha)_{h,\tilde H|\alpha}
\int_{{\cal G}_{\DS}}(Dg)_{|g}&\cr
\times&~K(t,g,h,\tilde H)
{1\over v(t,h,\tilde H)}J(t,h,\tilde H)
\Theta(h,\tilde H,A^*(t))&\cr
=&~\int_{(\Herm\times\Gau_{\DS})/{\cal G}_{\DS}(\ssans s_t)}
(D\tilde H)_{h|\tilde H}\otimes
(D\alpha)_{h,\tilde H|\alpha}
J(t,h,\tilde H)\Theta(h,\tilde H,A^*(t)).
&(A.25)\cr}
$$
Combining $(A.22)$ and $(A.25)$, one has
$$
{\cal J}_\Theta(h)
=\int_{{\cal M}_{\DS}}(Dt)_{|t}\int_{\Herm}(DH)_{h|H}
{V(h,H)\over v(t,h, H)}J(t,h,H)\Theta(h,H,A^*(t)).
\eqno(A.26)
$$
Gauge fixing is now easy. One simply deletes the infinite gauge
volume $V(h,H)$ in the above expression. The gauge fixed
functional integral is then
$$
{\cal J}^{\rm gf}_\Theta(h)
=\int_{{\cal M}_{\DS}}(Dt)_{|t}\int_{\Herm}(DH)_{h|H}
{1\over v(t,h, H)}J(t,h,H)\Theta(h,H,A^*(t)).
\eqno(A.27)
$$

$v(t,h,H)$ depends on $H$ and this is inconvenient. One can separate the
$H$ dependence from the group volume by the following method. Let
$\nu(g)$ be a left invariant positive $(q,q)$ form on ${\cal G}_{\DS}$. So
$L_f{}^*\nu(g)=\nu(g)$, for any $f\in{\cal G}_{\DS}$. Using $\nu(g)$, one
can define the group volume $v_\nu=\int_{{\cal G}_{\DS}}(Dg)_{|g}\nu(g)$
of ${\cal G}_{\DS}$. From the left ${\cal G}_{\DS}$ invariance of
$K(t,g,h,\tilde H)$, $(A.23)$ and the left ${\cal G}_{\DS}$ invariance of
$\nu$, it is easy to show the formal relation
$$
v(t,h,H)=v_\nu\nu(1)^{-1}K(t,1,h,H).
\eqno(A.28)
$$
Using $(A.28)$, $(A.27)$ can be cast
$$
{\cal J}^{\rm gf}_\Theta(h)
=\int_{{\cal M}_{\DS}}(Dt)_{|t}{\nu(1)\over v_\nu}
\int_{\Herm}(DH)_{h|H}{J(t,h,H)\over K(t,1,h,H)}\Theta(h,H,A^*(t)).
\eqno(A.29)
$$
This is the final form of the expression of ${\cal J}^{\rm gf}_\Theta(h)$.
Now, $(4.20)$ follows from $(A.29)$ by a straightforward calculation.

Using the isomorphisms
$(\Lie\Gau_{\DS})^\vee\cong\ECF^{\vee\hphantom{\rm S}1,0}_{\DS}$
and $\Lie\Gau_{\DS}\cong\ECF^{\phantom{\DS}0,0}_{\DS}$, one can
define real Hilbert structures on $(\Lie\Gau_{\DS})^\vee$
and $\Lie\Gau_{\DS}$. One simply views $(\Lie\Gau_{\DS})^\vee$ and
$\Lie\Gau_{\DS}$ as the real Hilbert manifolds
$\ECF^{\vee\hphantom{\rm S}1,0{\rm r}}_{\DS}$ and
$\ECF^{\phantom{\DS}0,0{\rm r}}_{\DS}$ with the Hilbert structure
$\langle\cdot,\cdot\rangle^{\vee{\rm r}}_{\DS h,H}$ and
$\langle\cdot,\cdot\rangle^{\rm r}_{\DS h,H}$, respectively.
This yields the ghost functional measures appearing in $(4.22)$.
\vskip.4cm
{\bf Appendix B.}
\vskip.4cm
\par
In this appendix, I shall provide some detail about the derivation of
$(6.7)$--$(6.8)$. To lighten the notation, I shall not indicate the
$\sans s$ dependence of the various objects. I also identify
$\varphi^{-1}(n)$ and $\varphi^{-1}(n)H_{\rm cl}(n)^{-1}$.

Let $n\in{\cal N}$. For any $\Phi\in\varphi^{-1}(n)$, the tangent space
$T_{\Phi}\varphi^{-1}(n)$ is the subspace of $\ECF^{0,0{\rm r}}$ spanned
by the $H_{\rm cl}(n)$--hermitian elements
$\exp(-\Phi/2)\delta\exp\Phi\exp(-\Phi/2)$ and is
equipped with the Hilbert
structure ${1\over 2}\langle\cdot,\cdot\rangle_{h,H_{\rm cl}(n)}^{\rm r}$.
Hence,
one has $\Vert\exp(-\Phi/2)\delta\exp\Phi$ $\times$ $\exp(-\Phi/2)
\Vert_{h,H_{\rm cl}(n)|\Phi}{}^2
=\Vert\exp(-\Phi/2)\delta\exp\Phi\exp(-\Phi/2)
\Vert_{h,H_{\rm cl}(n)}{}^2$,
where in the right hand side the norm is that of $\ECF^{0,0}$
In this way, $\varphi^{-1}(n)$ becomes a real Hilbert manifold.

Consider now $\cal N$. For any $n\in{\cal N}$, $T_n{\cal N}$ is just
$\Bbb R^r$, where $r=\dim{\cal N}$, with the standard euclidean inner
product $\langle\cdot,\cdot\rangle$. So, for $\delta n\in T_n{\cal N}$,
$\Vert\delta n\Vert_{|n}{}^2=|\delta n|^2$.

${\cal N}\times\varphi^{-1}(\cdot)$ can be given the structure of Hilbert
manifold as follows. For any $(n,\Phi)\in{\cal N}\times\varphi^{-1}(\cdot)$,
$T_{(n,\Phi)}{\cal N}\times\varphi^{-1}(\cdot)=
T_n{\cal N}\oplus T_{\Phi}\varphi^{-1}(n)$. The tangent vectors are of the
form $\delta n\oplus\exp(-\Phi/2)\delta_n\exp\Phi\exp(-\Phi/2)$,
where the notation $\delta_n$ means variation at fixed $n$. The norm is given
by
$$\eqalignno{
&~\Vert\delta n\oplus\exp(-\Phi/2)\delta_n\exp\Phi\exp(-\Phi/2)
\Vert_{h,H_{\rm cl}(n)|(n,\Phi)}{}^2&\cr
=&~\Vert\delta n\Vert_{|n}{}^2
+\Vert\exp(-\Phi/2)\delta_n\exp\Phi\exp(-\Phi/2)
\Vert_{h,H_{\rm cl}(n)|\Phi}{}^2.
&(B.1)\cr}
$$

The jacobian $M(h;n)$ of the map $(6.6)$ relating the measures
on $\Herm$ and ${\cal N}\times\varphi^{-1}(\cdot)$ is defined by
$$\eqalignno{
&~(D\delta H(\Phi;n)H(\Phi;n)^{-1})_{h|H(\Phi;n)}&\cr
=&~M(h;n)\Big[(D\delta n)_{|n}
\otimes(D\exp(-\Phi/2)\delta_n\exp\Phi
\exp(-\Phi/2))_{h,H_{\rm cl}(n)|\Phi}\Big].
&(B.2)\cr}
$$
By explicit calculations one finds
$$
M(h;n)=[\det J(h;n)]^{1\over 2},
\eqno(B.3)
$$
where $J(h;n)$ is given by $(6.9)$.

{\it Proof}. The tangent map of the parametrization $(6.6)$ is given by
$$\eqalignno{
\delta H(\Phi;n)H(\Phi;n)^{-1}
=&~\exp(\ad\Phi/2)
\big[\exp(-\Phi/2)\delta'_n\exp\Phi\exp(-\Phi/2)&\cr
&~\phantom{\exp(\ad\Phi/2)\big[}
+\delta'H_{\rm cl}(n)H_{\rm cl}(n)^{-1}\big].
&(B.4)\cr}
$$
The two terms in the right hand side are the components of
$\delta H(\Phi;n)H(\Phi;n)^{-1}$ on $T_{H(\Phi;n)}\varphi^{-1}(n)$
and ${\cal H}_{H(\Phi;n)}(n)$, respectively.
The notation $\delta'$ is used instead of $\delta$
since the decomposition does not follow by a straightforward variation of the
relation $(6.6)$. Then, by the orthogonality in $T_{H(\Phi;n)}\Herm$
of the two terms in the right hand side of $(B.4)$, one has
$$\eqalignno{
&~\Vert\delta H(\Phi;n)H(\Phi;n)^{-1}\Vert_{h|H(\Phi;n)}{}^2&\cr
=&~\Vert\exp(-\Phi/2)\delta'_n\exp\Phi
\exp(-\Phi/2)\Vert_{h,H_{\rm cl}(n)|\Phi}{}^2
+\Vert\delta'H_{\rm cl}(n)H_{\rm cl}(n)^{-1}\Vert_{h|H_{\rm cl}(n)}{}^2.
&(B.5)\cr}
$$
Using the jacobian relation $(B.2)$, the normalization
condition of the measures and $(B.5)$, it is straightforward to obtain
$(B.3)$. {\it QED}

{}From $(B.2)$--$(B.3)$, $(6.7)$--$(6.8)$ follows readily.
\vskip.6cm
\centerline{\bf REFERENCES}
\def\NP#1{Nucl.~Phys.~{\bf #1}}
\def\PL#1{Phys.~Lett.~{\bf #1}}

\def\CMP#1{Commun.~Math.~Phys.~{\bf #1}}

\def\IJMP#1{Int.~J.~Mod.~Phys.~{\bf #1}}
\def\PREP#1{Phys.~Rep.~{\bf #1}}

\def\AP#1{Ann.~Phys.~{\bf #1}}
\def\CQG#1{Class.~Quantum~Grav.~{\bf #1}}
\vskip.4cm
\par\noindent

\item{\ref{1}}
P. Bouwknegt and K. Schoutens, \PREP{223} (1993) 183 and references
therein.

\item{\ref{2}}
E. Bergshoeff, A. Bilal and K. S. Stelle, \IJMP{A6} (1991) 4959 and
references therein.

\item{\ref{3}}
C. M. Hull, proc. of {\it Strings and Symmetries 1991}, Stony Brook,
May 1991, ed. N. Berkovits {\it et al.}, World Scientific
(1992) and references therein.

\item{\ref{4}}
K. Schoutens. A. Sevrin and P. van Nieuwenhuizen, \PL{B243} (1990) 245,
\PL{B251} (1990) 355, \NP{B349} (1991) 791, \NP{B364} (1991) 584,
\NP{B371} (1992) 315, proc. of {\it Strings and Symmetries 1991},
Stony Brook, May 1991, ed. N. Berkovits {\it et al.} World Scientific
(1992) and references therein.

\item{\ref{5}}
P. West, lecture at {\it Salamfest} (1993), hep--th/9309095
and references therein.

\item{\ref{6}}
P. Forg\'acs, A. Wipf, J. Balog, L. Feh\'er, and L. O'Raifeartaigh,
\PL{B227} (1989) 214.

\item{\ref{7}}
J. Balog, L. Feh\'er, P. Forg\'acs, L. O'Raifeartaigh and A. Wipf,
\PL{B244} (1990) 435 and \AP{203} (1990) 76.

\item{\ref{8}}
L. O'Raifertaigh and A. Wipf, \PL{B251} (1990) 361.

\item{\ref{9}}
L. O'Raifertaigh, P. Ruelle, I. Tsutsui and A. Wipf, \CMP{143} (1992) 333.

\item{\ref{10}}
L. Feh\'er, L. O'Raifertaigh, P. Ruelle, I. Tsutsui and A. Wipf,
\AP{213} (1992) 1. and \PREP{222} no. 1, (1992) 1.

\item{\ref{11}}
F. A. Bais, T. Tjin and P. van Driel, \NP{B357} (1991) 632.

\item{\ref{12}}
T. Tjin and P. van Driel, preprint ITFA--91--04.

\item{\ref{13}}
M. Bershadsky and H. Ooguri, \CMP{126} (1989) 49.

\item{\ref{14}}
R. Zucchini, \PL{B323} (1994) 322 and J. Geom. Phys. {\bf 16} (1995) 237.

\item{\ref{15}}
V. G. Drinfeld and V. V. Sokolov, J. Sov. Math. {\bf 30} (1985) 1975.

\item{\ref{16}}
A. M. Polyakov, \PL{B103} (1981) 207, \PL{B103} (1981) 211.

\item{\ref{17}}
D. Friedan, in {\it Recent Developments in Field Theory and Statistical
Mechanics}, North--Holland (1984).

\item{\ref{18}}
O. Alvarez, \NP{B216} (1983) 125.

\item{\ref{19}}
G. Moore and P. Nelson, \NP{B266} (1986) 58.

\item{\ref{20}}
K. Kodaira, {\it Complex Manifolds and Deformations of Complex Structures},
Grund\-lehr\-en Math. Wiss. 283, Springer--Verlag (1985).

\item{\ref{21}}
S. K. Donaldson, Proc. London Math. Soc. {\bf 50} (1985) 1 and
Duke Math. J. {\bf 54} (1987) 231.

\item{\ref{22}}
A. M. Polyakov, \IJMP{A5} (1990) 833.

\item{\ref{23}}
J. De Boer and J. Goeree, \PL{B274} (1992) 289, \NP{B401} (1993) 369.

\item{\ref{24}}
N. J. Hitchin, Topology {\bf 31} no. 3, (1992) 449.

\item{\ref{25}}
S. Govindarajan and T. Jayaraman, \PL{B345} (1995) 211.

\item{\ref{26}}
R. Gunning, {\it Lectures on Riemann surfaces}, Princeton University Press
(1966) and  {\it Lectures on Vector Bundles on Riemann surfaces}, Princeton
University Press (1967).

\item{\ref{27}}
R. O. Wells, {\it Differential Analysis on Complex Manifolds},
Graduate Texts in Mathematics, Springer Verlag (1980).

\item{\ref{28}}
S. Kobayashi, {\it Differential Geometry of Complex Vector Bundles}
Iwanami Shoten Publishers and Princeton University Press (1987).

\item{\ref{29}}
R. Zucchini, preprint DFUB 95--2, hep--th/9505056.

\item{\ref{30}}
M. Knecht, S. Lazzarini and R. Stora, \PL{B273} (1991) 63.

\item{\ref{31}}
R. Zucchini, \CQG{11} (1994) 1697.

\item{\ref{32}}
H. Kawai and R. Nakayama, \PL{B306} (1993) 224.

\item{\ref{33}}
S. Ichinose, N. Tsuda and T. Yukawa, preprint hep-th/9502101.

\item{\ref{34}}
K. K. Uhlenbeck and S. T. Yau, Comm. Pure and Appl. Math. {\bf 39-S}
(1986) 257

\item{\ref{35}}
C. T. Simpson, J. Am. Math. Soc. {\bf 1} no. 4, (1988) 867.

\item{\ref{36}}
E. Aldrovandi and G. Falqui, hep-th/9312093 to appear in J. Geom. Phys.
and hep-th/9411184.

\item{\ref{37}}
E. Aldrovandi and L. Bonora, J. Geom. Phys. {\bf 14} (1994) 65.

\item{\ref{38}}
A. Bilal, V. V. Fock and I. I. Kogan, \NP{B359} (1991) 635.

\item{\ref{39}}
R. Zucchini, \CQG{10} (1993) 253.
\bye